\documentclass[12pt]{article}
\usepackage{graphicx}
\usepackage{epsfig}
\usepackage{hyperref}
\def\lsim{\raise0.3ex\hbox{$<$\kern-0.75em\raise-1.1ex\hbox{$\sim$}}}
\def\gsim{\raise0.3ex\hbox{$>$\kern-0.75em\raise-1.1ex\hbox{$\sim$}}}

\def\mean#1{\left<#1\right>}
\def\Journal#1#2#3#4{{#1}{\bf #2} (#4) #3}
\def\IJMPE{{Int. J. Mod. Phys. E}}
\def\EPJC{{Eur. Phys. J. C}}
\def\JPG{{J. Phys. G}}
\def\JPCS{{J. Phys: Conf. Series\ }}

\def\NIMA{{Nucl. Instrum. Methods A}}
\def\NPA{{Nucl. Phys. A}}
\def\NPB{{Nucl. Phys. B}}
\def\PLB{{Phys. Lett. B}}
\def\PLC{Phys. Repts.\ }

\def\PRL{Phys. Rev. Lett.\ }
\def\PRD{{Phys. Rev. D}}
\def\PRC{{Phys. Rev. C}}
\def\PR{Phys. Rev.\ }

\def\ARNPS{{Ann. Rev. Nucl. Part. Sci.\ }} 
\def\RMP{Rev. Mod. Phys.\ }

\def\RPP{Rep. Prog. Phys.\ }
\normalsize
\begin{document}
\title{Results from PHENIX at RHIC\\ with Implications for LHC}
\author{M.~J.~Tannenbaum % etc
\thanks{Research supported by U.S. Department of Energy, DE-AC02-98CH10886.}
%\\ Brookhaven National Laboratory\\Upton, NY 11973-5000 USA}
\\ Physics Department, 510c,\\
Brookhaven National Laboratory,\\
Upton, NY 11973-5000, USA\\
mjt@bnl.gov}\maketitle
\begin{abstract}
This article is based on my Proceedings for the 47th Course of the International School of Subnuclear Physics on the Most Unexpected at LHC and the Status of High Energy Frontier, Erice, Sicily, Italy, 2009. Results from the PHENIX experiment at the Relativistic Heavy Ion Collider (RHIC) in nucleus-nucleus and proton-proton collisions at c.m. energy $\sqrt{s_{NN}}=200$ GeV are presented in the context of the methods of single and two-particle inclusive Êreactions which were used in the discovery of hard-scattering in p-p collisions at the CERN ISR in the 1970's. ÊThese techniques are used at RHIC in A+A collisions because of the huge combinatoric background from the large particle multiplicity. Topics include $J/\Psi$ suppression, jet quenching in the dense medium (sQGP) as observed with $\pi^0$ at large transverse momentum, thermal photons, collective flow, two-particle correlations, suppression of heavy quarks at large $p_T$ and its possible relation to Higgs searches at the LHC. The differences and similarities of the measurements in p-p and A+A collisions are presented. The two discussion sessions which followed the lectures on which this article is based are included at the end.
\end{abstract}
\maketitle
\thispagestyle{empty}
\tableofcontents
\section{Introduction}\label{sec:introduction}
High energy nucleus-nucleus collisions provide the means of creating nuclear matter in conditions of extreme temperature and density~\cite{BearMountain,seeMJTROP,MJTROP}.  
 The kinetic energy of the incident projectiles would be dissipated in the large 
volume of nuclear matter involved in the reaction.  At large energy or baryon density, a phase transition is expected from a state of nucleons containing confined quarks and gluons to a state of ``deconfined'' (from their individual nucleons) quarks and gluons, in chemical and thermal equilibrium, covering a volume that is many units of the confining length scale. This state of nuclear matter was originally given the name Quark Gluon Plasma(QGP)~\cite{Shuryak80}, a plasma being an ionized gas. However the results at RHIC~\cite{seeMJTROP} indicated that instead of behaving like a gas of free quarks and gluons, the matter created in heavy ion collisions at nucleon-nucleon c.m. energy $\sqrt{s_{NN}}=200$ GeV appears to be more like a {\em liquid}. This matter interacts much more strongly than originally expected, as elaborated in peer reviewed articles by the 4 RHIC experiments~\cite{BRWP,PHWP,STWP,PXWP}, which inspired the theorists~\cite{THWPS} to give it the new name ``sQGP" (strongly interacting QGP).  

	In the terminology of high energy physics, 
the QGP or sQGP is called a ``soft'' process, related to the QCD confinement scale 
\begin{equation}
\Lambda^{-1}_{\rm QCD} \simeq {\rm (0.2\ GeV)}^{-1} \simeq 1 \, 
\mbox{fm}\qquad .
\label{eq:LambdaQCD}
\end{equation}
   With increasing temperature, $T$, in analogy to increasing $Q^2$, the strong coupling constant $\alpha_{s}(T)$ becomes smaller, reducing the binding energy,  and the string tension, $\sigma(T)$, becomes smaller, increasing the confining radius, effectively screening the potential\cite{SatzRPP63}: 
  \begin{equation}
  V(r)=-{4\over 3}{\alpha_{s}\over r}+\sigma\,r \rightarrow 
-{4\over 3}{\alpha_{s}\over r} e^{-\mu\,r}+\sigma\,{{(1-e^{-\mu\,r})}\over \mu}
\label{eq:VrT}
\end{equation} 
where $\mu=\mu(T)=1/r_D$ is the Debye screening mass~\cite{SatzRPP63}. For $r< 1/\mu$ a quark feels the full color charge, but for $r>1/\mu$, the quark is free of the potential and the string tension, effectively deconfined. 

\begin{figure}[!thb]
\begin{center}
%\begin{minipage}{1.0\textwidth}
%\begin{center}
\includegraphics[width=0.6\textwidth]{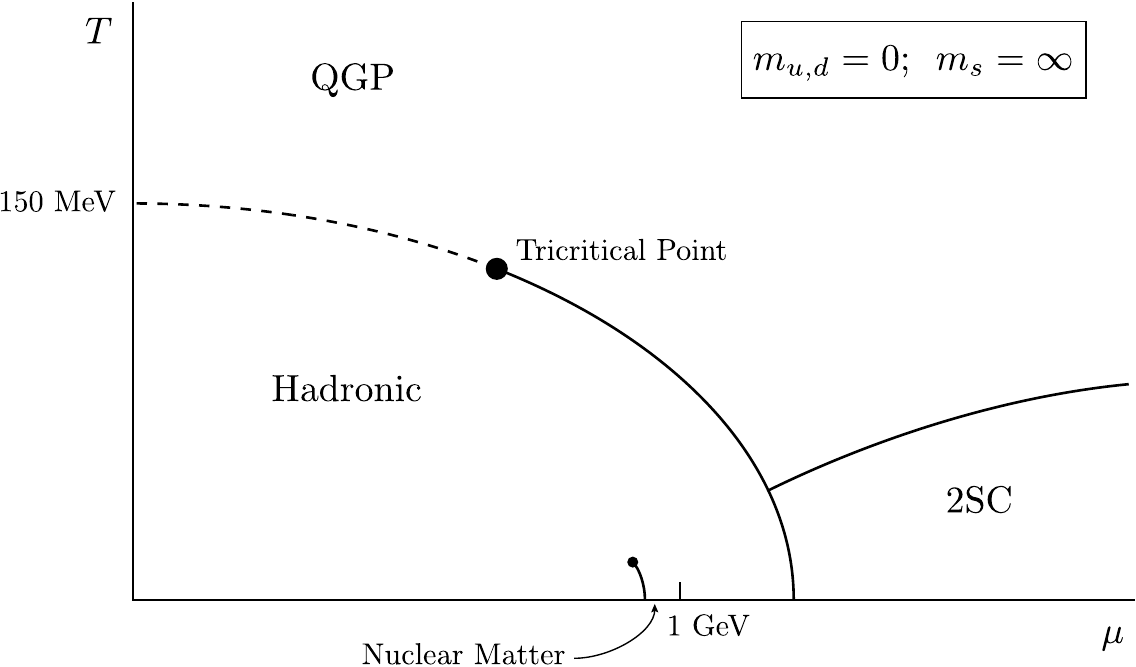}
\end{center}\vspace*{-1.0pc}
\caption[]{(left) A proposed phase diagram for nuclear matter~\cite{Krishna99}: Temperature,  $T$,  vs Baryon Chemical Potential, $\mu$. \label{fig:phaselat}}
%\end{minipage}
%\end{center}
\end{figure}

   There has been considerable work over the past 
three decades in making quantitative predictions for the QGP~\cite{seeMJTROP}. The predicted transition temperature from a state of hadrons to the QGP varies, from $T_c\sim 150$ MeV at zero baryon density, to zero temperature at a critical baryon density roughly 1 GeV/fm$^3$, 
$\sim$ 6.5 times the normal density of cold nuclear matter   
($\rho_0 = 0.14\,  {\rm nucleons}/ {\rm fm}^3$, $\mu_B\simeq 930$ MeV), 
where $\mu_B$ is the Baryon chemical potential. A typical expected phase diagram of nuclear matter~\cite{Krishna99} is shown in Fig.~\ref{fig:phaselat}. Not distinguished on Fig.~\ref{fig:phaselat} in the hadronic phase are the liquid self-bound ground state of nuclear matter and the gas of free nucleons~\cite{DAgostino05}. 

	   A nice feature of the search for the QGP is that it requires the integrated use of many disciplines in Physics: High Energy Particle Physics, Nuclear Physics, Relativistic Mechanics, Quantum Statistical Mechanics, and, recently, AdS/CFT string theory~\cite{Policastro,Nastase-BlackHole}. From the point of view of an experimentalist there are two major questions in this field. The first is how to relate the thermodynamical properties (temperature, energy density, entropy, viscosity ...) of the QGP or hot nuclear matter to properties that can be measured in the laboratory. The second question is how the QGP can be detected. 

    One of the
major challenges in this field is to find signatures that are unique to the QGP so that this new state of matter can be distinguished from the ``ordinary physics" of relativistic nuclear collisions.  Another more general challenge is to find effects which are specific to A+A collisions, such as collective or coherent phenomena, in distinction to cases for which  A+A collisions can be considered as merely an incoherent superposition of nucleon-nucleon collisions~\cite{specificity,Weiner05,Alexopoulos02}. 

\section{Issues in Relativistic Heavy Ion Physics}
\subsection{$J/\Psi$ suppression---the original ``gold-plated" QGP signature}

   Since 1986, the `gold-plated' signature of deconfinement was thought to be $J/\Psi$ suppression. Matsui and Satz~\cite{MatsuiSatz86} proposed that $J/\Psi$ production in A+A collisions will be suppressed by Debye screening of the quark
color charge in the QGP. The $J/\Psi$ is produced when two gluons
interact to produce a $c, \bar c$ pair which then resonates to form the
$J/\Psi$. In the plasma the $c, \bar c$ interaction is screened so that the 
$c, \bar c$ go their separate ways and eventually pick up other quarks at
the periphery to become {\it open charm}. ``Anomalous suppression'' of $J/\Psi$ was found in Pb+Pb collisions at the CERN SpS $\sqrt{s_{NN}}=17.2$ GeV~\cite{NA50EPJC39} (e.g. see Fig.~\ref{fig:JPsiAB} below) . This is the CERN fixed target heavy ion program's main claim to fame: but the situation is complicated because $J/\Psi$ are suppressed in p+A collisions~\cite{E772}. 

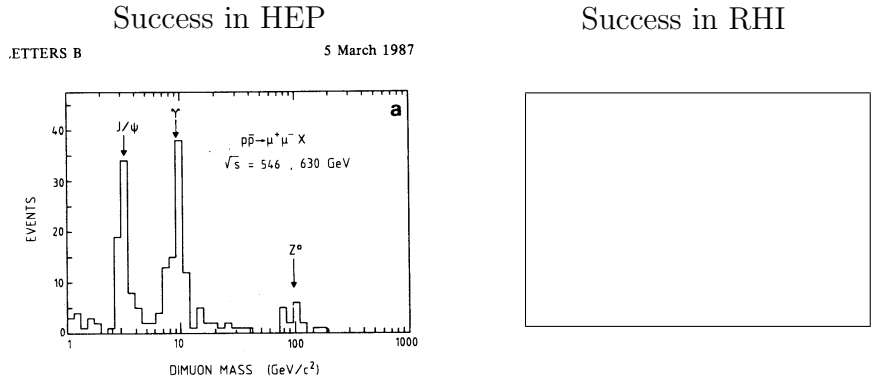
\begin{figure}[ht]
\begin{center}
\begin{minipage}[b]{2.4in}
\centerline{Success in HEP}
\centerline{\psfig{file=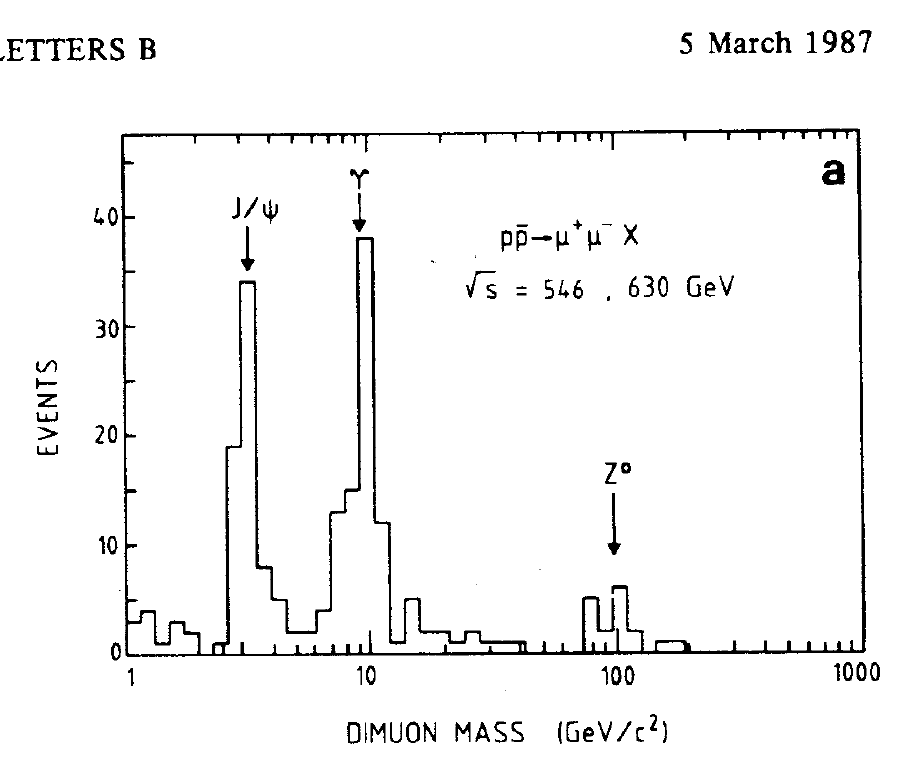,width=2.2in}}
\end{minipage}
\begin{minipage}[b]{2.4in}
\centerline{Success in RHI\hspace*{0.05in}}
%\centerline{\psfig{file=figs9/PLBsuccessr,width=2.2in}}
\vspace*{0.325in}
\hspace*{0.35in}\begin{picture}(130,88)
\put(0,0){\framebox(130,88){}}
\end{picture}
\vspace*{0.35in}
\end{minipage}
\end{center}\vspace*{-0.12in}
\caption[]{``The road to success": In High Energy Physics (left) a UA1 measurement\cite{UA1} of pairs of muons each with $p_T\geq 3$ GeV/c shows two Nobel prize winning dimuon peaks and one which won the Wolf prize. Success for measuring these peaks in RHI physics is shown schematically on the right. }
\label{fig:success}
\end{figure}

	The search for $J/\Psi$ suppression and thermal photon/dilepton radiation from the QGP drove the design of the RHIC experiments. My summary of the different views of dilepton resonances in the High Energy\cite{UA1} and Relativistic Heavy Ion\cite{MatsuiSatz86} Physics communities since the mid 1980's is shown in Fig.~\ref{fig:success}. 

\subsection{Detector issues in A+A compared to p-p collisions} 

\begin{figure}[!bht]
\begin{center}
\begin{tabular}{cc}
%\hspace*{-4cm}
\psfig{file=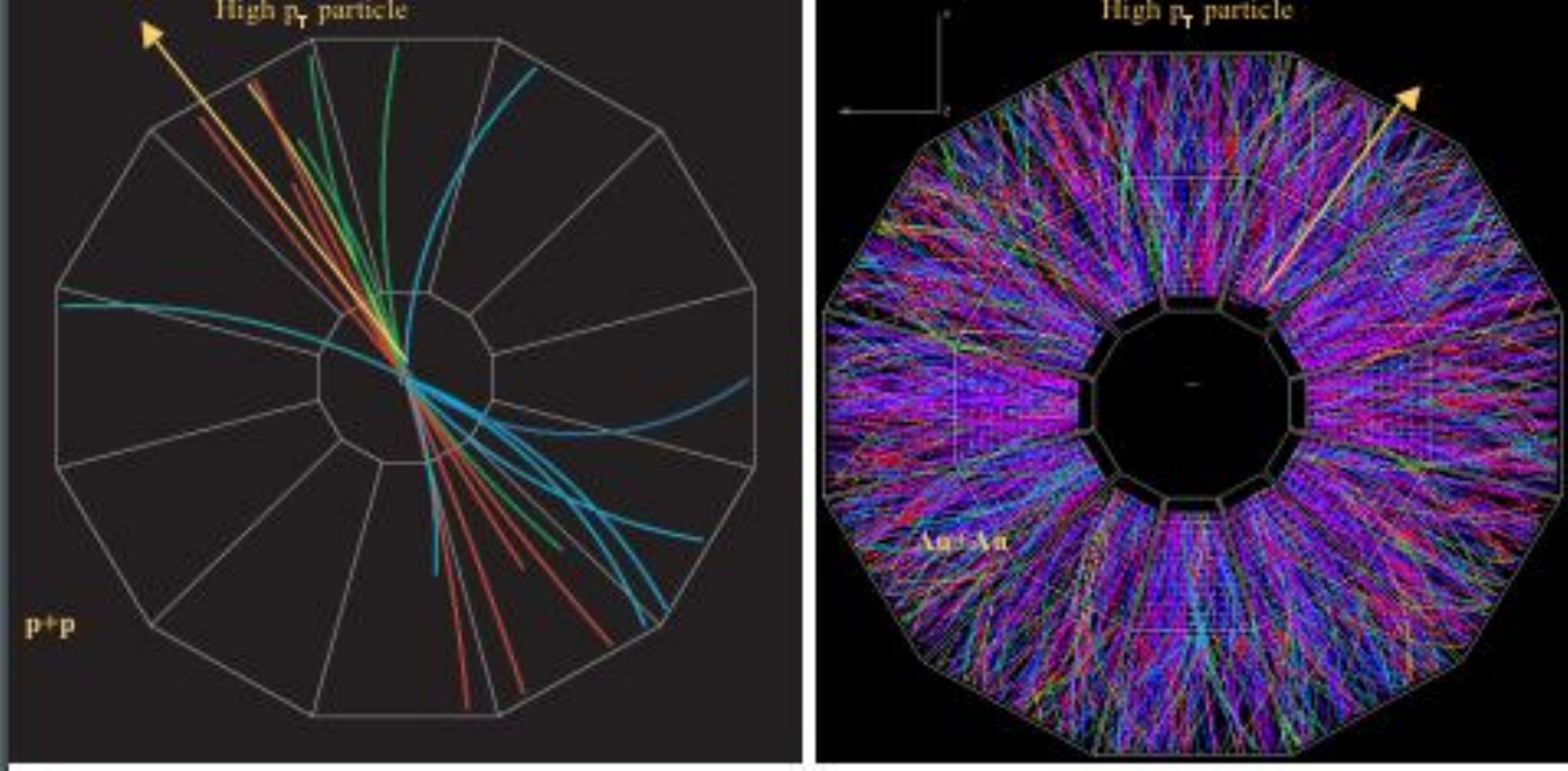,width=0.64\linewidth}&\hspace*{-0.025\linewidth}
\psfig{file=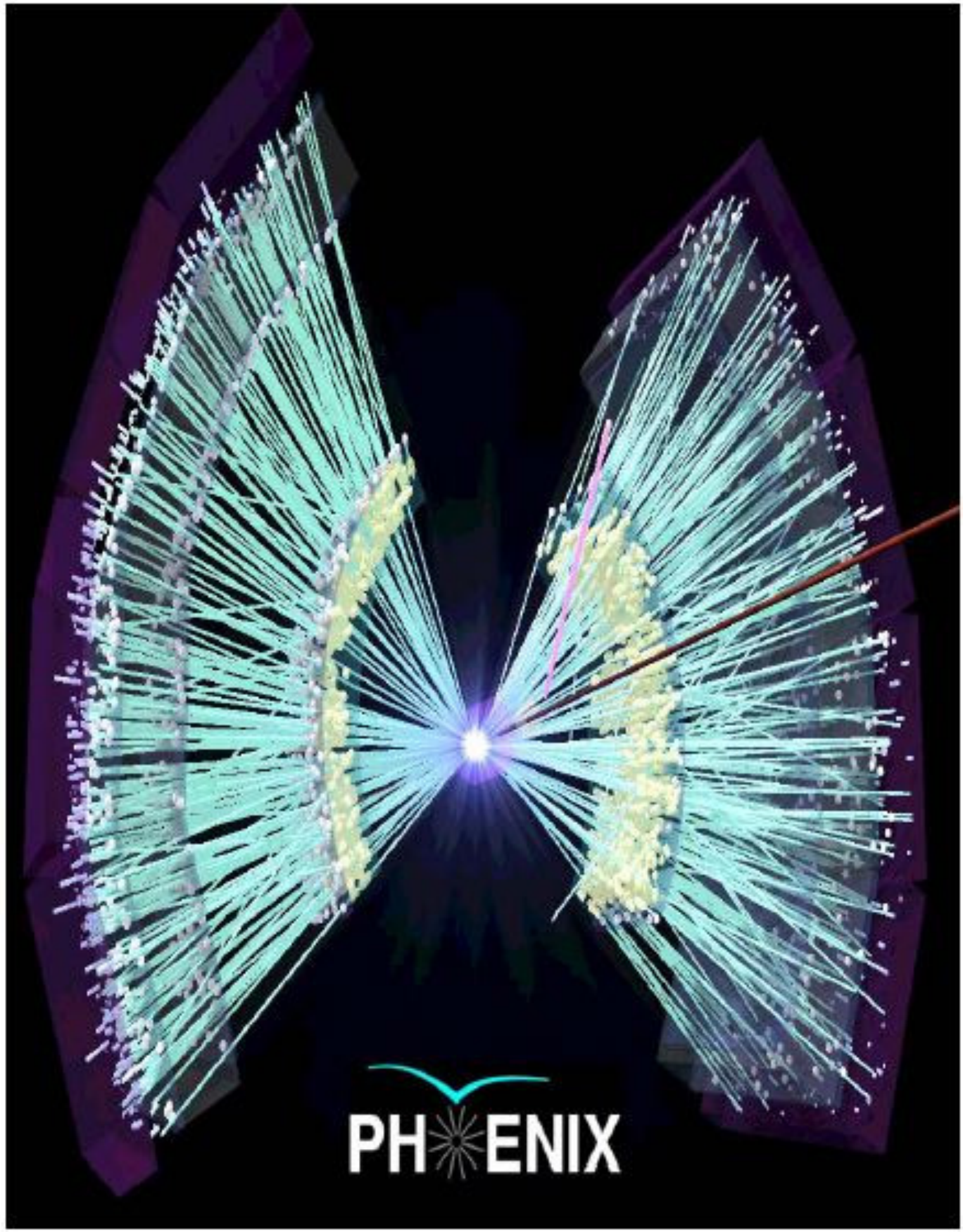,width=0.315\linewidth,height=0.315\linewidth}
\end{tabular}
\end{center}\vspace*{-0.15in}
\caption[]{a) (left) A p-p collision in the STAR detector viewed along the collision axis; b) (center) Au+Au central collision at $\sqrt{s_{NN}}=200$ GeV in the STAR detector;  c) (right) Au+Au central collision at $\sqrt{s_{NN}}=200$ GeV in the PHENIX detector.  
\label{fig:collstar}}

\end{figure}

 	Another main concern of experimental design in RHI collisions is the huge multiplicity in A+A central collisions compared to  p-p collisions. 
\begin{figure}[!hbt]
\begin{center}
\begin{tabular}{cc}
\psfig{file=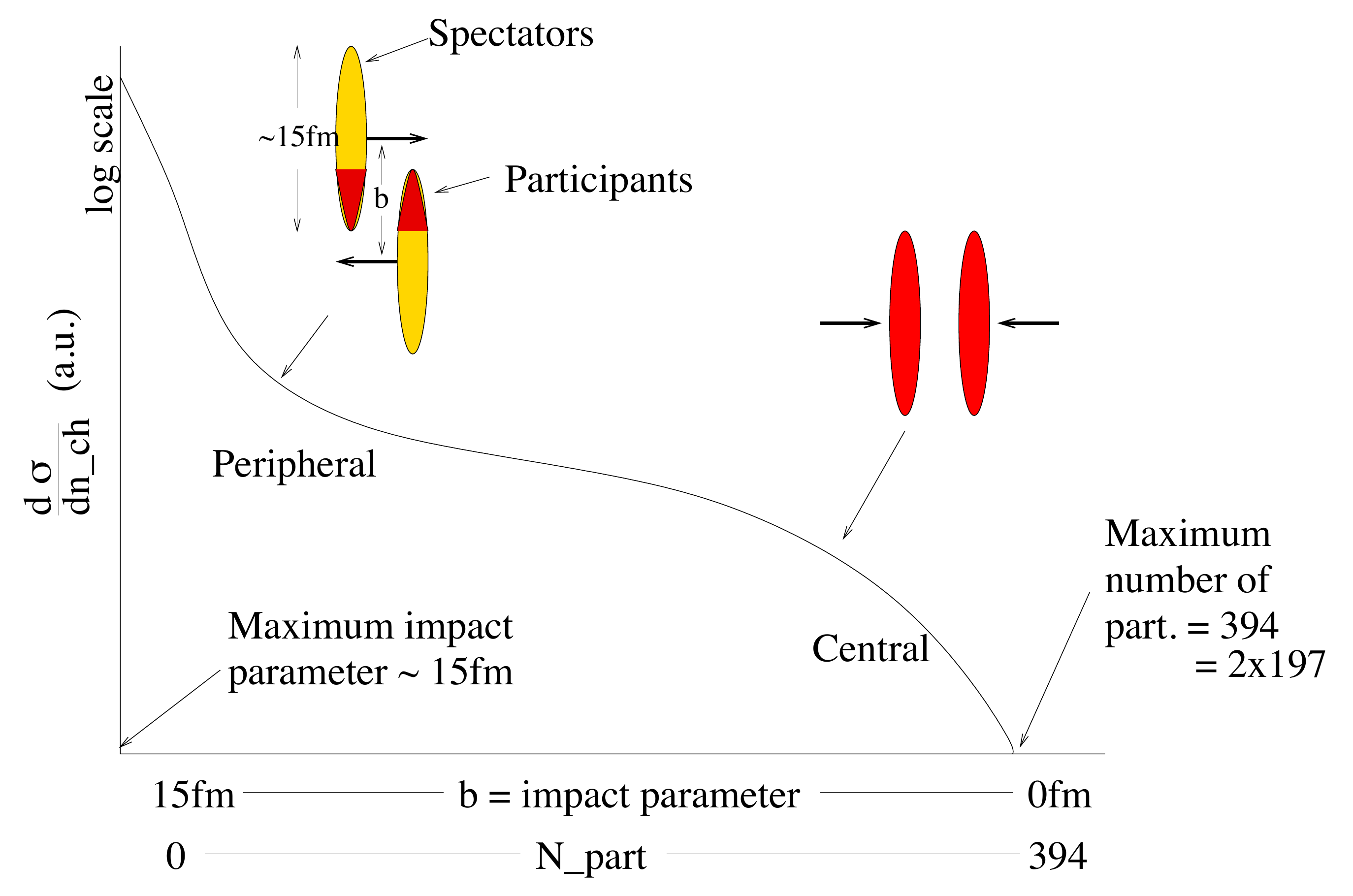,width=0.50\linewidth}
\psfig{file=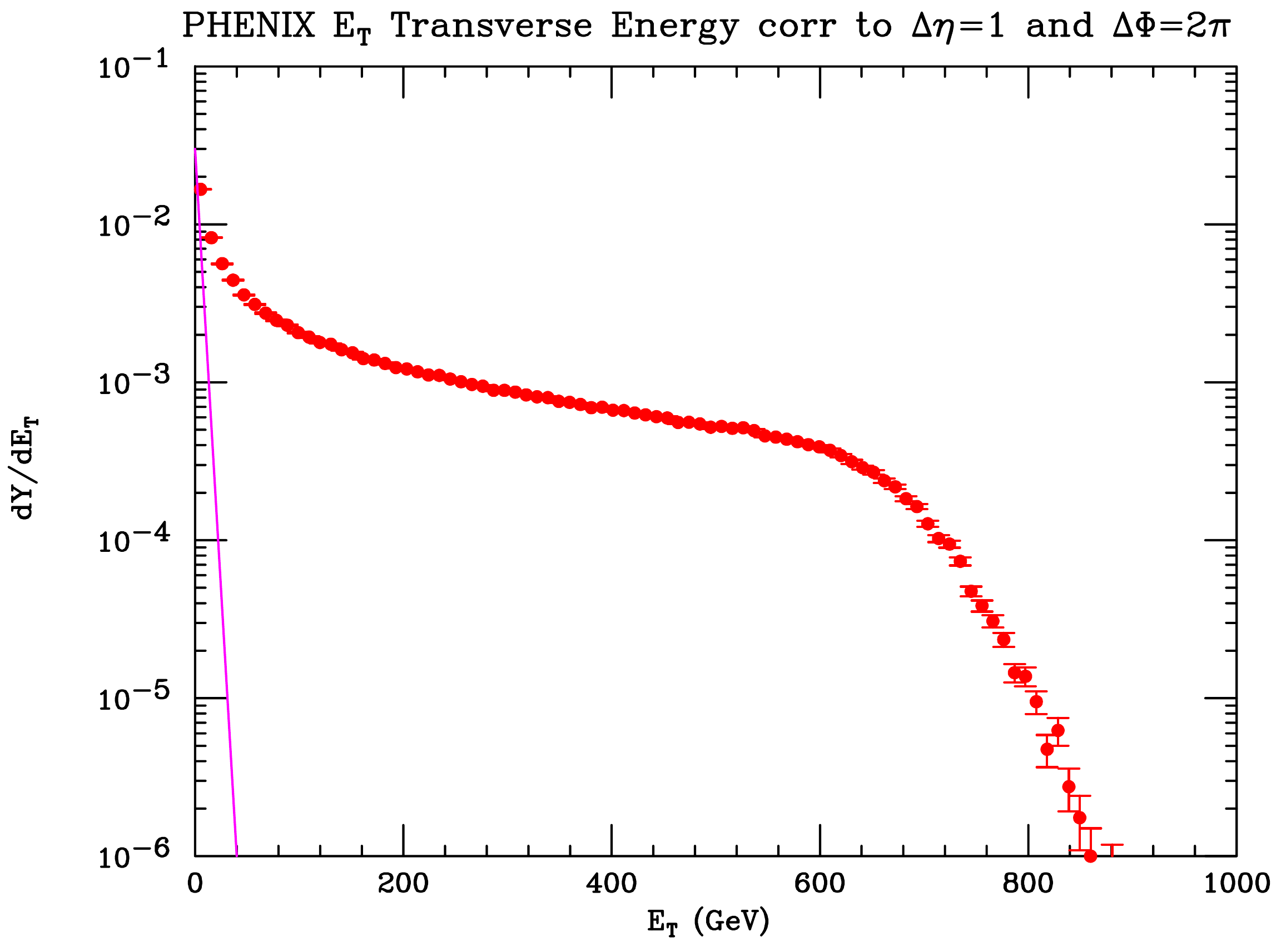,width=0.45\linewidth,angle=0}\end{tabular}
\end{center}\vspace*{-0.15in}
%-90-->0 height-->width
\caption[]{a) (left) Schematic of collision of two nuclei with radius $R$ and impact parameter $b$. The curve with the ordinate labeled $d\sigma/d n_{\rm ch}$ represents the relative probability of charged particle  multiplicity $n_{\rm ch}$ which is directly proportional to the number of participating nucleons, $N_{\rm part}$. b)(right) Transverse energy ($E_T$) distribution in Au+Au and p-p collisions at $\sqrt{s_{NN}}=200$ GeV from PHENIX~\cite{ppg019}.  
\label{fig:nuclcoll}}

\end{figure}
A schematic drawing of a collision of two relativistic Au nuclei is shown in Fig.~\ref{fig:nuclcoll}a. In the center of mass system of the nucleus-nucleus collision, the two Lorentz-contracted nuclei of radius $R$ approach each other with impact parameter $b$. In the region of overlap, the ``participating" nucleons interact with each other, while in the non-overlap region, the ``spectator" nucleons simply continue on their original trajectories and can be measured in Zero Degree Calorimeters (ZDC), so that the number of participants can be determined. The degree of overlap is called the centrality of the collision, with $b\sim 0$, being the most central and $b\sim 2R$, the most peripheral. The maximum time of overlap is $\tau_\circ=2R/\gamma\,c$ where $\gamma$ is the Lorentz factor and $c$ is the velocity of light. 
The energy of the inelastic collision is predominantly dissipated by multiple particle production, where $n_{\rm ch}$, the number of charged particles produced, is directly proportional~\cite{PXWP} to the number of participating nucleons ($N_{\rm part}$) as sketched on Fig.~\ref{fig:nuclcoll}a. Thus, $n_{\rm ch}$ or the total transverse energy $E_T$ in central Au+Au collisions is roughly $A$ times larger than in a p-p collision, as shown in the measured transverse energy spectrum in the PHENIX detector for Au+Au compared to p-p (Fig.~\ref{fig:nuclcoll}b) and in actual events from the STAR and PHENIX detectors at RHIC in Fig.~\ref{fig:collstar}.

	As it is a daunting task to reconstruct all the particles produced in such events, the initial detectors at RHIC~\cite{RHICNIM} concentrated on the measurement of single-particle or multi-particle inclusive variables to analyze RHI collisions, with inspiration from the CERN ISR which emphasized those techniques before the era of jet reconstruction. There are two major detectors in operation at RHIC, STAR and PHENIX, and there were also two smaller detectors, BRAHMS and PHOBOS, which have completed their program. As may be surmised from Fig.~\ref{fig:collstar}, STAR, which emphasizes hadron physics, is most like a conventional general purpose collider detector, a TPC to detect all charged particles over the full azimuth ($\Delta\phi=2\pi$) and  $\pm 1$ units of pseudo-rapidity ($\eta$), while PHENIX is a very high granularity high resolution special purpose detector covering a smaller solid angle at mid-rapidity, together with a muon-detector at forward rapidity~\cite{egseePT}. PHENIX is designed to measure and trigger on rare processes involving leptons, photons and identified hadrons at the highest luminosities with the special features: i) a minimum of material (0.4\% $X_\circ$) in the aperture to avoid photon conversions; ii) possibility of zero magnetic field on axis to prevent de-correlation of $e^+ e^-$ pairs from photon conversions; iii) Electro-Magnetic Calorimeter (EMCal) and Ring Imaging Cherenkov Counter (RICH) for $e^{\pm}$ identification and level-1 $e^{\pm}$ trigger; iv) a finely segmented EMCal ($\delta\eta$, $\delta\phi=0.01 \times$ 0.01) to avoid overlapping showers due to the high multiplicity and for separation of single-$\gamma$ and $\pi^0$ up to $p_T\sim 25$ GeV/c; v) EMCal and precison Time of Flight measurement for particle identification.

	In addition to the large multiplicity, there are two other issues in RHI physics which are different from p-p physics: i) space-time issues, both in momentum space and coordinate space---for instance what is the spatial extent of fragmentation? is there a formation time/distance?; ii) huge azimuthal anisotropies of particle production in non-central collisions (colloquially collective flow) which are interesting in their own right but can be troublesome. 
	\subsection{Collective Flow} 
   A distinguishing feature of A+A collisions compared to either p-p or p+A collisions is the collective flow observed. This effect is seen over the full range of energies studied in heavy ion collisions, from incident kinetic energy of $100A$ MeV to c.m. energy of $\sqrt{s_{NN}}=200$ GeV~\cite{LaceyQM05}. Collective flow, or simply flow, is a collective effect which can not be obtained from a superposition of independent N-N collisions.

   \begin{figure}[!thb]
   \begin{center}
\includegraphics[width=0.45\linewidth]{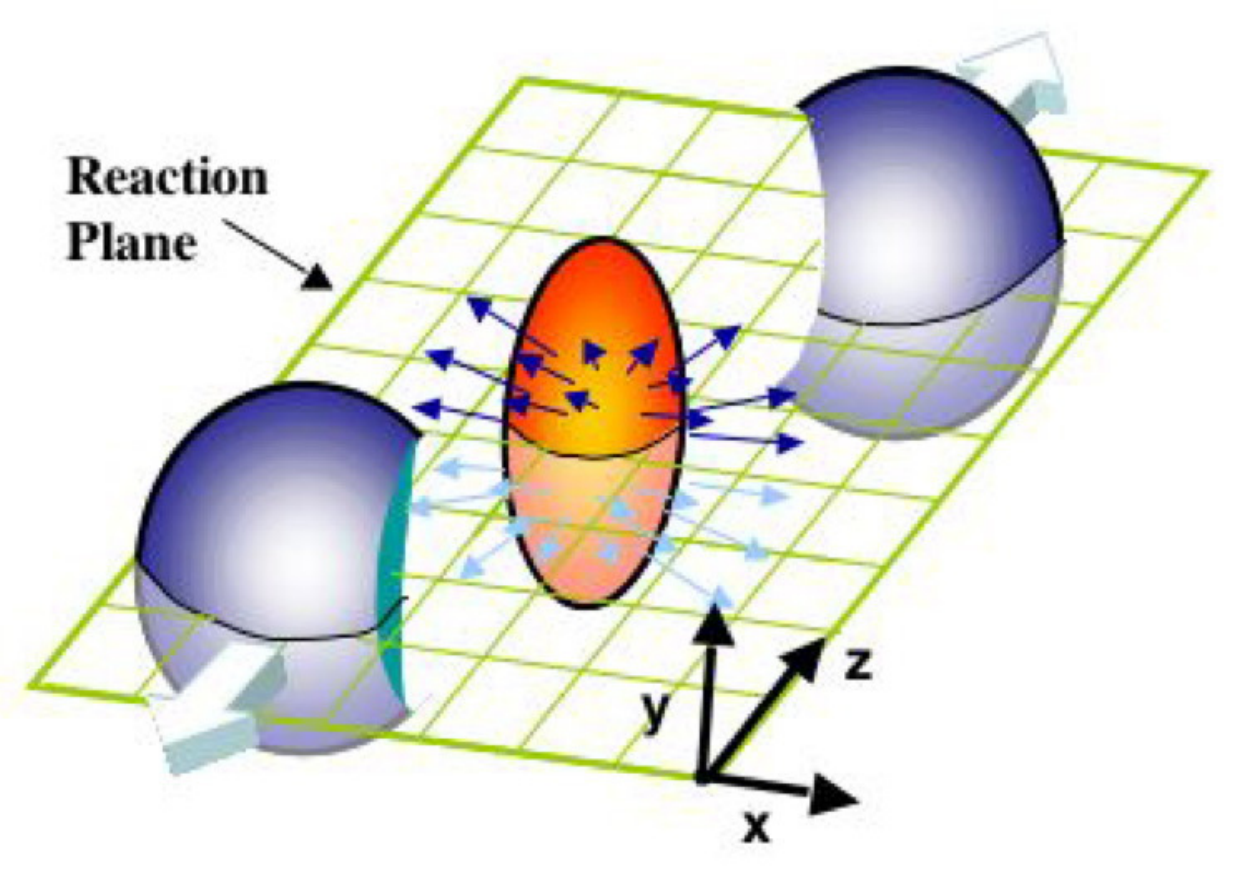}
\includegraphics[width=0.54\linewidth]{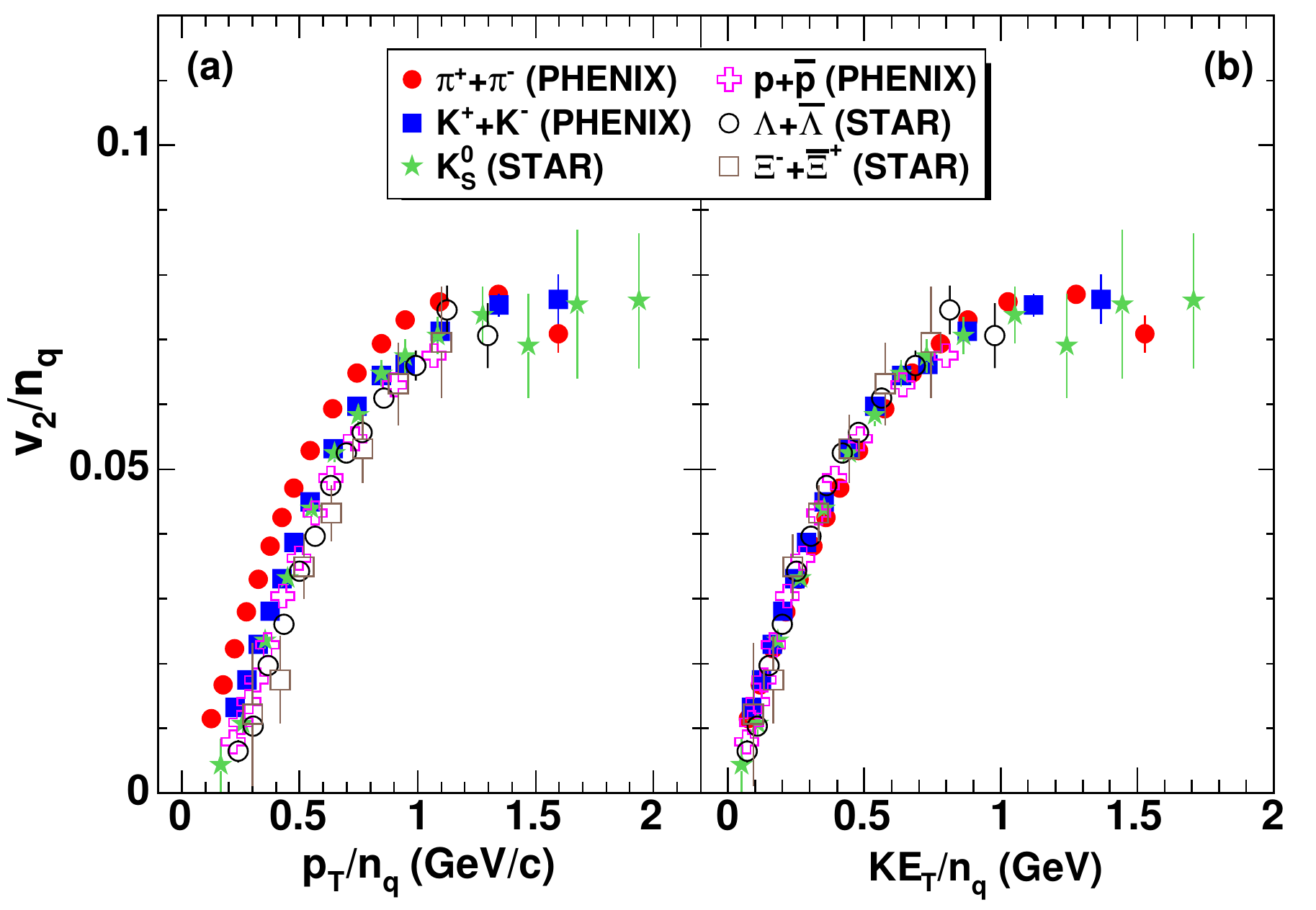}
\end{center}\vspace*{-0.25in}
\caption[]{(left) Almond shaped overlap zone generated just after an A+A collision where the incident nuclei are moving along the $\pm z$ axis. The reaction plane by definition contains the impact parameter vector (along the $x$ axis)~\cite{KanetaQM04} (see discussion session). (right) Measurements of elliptical-flow ($v_2$)  for identified hadrons plotted as $v_2$ divided by the number of constituent quarks $n_q$ in the hadron as a function of (a) $p_T/n_q$, (b) $KE_T/n_q$~\cite{PXArkadyQM06}.   
\label{fig:MasashiFlow}}
\end{figure}
   Immediately after an A+A collision, the overlap region defined by the nuclear geometry is almond shaped (see Fig~\ref{fig:MasashiFlow}) with the shortest axis along the impact parameter vector. Due to the reaction plane breaking the $\phi$ symmetry of the problem, the semi-inclusive single particle spectrum is modified by an expansion in harmonics~\cite{Ollitrault} of the azimuthal angle of the particle with respect to the reaction plane, $\phi-\Phi_R$~\cite{HeiselbergLevy}, where the angle of the reaction plane $\Phi_R$ is defined to be along the impact parameter vector, the $x$ axis in Fig.~\ref{fig:MasashiFlow}: 
  \begin{equation}
{Ed^3 N \over dp^3}={d^3 N\over p_T dp_T dy d\phi}
={d^3 N\over 2\pi\, p_T dp_T dy} \left[ 1+\sum_n 2 v_n \cos n(\phi-\Phi_R)\right] .
\label{eq:siginv2} 
\end{equation} 
The expansion parameter $v_2$, called elliptical flow, is predominant at mid-rapidity. In general, the fact that flow is observed in final state hadrons  shows that thermalization is rapid so that hydrodynamics comes into play before the spatial anisotropy of the overlap almond dissipates. At this early stage hadrons have not formed and it has been proposed that the constituent quarks flow~\cite{VoloshinQM02}, so that the flow should be proportional to the number of constituent quarks $n_q$, in which case $v_2/n_q$ as a function of $p_T/n_q$ would represent the constituent quark flow as a function of constituent quark transverse momentum and would be universal. However, in relativistic hydrodynamics, at mid-rapidity, the transverse kinetic energy, $m_T-m_0=(\gamma_T-1) m_0\equiv KE_T$, rather than $p_T$ is the relevant variable, and in fact $v_2/n_q$ as a function of $KE_T/n_q$ seems to exhibit nearly perfect scaling~\cite{PXArkadyQM06} (Fig.~\ref{fig:MasashiFlow}b). 

    The fact that the flow persists for $p_T>1$ GeV/c implies that the viscosity is small~\cite{TeaneyPRC68}, perhaps as small as a quantum viscosity bound from string theory~\cite{Kovtun05}, $\eta/s=1/(4\pi)$ where $\eta$ is the shear viscosity and $s$ the entropy density per unit volume.  This has led to the description of the ``sQGP'' produced at RHIC as ``the perfect fluid''~\cite{THWPS}. 
\subsection{Triangular flow, odd harmonics}
For the first 10 years of RHIC running, and dating back to the Bevalac, all the experts thought that the odd harmonics in Eq.~\ref{eq:siginv2} would vanish by the symmetry $\phi\rightarrow \phi+\pi$ of the almond shaped overlap region~\cite{AlverOllitrault} (Fig.~\ref{fig:MasashiFlow}). However, in 2010, an MIT graduate student and his Professor in experimental physics, seeking (at least since 2006) how to measure the fluctuations of $v_2$ in the PHOBOS experiment at RHIC,  realized that fluctuations in the collision geometry on an event-by-event basis, i.e. the distribution of participants from event-to-event, did not respect the average symmetry. This resulted in what they called ``participant triangularity'' and ``triangular flow'', or $v_3$ in Eq.~\ref{eq:siginv2}, which they measured using both PHOBOS and STAR data~\cite{AlverRoland}. A Brazilian group had shown in 2009 that $v_3$, does appear in an event-by-event hydrodynamics calculation without jets~\cite{BrazilNuXuv3}, but the MIT group~\cite{AlverRoland} was the first to show it with real data.  

\begin{figure}[!h]
\begin{center}
\includegraphics[width=0.75\linewidth]{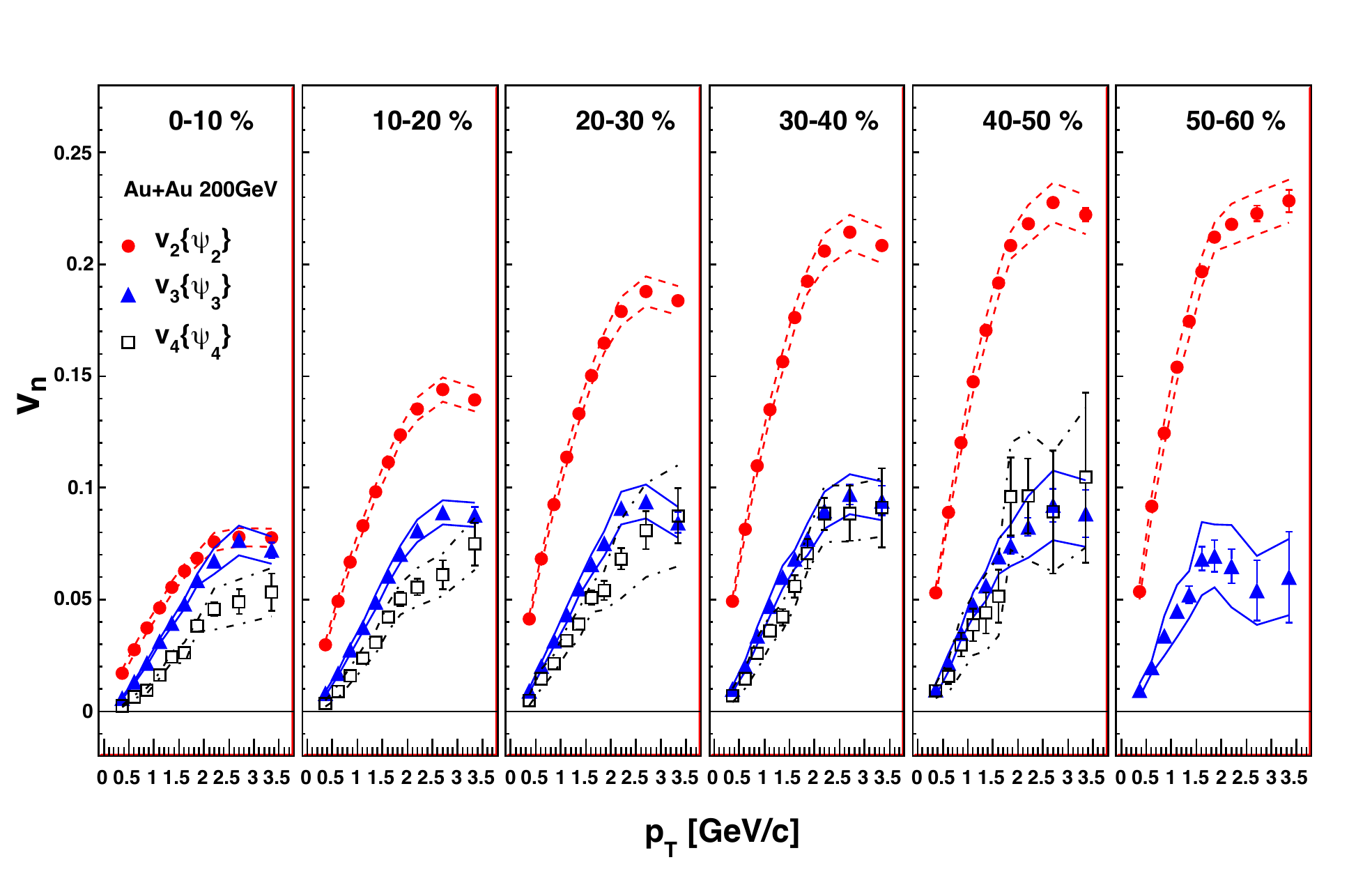}
\end{center}
%\vspace*{-0.28in}
\caption[]
{PHENIX~\cite{EsumiQM11} measurements of the $v_n$ parameters using Eq.~\ref{eq:siginv2} (with the appropriate reaction plane) as a function of $p_T$ for different centrality slices in $\sqrt{s_{NN}}=200$ GeV Au+Au collisions. \label{fig:PXv3} }
\end{figure}

Many experiments presented measurements of $v_3$ at Quark Matter 2011 this year, e.g. Fig.~\ref{fig:PXv3}~\cite{EsumiQM11},   and it was one of the most exciting results of this past year. 
There are two striking observations from Fig.~\ref{fig:PXv3} which indicate that fluctuations of the initial collision geometry are driving the observed $v_3$: i) the centrality dependence of $v_3(p_T)$ is weak as one would expect from fluctuations, but $v_2(p_T)$ which is most sensitive to the geometry of the ``almond''-shaped overlap region tracks the change in eccentricity with centrality; ii) for the most central collisions (0-10\%), where the overlap region is nearly circular so that all the $v_n$ are driven by fluctuations, $v_2(p_T)$, $v_3(p_T)$, $v_4(p_T)$ are comparable.  The fact that the observed collective flow of final state particles follows the fluctuations in the initial state geometry points to real hydrodynamic flow of a nearly perfect fluid (and convinces this author of the validity of hydrodynamics in RHI collisions, of which he had been quite skeptical). 
\section{Measurements in p-p collisions at RHIC}

     In addition to being the first heavy ion collider, RHIC is also the first polarized proton collider. Proton-proton collisions are performed with both beams either longitudinally or transversely polarized~\cite{RHICNIM,alsoMJT95}.  The bunch-by-bunch polarization is arranged so that the spin averaged cross section is obtained to high accuracy if polarization information is ignored.  The emphasis on precision EM calorimetry allows PHENIX to excel in the measurement of reactions producing photons, such as direct-single-photon production, or particles which decay to photons,  $\pi^0\rightarrow \gamma+\gamma$, $\eta\rightarrow \gamma+\gamma$, etc. 
     
\begin{figure}[!t]
\begin{center}
\includegraphics[width=0.48\linewidth]{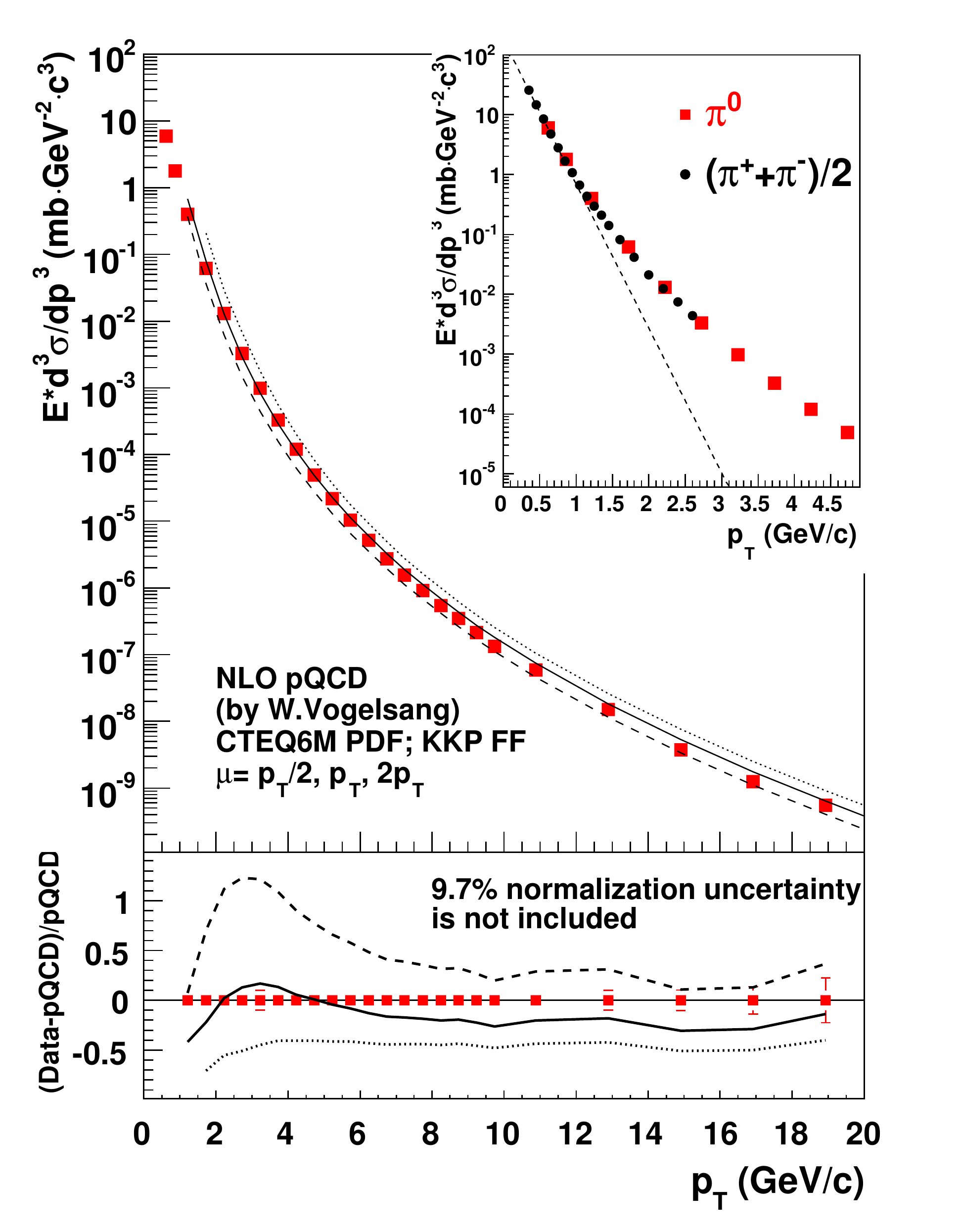}
\hspace*{-0.03\linewidth}\includegraphics[width=0.53\linewidth]{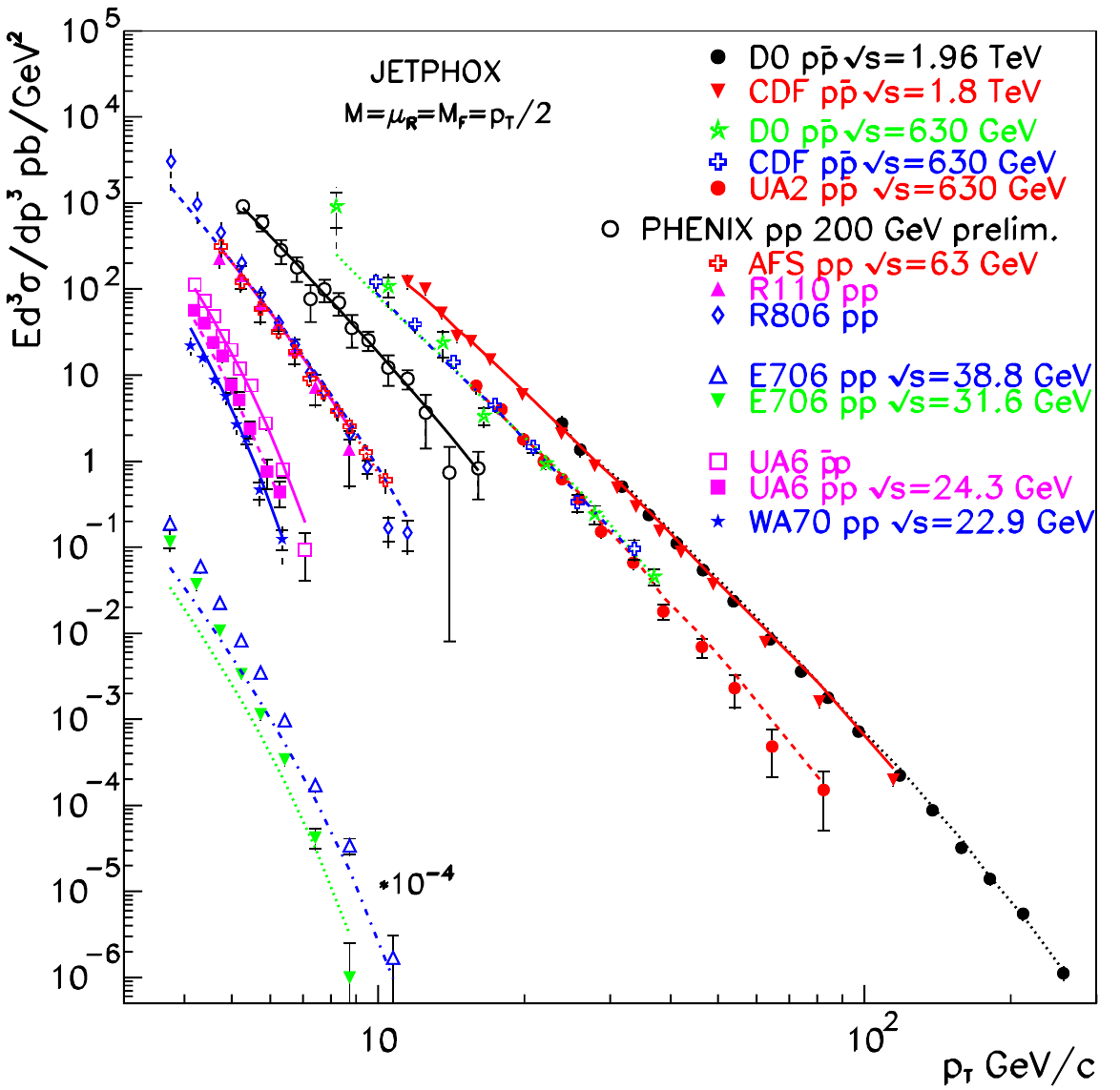}
\end{center}
\caption[]{a) (left) PHENIX measurement of invariant cross section of $\pi^0$ vs. $p_T$ at mid-rapidity in p-p collisons at $\sqrt{s}=200$ GeV.~\cite{PXpi0PRD}. b) (right) PHENIX measurement of inclusive direct-single $\gamma$ in p-p collisons at $\sqrt{s}=200$ GeV ($\circ$), together with all previous data compared to the theory.~\cite{Aurenche} }
\label{fig:pi0GamRHIC}
\end{figure}

     In order to understand whether an effect observed in A+A collisions exhibits a sensitivity to collective effects or to the presence of a medium such as the QGP it is important to establish a precise baseline measurement in p-p collisions at the same value of nucleon-nucleon c.m. energy $\sqrt{s_{NN}}$. PHENIX measurements of the invariant cross section, $E d^3\sigma/dp^3$, for $\pi^0$ and direct-single-$\gamma$ production in p-p collisions at $\sqrt{s}=200$ GeV are shown in Fig.~\ref{fig:pi0GamRHIC}a~\cite{PXpi0PRD} and Fig.~\ref{fig:pi0GamRHIC}b~\cite{Aurenche}, respectively.       
The inset on Fig.~\ref{fig:pi0GamRHIC}a shows that the $\pi^0$ cross section is exponential $\sim e^{-6p_T}$ for $p_T< 2$ GeV/c, as originally paramaterized by Cocconi~\cite{Cocconi1961,BBK}, which is the region of soft-multiparticle physics. For $p_T > 2$ GeV/c the spectrum is a power law which is indicative of the hard-scattering of the quark and gluon constituents of the proton. 
The excellent agreement of the measurements with theory is rewarding, although not surprising, since, after all, the discovery of $\pi^0$ production at large transverse momentum at the CERN-ISR proved that the partons of deeply inelastic scattering (DIS) interacted strongly with each other~\cite{BBK,CCR}.

\subsection{The influence of the CERN-ISR}

\begin{figure}[!h]
\begin{center}
\includegraphics[width=0.50\linewidth]{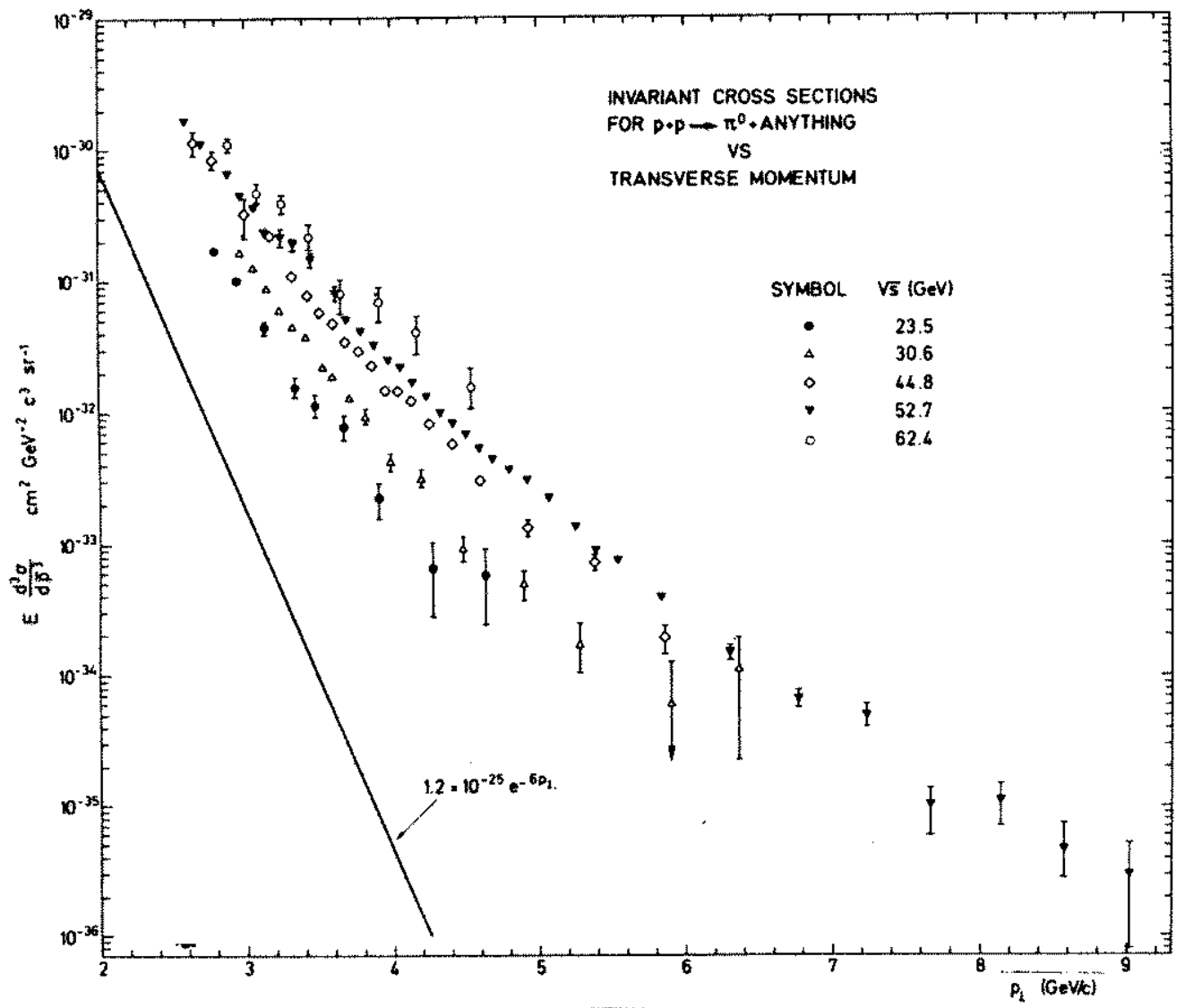}
\includegraphics[width=0.40\linewidth]{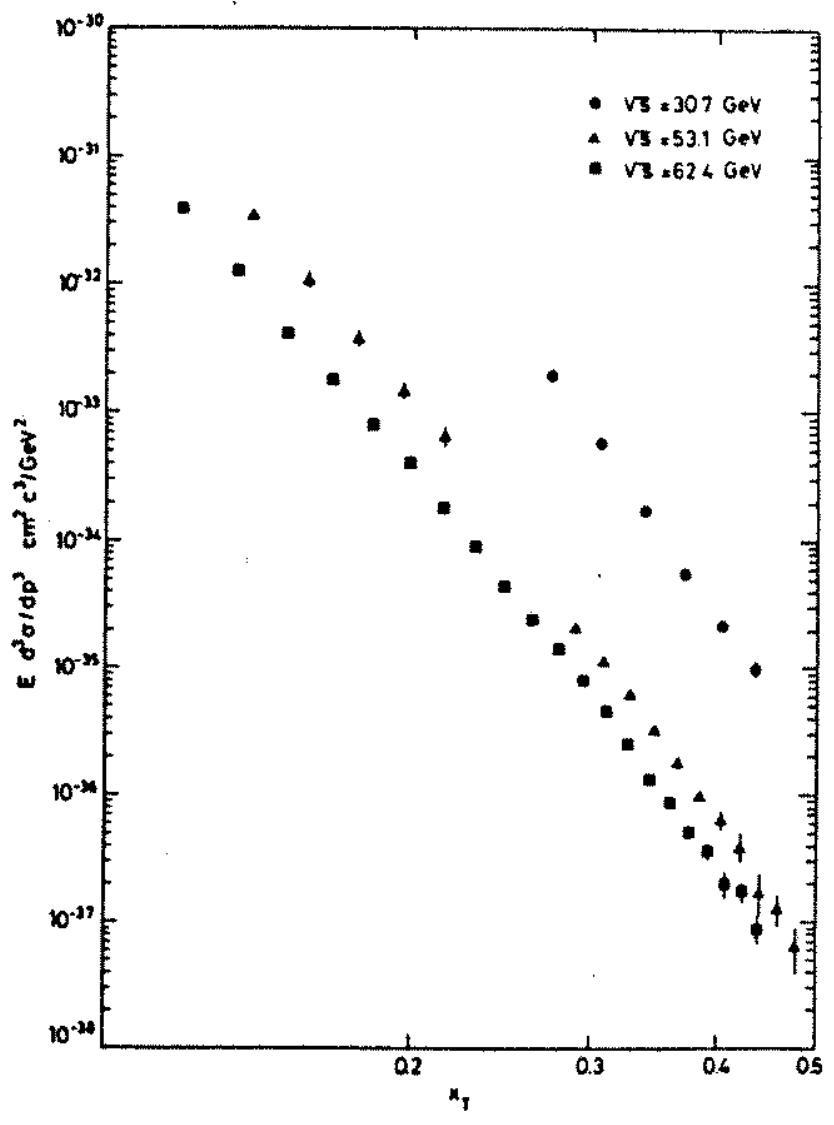}
\end{center}
\caption[]{a) (left) CCR~\cite{CCR} measurement of the invariant cross section of $\pi^0$ vs. $p_T$ at mid-rapidity in p-p collisons for 5 values of $\sqrt{s}$. b) (right) Later ISR measurement of invariant cross section of $\pi^0$ vs. $x_T=2p_T/\sqrt{s}$ at mid-rapidity in p-p collisons for 3 values of $\sqrt{s}$~\cite{CCOR} }
\label{fig:pizeroISR}
\end{figure}

 The ISR discovery~\cite{CCR} (Fig.~\ref{fig:pizeroISR}a) showed that the $e^{-6p_T}$ dependence at low $p_T$ breaks to a power law with characteristic $\sqrt{s}$ dependence for $p_T > 2$ GeV/c, which is more evident from the log-log plot of subsequent data~\cite{CCOR} (Fig.~\ref{fig:pizeroISR}b) as a function of $x_T=2p_T/\sqrt{s}$. 
This plot exhibits that the cross section for hard-processes obeys the scaling law:
\begin{equation}
E{{ d^3\sigma} \over {d^3p}}={1 \over p_T^{\,n_{\rm eff}} } F ({p_T \over \sqrt{s} })={1 \over \sqrt{s}^{\,n_{\rm eff}} } G({x_T} ) 
\end{equation}
where $n_{\rm eff}(x_T,\sqrt{s})\sim 4-6$ gives the form of the force-law between constituents as later predicted by Quantum Chromodynamics (QCD) with non-scaling structure and fragmentation functions and running coupling constant~\cite{BBKBBGCGKS}.  The more familiar equation for the constituent reaction $a+b\rightarrow c +d$ 
(e.g. $g+q\rightarrow g+q$) at parton-parton center-of-mass (c.m.) energy $\sqrt{\hat{s}}$ in ``leading logarithm'' pQCD~\cite{Owens} is:   
\begin{equation}
\frac{d^3\sigma}{dx_1 dx_2 d\cos\theta^*}=
\frac{s d^3\sigma}{d\hat{s} d\hat{y} d\cos\theta^*}=
\frac{1}{s}\sum_{ab} f_a(x_1) f_b(x_2) 
\frac{\pi\alpha_s^2(Q^2)}{2x_1 x_2} \Sigma^{ab}(\cos\theta^*)
\label{eq:QCDabscat}
\end{equation} 
where $f_a(x_1)$, $f_b(x_2)$, are parton distribution functions, 
the differential probabilities for partons
$a$ and $b$ to carry momentum fractions $x_1$ and $x_2$ of their respective 
protons (e.g. $u(x_2)$), and where $\theta^*$ is the scattering angle in the parton-parton c.m. system. 
The parton-parton c.m. energy squared is $\hat{s}=x_1 x_2 s$,
where $\sqrt{s}$ is the c.m. energy of the p-p collision. The parton-parton 
c.m. system moves with rapidity $\hat{y}=1/2 \ln (x_1/x_2)$ in the p-p c.m. system and the transverse momentum of a scattered parton is 
$p_T = p_T^* = { \sqrt{\hat{s}} \over 2 } \; \sin\theta^*$. 
Only the characteristic subprocess angular distributions,
{\bf $\Sigma^{ab}(\cos\theta^*)$} 
and the coupling constant,
$\alpha_s(Q^2)={12\pi}/({25\, \ln(Q^2/\Lambda^2))}$,
are fundamental predictions of QCD~\cite{CutlerSivers,Combridge:1977dm}. 

	Subsequent ISR measurements utilizing inclusive single or pairs of hadrons established that high $p_T$ particles in p-p collisions are produced from states with two roughly back-to-back jets which are  the result of scattering of constituents of the nucleons as described by Quantum Chromodynamics (QCD), which was developed during the course of those measurements. These techniques have been used extensively and further developed at RHIC since they are the only practical method to study hard-scattering and jet phenomena in Au+Au central collisions at RHIC energies. 

\begin{figure}[!t]
\begin{center}
%\begin{tabular}{cc}
\includegraphics[width=0.45\linewidth]{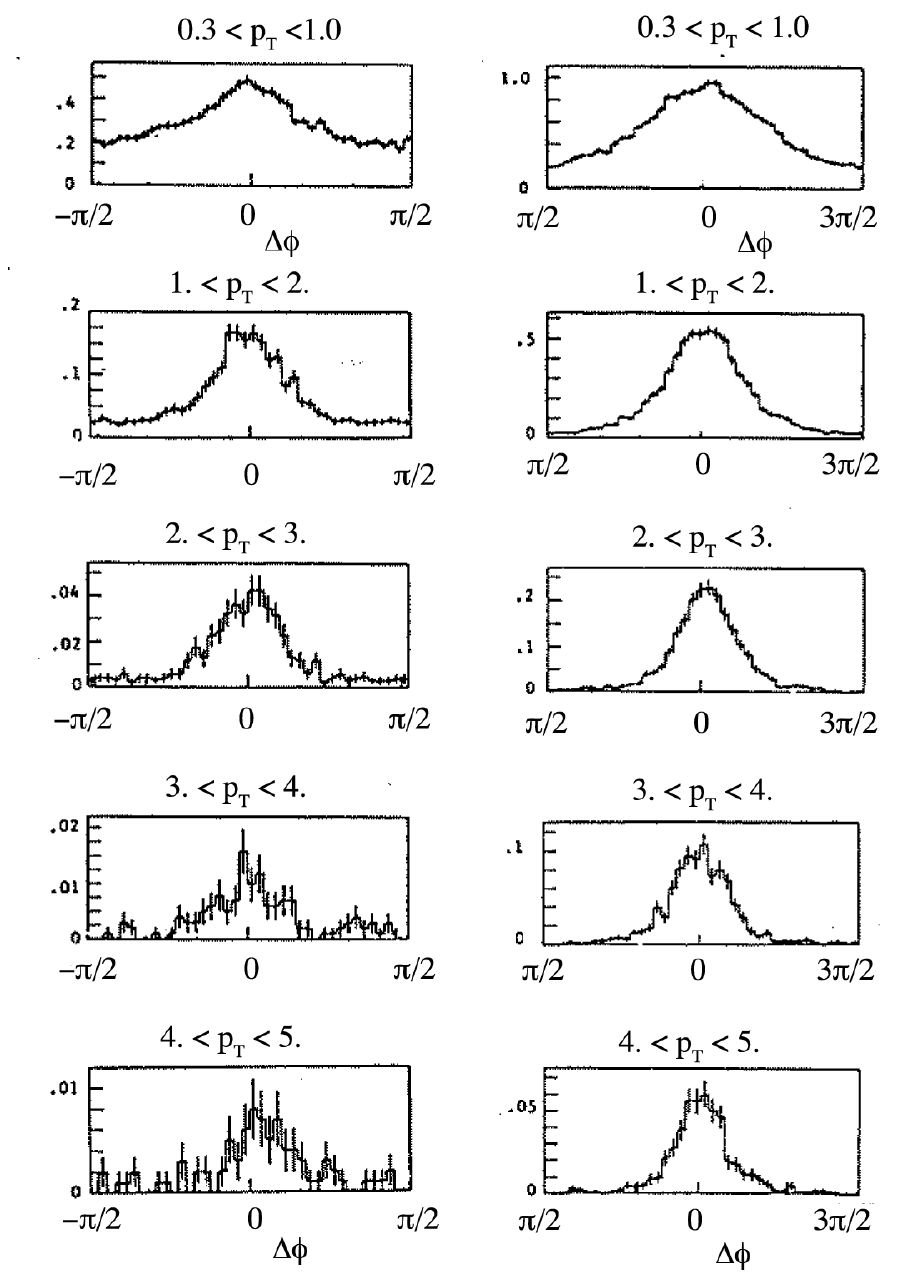} %&
\hspace*{0.25in}\includegraphics[width=0.45\linewidth]{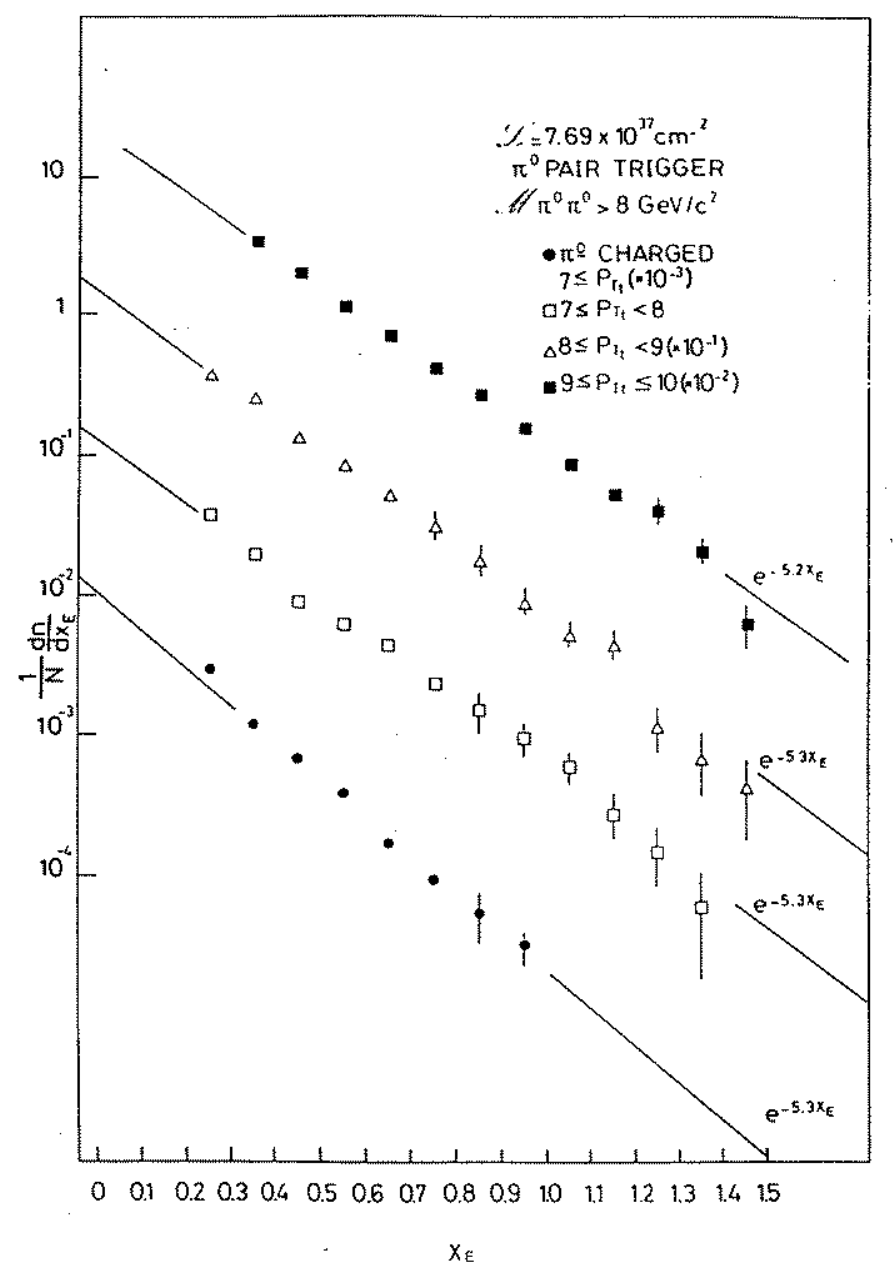}
%\end{tabular}
\end{center}
\vspace*{-2.0pc}
\caption[]
{CCOR~\cite{Angelis79,JacobEPS79} measurements at $\sqrt{s}=62.4$ GeV. a,b)Distributions of azimuthal angle ($\Delta \phi)$ of associated charged particles of transverse momentum $p_{T_a}$, with respect to a trigger $\pi^0$ with $p_{T_t}\geq 7$ GeV/c, for 5 intervals of $p_{T_{(a)}}$: a) (left-most panel) for $\Delta\phi=\pm \pi/2$ rad about the trigger particle, and b) (middle panel) for $\Delta\phi=\pm \pi/2$ about $\pi$ radians (i.e. directly opposite in azimuth) to the trigger. The trigger particle is restricted to $|\eta|<0.4$, while the associated charged particles are in the range $|\eta|\leq 0.7$. c) (right panel) $x_E$ distributions (see text) corresponding to the data of the center panel.   
\label{fig:mjt-ccorazi} }\vspace*{-1.0pc}
\end{figure}

The di-jet structure of events triggered by a high $p_T$ $\pi^0$, measured via two-particle correlations at the ISR, is shown in Fig~\ref{fig:mjt-ccorazi}~\cite{Angelis79,JacobEPS79}. 
The peaks on both the same side (Fig.~\ref{fig:mjt-ccorazi}a) as the trigger $\pi^0$ and opposite in azimuth (Fig.~\ref{fig:mjt-ccorazi}b) are due to the correlated charged particles from jets. 
The integrated (in $\Delta\phi$) yield of the away side-particles as a function of the variable $x_E\equiv -p_{T_a} \cos(\Delta\phi)/p_{T_t}\approx z_{a}/z_{t}$, where $z_t=p_{T_t}/\hat{p}_{T_t}$ is the fragmentation variable of the trigger jet (with $\hat{p}_{T_t}$) and $z_a=p_{T_a}/\hat{p}_{T_a}$ is the fragmentation variable of the away  jet (with $\hat{p}_{T_a}$),    
was thought in the ISR era to measure the fragmentation function of the away jet (Fig.~\ref{fig:mjt-ccorazi}c) but was found at RHIC to be sensitive, instead, to the ratio of the transverse momenta of the away-jet to the trigger jet, $\hat{x}_h\equiv \hat{p}_{T_a}/\hat{p}_{T_t}$~\cite{ppg029}. 

\begin{figure}[ht]
\begin{center}
\begin{tabular}{cc}
\hspace*{-0.1in}\includegraphics[width=0.75\linewidth]{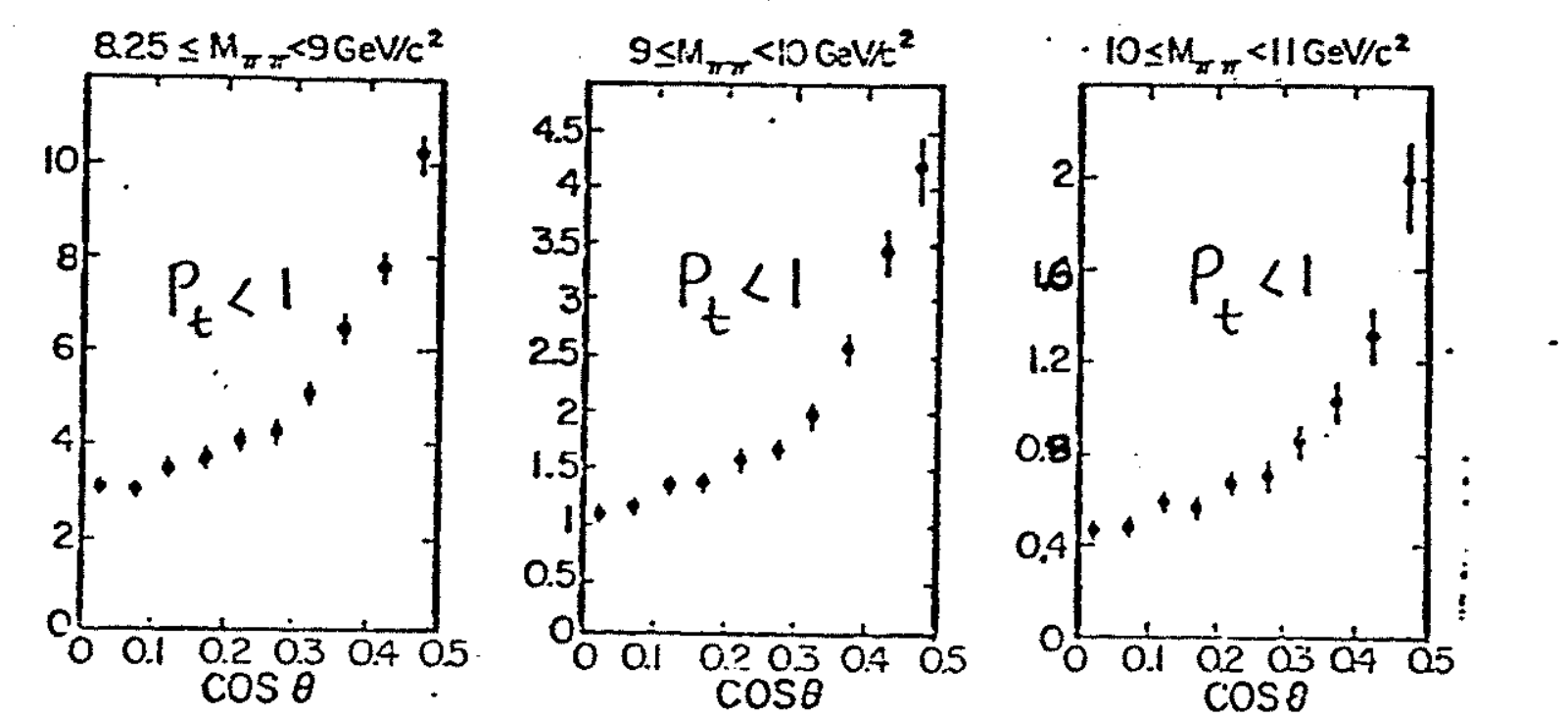} &
%\hspace*{-0.4in}\includegraphics[width=0.31\linewidth]{figs9/q_scat}
\hspace*{-0.35in}\includegraphics[width=0.288\linewidth]{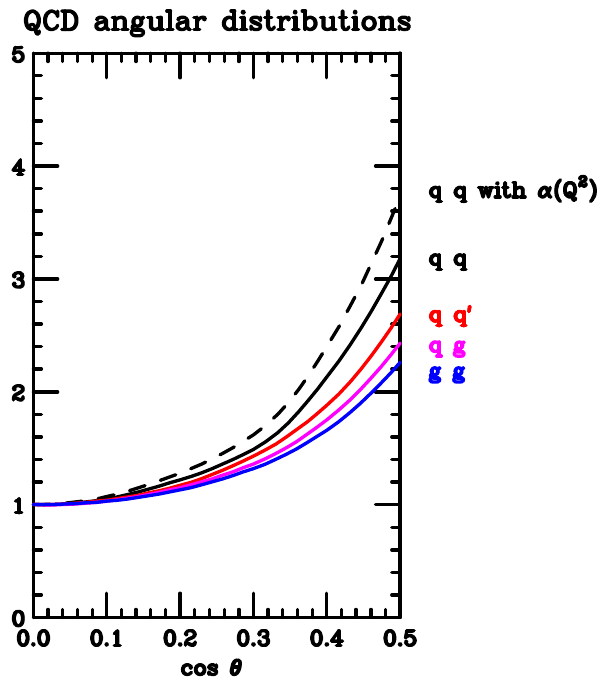}
\end{tabular}
\end{center}
%\vspace*{-10mm}
\caption[]
{a) (left 3 panels) CCOR measurement~\cite{Paris82,CCOR82NPB} of polar angular distributions of $\pi^0$ pairs with net $p_T < 1$ GeV/c at mid-rapidity in p-p collisions with $\sqrt{s}=62.4$ GeV for 3 different values of $\pi\pi$ invariant mass $M_{\pi \pi}$. b) (rightmost panel) QCD predictions for $\Sigma^{ab}(\cos\theta^*)$ for the elastic scattering of $gg$, $qg$, $qq'$, $qq$, and $qq$ with $\alpha_s(Q^2)$ evolution.    
\label{fig:mjt-ccorqq} }
\end{figure}

The QCD subprocess angular distribution {\bf $\Sigma^{ab}(\cos\theta^*)$} was also first measured with two-particle correlations of $\pi^0$ pairs of large invariant mass at the CERN-ISR~\cite{Paris82,CCOR82NPB} (Fig.~\ref{fig:mjt-ccorqq}), in agreement with QCD~\cite{CutlerSivers,Combridge:1977dm} at a fundamental level.   

\subsection{Other ISR discoveries important at RHIC}
Two other ISR discoveries, direct single-$\gamma$ production and direct-single $e^{\pm}$ production, and one near miss, $J/\Psi$ production, are important components of physics at RHIC. 

\begin{figure}[!h]
\begin{center}
\begin{tabular}{cc}
\includegraphics[width=0.44\linewidth,height=0.50\linewidth]{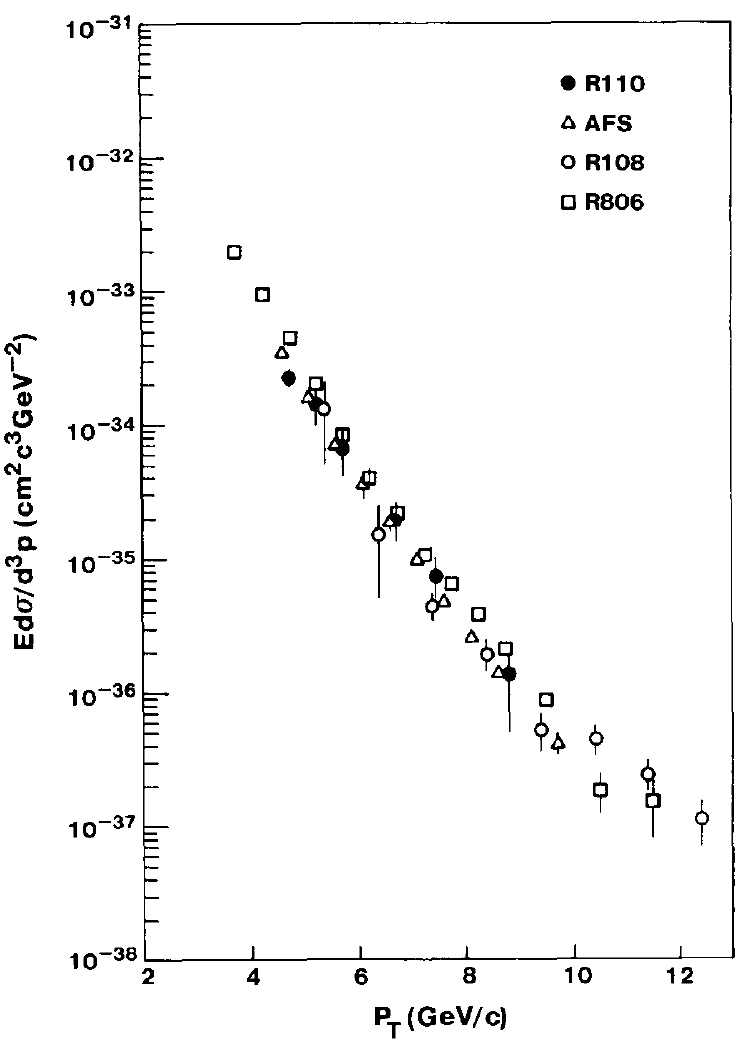}&\hspace*{-0.02\linewidth}  
\includegraphics[width=0.44\linewidth]{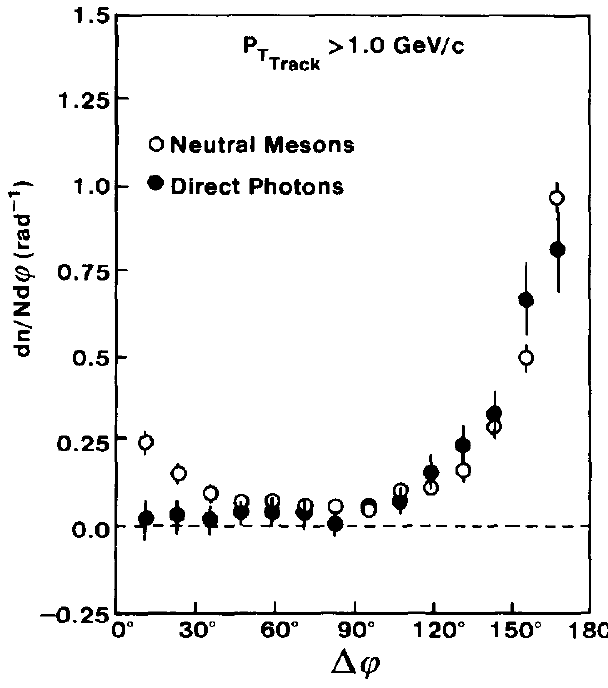} 
\end{tabular}
\end{center}
\caption[]{a)(left) Compilation of invariant cross sections of direct-$\gamma$ production at ISR~\cite{CMOR}; (right) azimuthal correlations of neutral mesons and direct-$\gamma$ with $h^{\pm}$~\cite{CMOR}. }
\label{fig:I4}
\end{figure}

Direct single-$\gamma$ production via the inverse QCD-compton process~\cite{QCDCompton} $g+q \rightarrow \gamma+q$ is an important probe in A+A collisions because the $\gamma$ is a direct participant in the reaction (at the constituent level), which emerges from the medium without interacting and can be measured precisely.   The cross sections for direct single-$\gamma$ production at $\sqrt{s}=62.4$ GeV~\cite{CMOR} are shown in Fig.~\ref{fig:I4}a. Two-particle azimuthal correlations of charged hadrons with neutral mesons ($\pi^0$), compared to direct-$\gamma$ (Fig.~\ref{fig:I4}b), show that direct-$\gamma$ are isolated, with no accompanying same-side particles, while $\pi^0$ have accompanying particles since they are fragments of jets from high $p_T$ partons.  

\begin{figure}[!h]
\vspace*{-1pc}
\begin{center}
\hspace*{-0.03\linewidth}\includegraphics[width=1.13\linewidth]{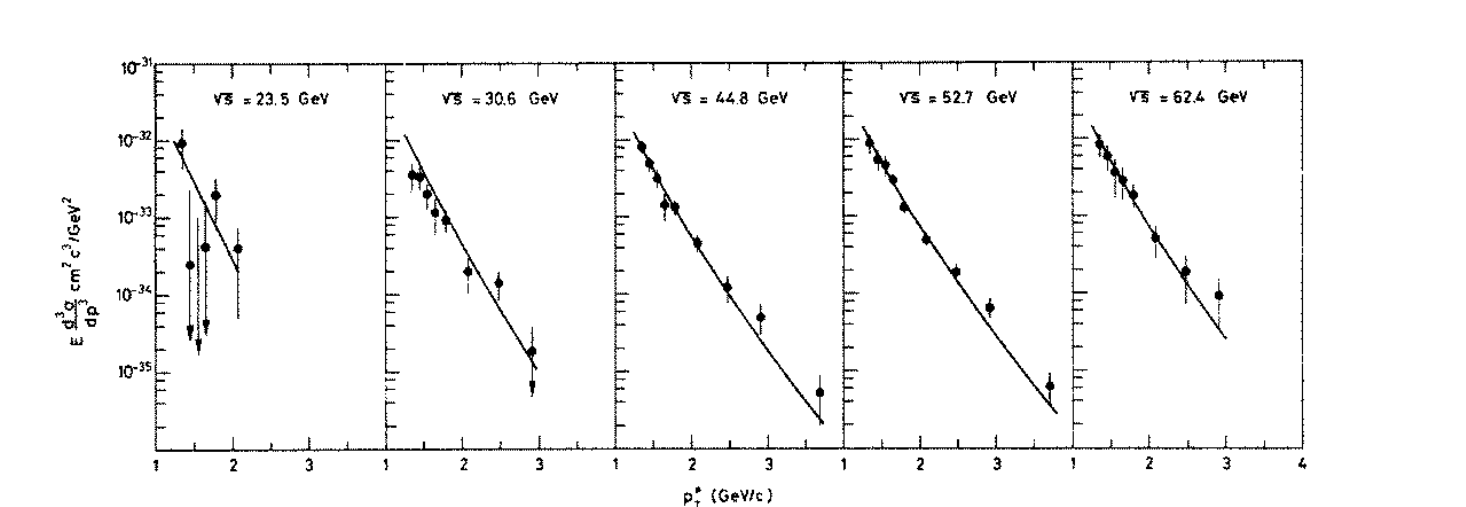}
\end{center}
\vspace*{-1.5pc}
\caption[]{Invariant cross sections at mid-rapidity: $(e^+ + e^-)/2$ (points); $10^{-4}\times (\pi^+ +\pi^-)/2$ (lines)~\cite{CCRS}.}
\label{fig:I2}
\end{figure}

Direct single-$e^{\pm}$ at a level of $e^{\pm}/\pi^{\pm}\approx 10^{-4}$ for all values of $\sqrt{s}$ at the CERN-ISR were discovered before either the $J/\Psi$ or open-charm~\cite{CCRS} (Fig.~\ref{fig:I2}). After the discovery of the $J/\Psi$ in 1974,
\begin{figure}[!ht]
\begin{center}
\begin{tabular}{ccc}
\hspace*{-0.02\linewidth}\includegraphics[width=0.31\linewidth]{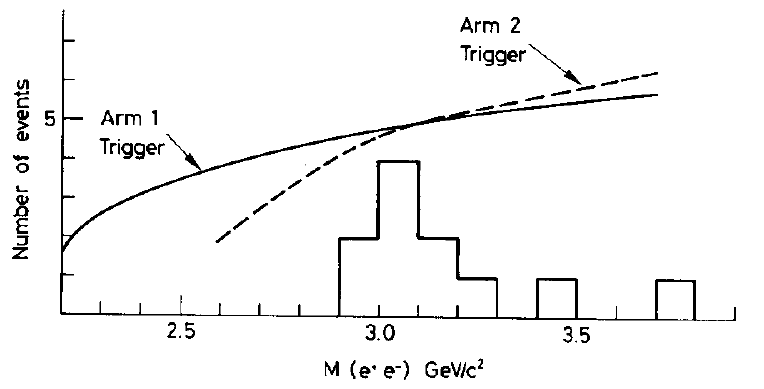}&
\hspace*{-0.04\linewidth}\includegraphics[width=0.33\linewidth]{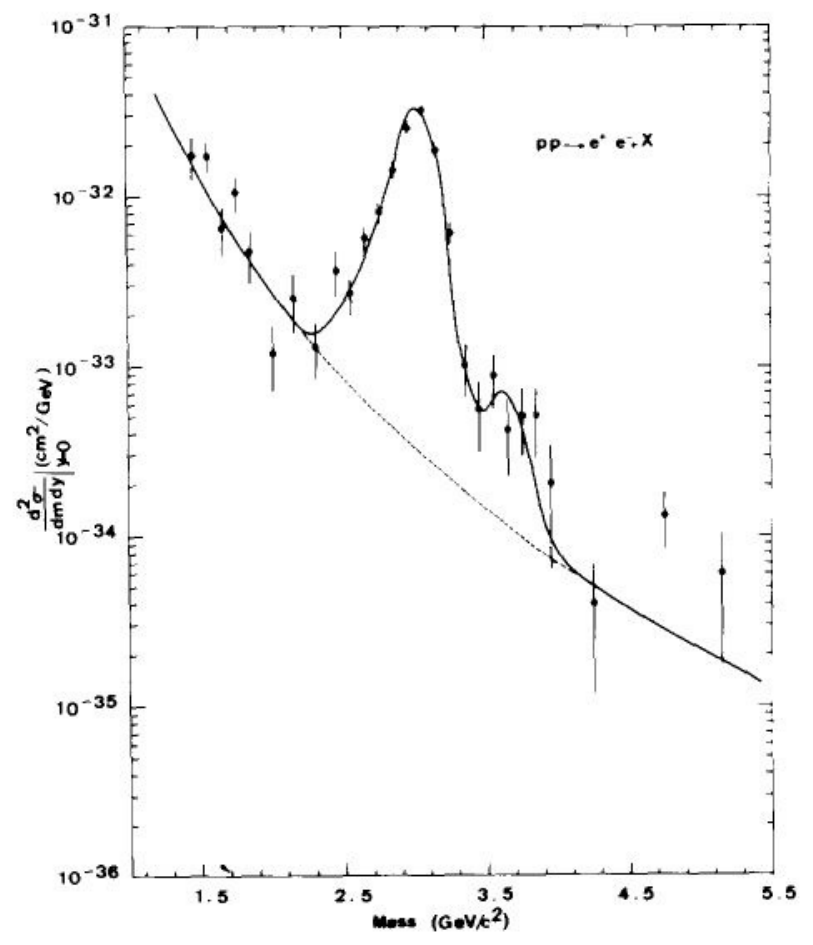}&
\hspace*{-0.02\linewidth}\includegraphics[width=0.33\linewidth]{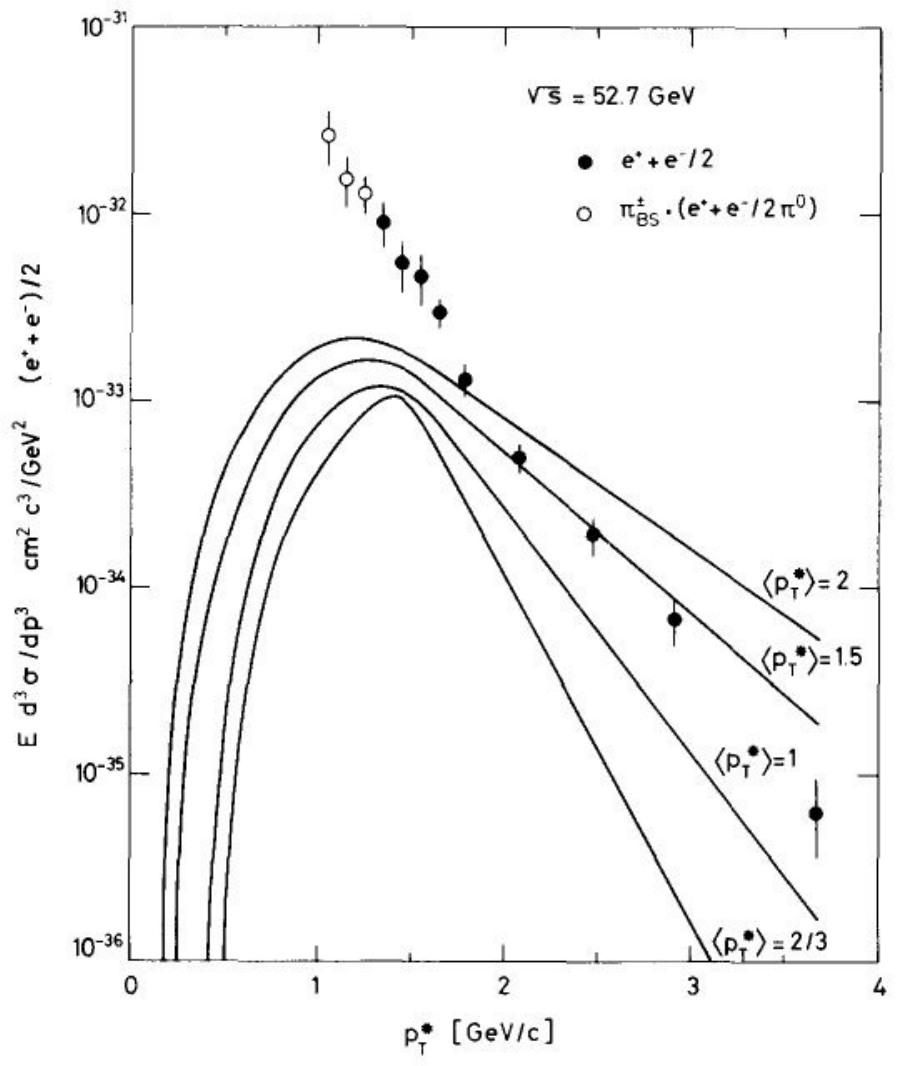} 
\end{tabular}
\end{center}
\caption[]{a)(left) First $J/\Psi$ at ISR~\cite{CCRSPLB56}; b) (center) Best $d\sigma_{ee}/dm_{ee} dy|_{y=0}$~\cite{Clark}; c)(right) direct-$e^{\pm}$ data at $\sqrt{s}=52.7$ GeV (Fig.~\ref{fig:I2}) with calculated $e^{\pm}$ spectrum for $J/\Psi$ for several values of $\mean{p_T}$~\cite{CCRS}. }
\label{fig:I3}
\end{figure}
it was demonstrated that the $J/\Psi$ was not the source of the single-$e^{\pm}$ (Fig.~\ref{fig:I3})  and two years later, when open charm was discovered, it was shown that the direct $e^{\pm}$ were due to the semi-leptonic decay of charm mesons~\cite{HLLS}. Fig.~\ref{fig:I3}a~\cite{CCRSPLB56} shows the first $J/\Psi$ at the ISR~\cite{CCRSPLB56}, Fig.~\ref{fig:I3}b shows the best $J/\Psi$ measurement at the ISR~\cite{Clark} while Fig.~\ref{fig:I3}c~\cite{CCRS} shows that the direct electrons (Fig.~\ref{fig:I2}) are not the result of $J/\Psi$ decay since $\mean{ p_T}=1.1\pm 0.05$ GeV/c~\cite{Clark}.  

%%test
\pagebreak
\section{From ISR p-p to RHIC A+A physics}

   Since hard-scattering at high $p_T >2$ GeV/c is point-like, with distance scale $1/p_T < 0.1$ fm, the cross section in p+A (A+A) collisions, compared to p-p, should be larger by the relative number of possible point-like encounters, a factor of $A$ ($A^2$) for p+A (A+A) minimum bias collisions. When the impact parameter or centrality of the collision is defined, the proportionality factor becomes $\mean{T_{AA}}$, the average overlap integral of the nuclear thickness functions. 
 
\subsection{Jet quenching from inclusive $\pi^0$ production}   
\begin{figure}[!h]
\begin{center}
%\begin{tabular}{cc}
\includegraphics[width=0.45\linewidth]{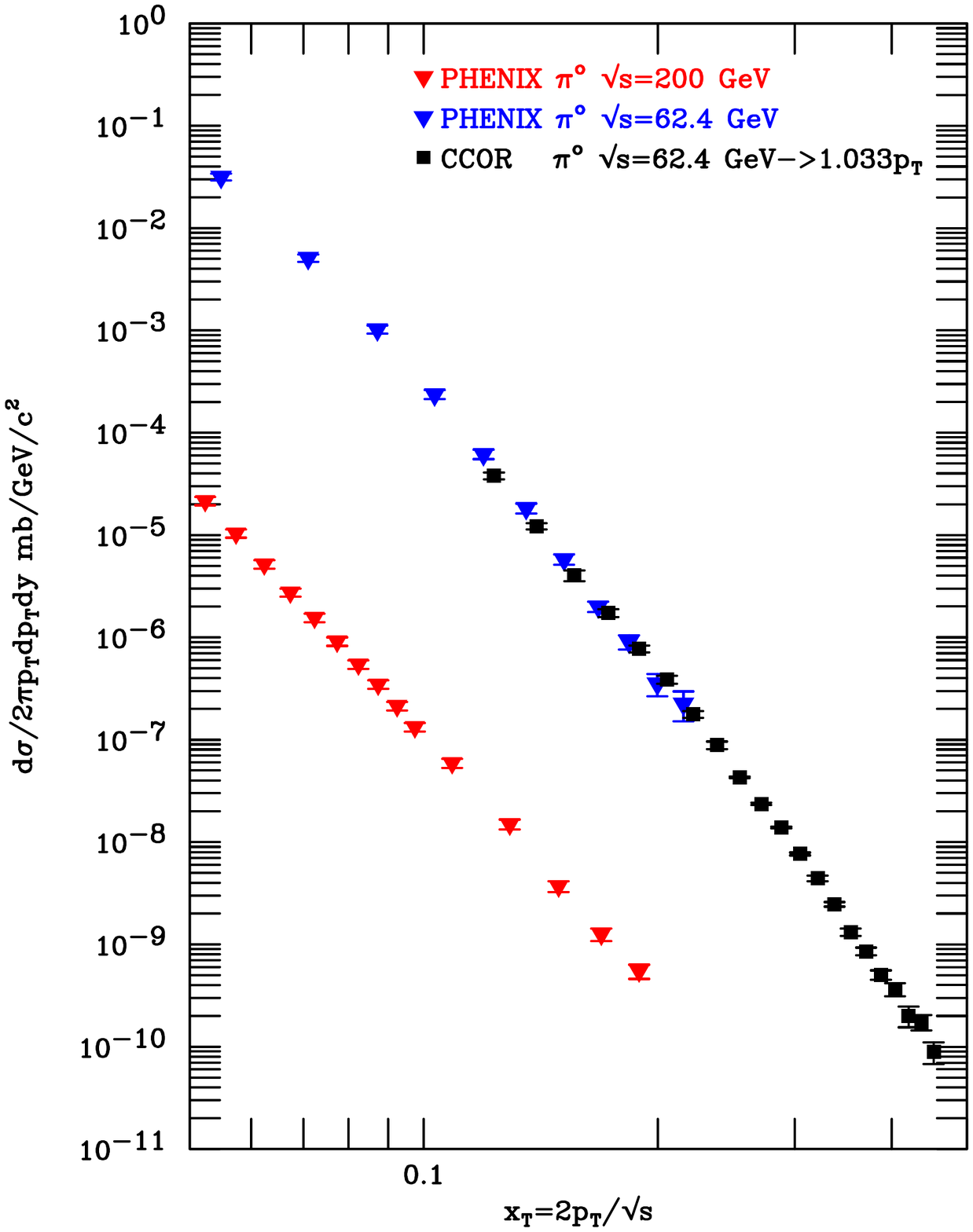} %&
\hspace*{-0.04\linewidth} \raisebox{1.0pc}{\includegraphics[width=0.51\linewidth]{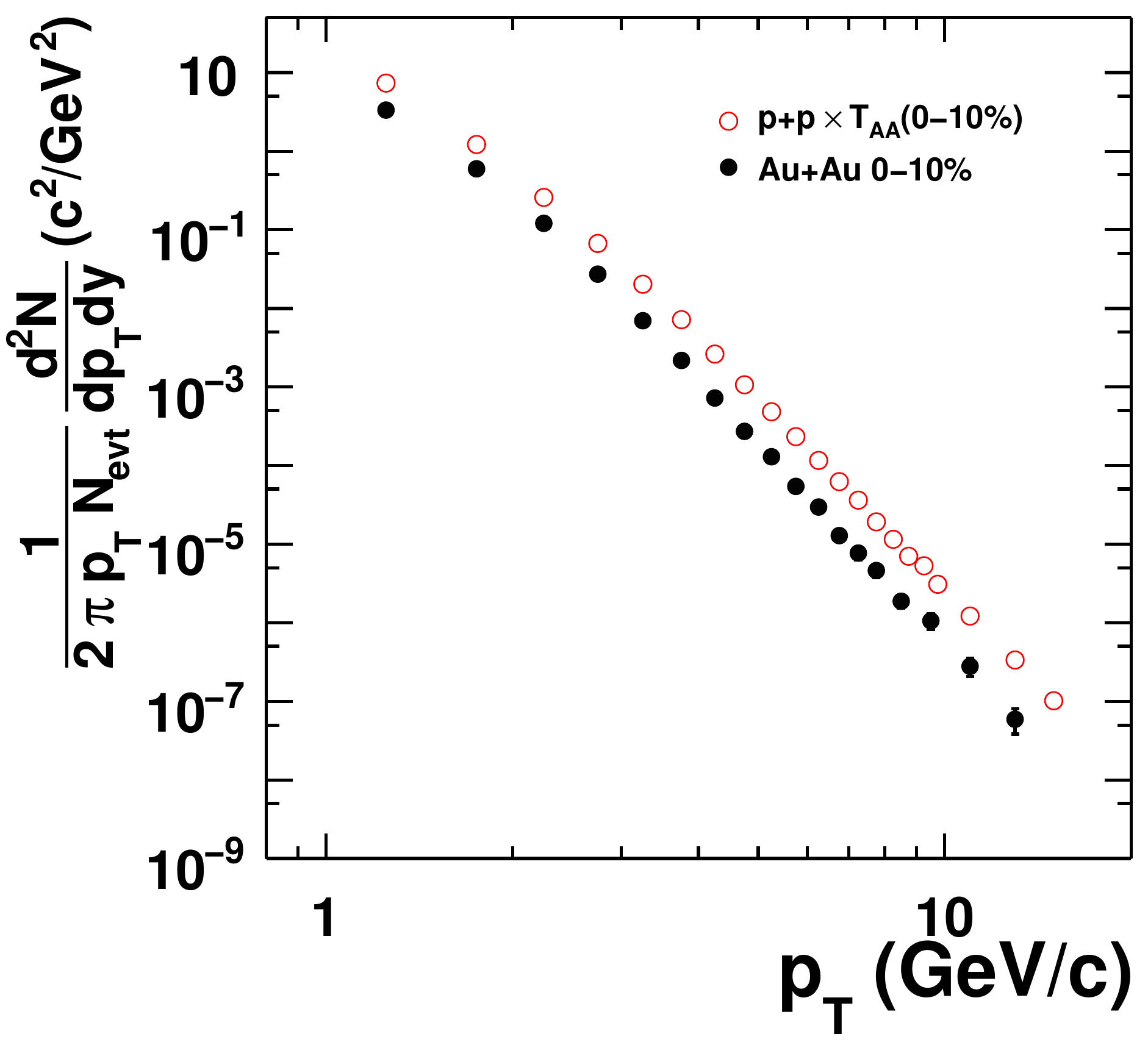}}
%\end{tabular}
\end{center}\vspace*{-1.5pc}
\caption[]{a) (left) $Ed^3\sigma/dp^3$ vs. $x_T$ for PHENIX mid-rapidity $\pi^0$ at $\sqrt{s}=200$ GeV in p-p collisions~\cite{PXpi0PRD} plus PHENIX~\cite{PXpi062pp} and CCOR-ISR~\cite{CCOR} measurements at $\sqrt{s}=62.4$ GeV, where the absolute $p_T$ scale of the ISR measurement has been corrected upwards by 3\% to agree with the PHENIX data. b) (right) $\pi^0$ p-p data vs. $p_T$ at $\sqrt{s}=200$ GeV from a) multiplied by $\mean{T_{AA}}$ for Au+Au central (0-10\%) collisions compared to semi-inclusive $\pi^0$ invariant yield in Au+Au central (0-10\%) collisions at $\sqrt{s_{NN}}=200$ GeV. } 
\label{fig:f1}\vspace*{-1.0pc}
\end{figure}
   The discovery, at RHIC, that $\pi^0$ are suppressed by roughly a factor of 5 compared to point-like scaling of hard-scattering in central Au+Au collisions is arguably {\em the}  major discovery in Relativistic Heavy Ion Physics. In Fig.~\ref{fig:f1}a), the PHENIX measurement of $E d^3 \sigma/dp^3$ for $\pi^0$ 
production in p-p collisions at $\sqrt{s}=62.4$ GeV~\cite{PXpi062pp} is in excellent agreement with the ISR data and the PHENIX $\pi^0$ data follow the same trend as the lower energy data, with a pure power law, $E d^3 \sigma/dp^3\propto p_T^{-8.1\pm 0.1}$ for $p_T > 3$ GeV/c at $\sqrt{s}=200$ GeV. In Fig.~\ref{fig:f1}b), the 200 GeV p-p data, multiplied by the point-like scaling factor $\mean{T_{AA}}$ for (0-10\%) central Au+Au collisions are compared to the semi-inclusive invariant $\pi^0$ yield in central (0-10\%) Au+Au collisions at $\sqrt{s_{NN}}=200$ GeV and,  amazingly, the Au+Au data follow the same power-law as the p-p data but are suppressed from the point-like scaled p-p data by a factor of $\sim 5$, independent of $p_T$. The suppression is represented quantitatively by the ``nuclear modification factor'', $R_{AA}(p_T)$, the ratio of the measured semi-inclusive yield in A+A collisions to the point-like scaled p-p cross section at a given $p_T$: 
         \begin{equation}
  R_{AA}(p_T)={{d^2N^{\pi}_{AA}/dp_T dy N_{AA}}\over {\langle T_{AA}\rangle d^2\sigma^{\pi}_{pp}/dp_T dy}} \quad . 
  \label{eq:RAA}
  \end{equation}

\begin{figure}[!b]
\includegraphics[width=0.50\textwidth]{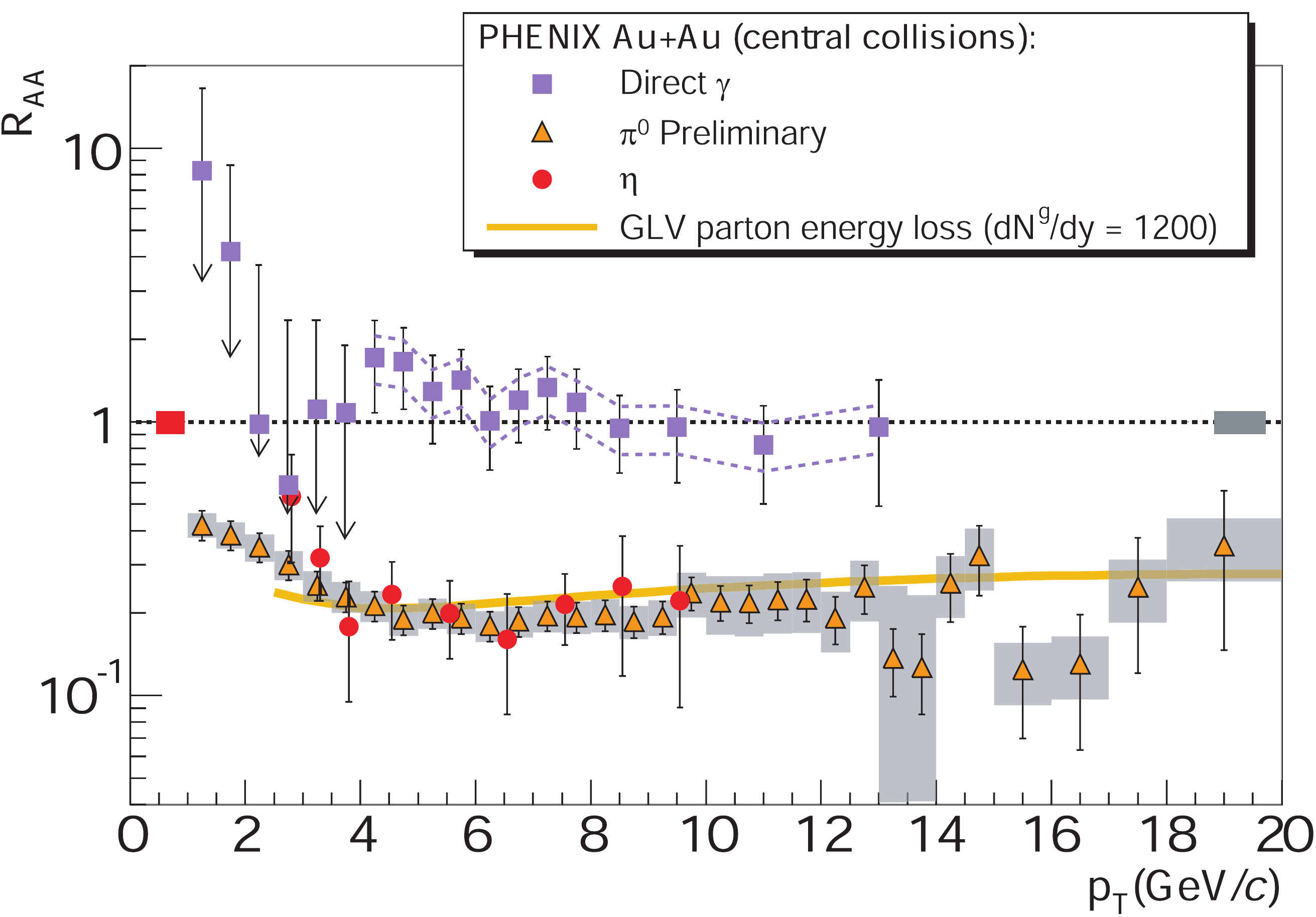} 
\includegraphics[width=0.50\textwidth]{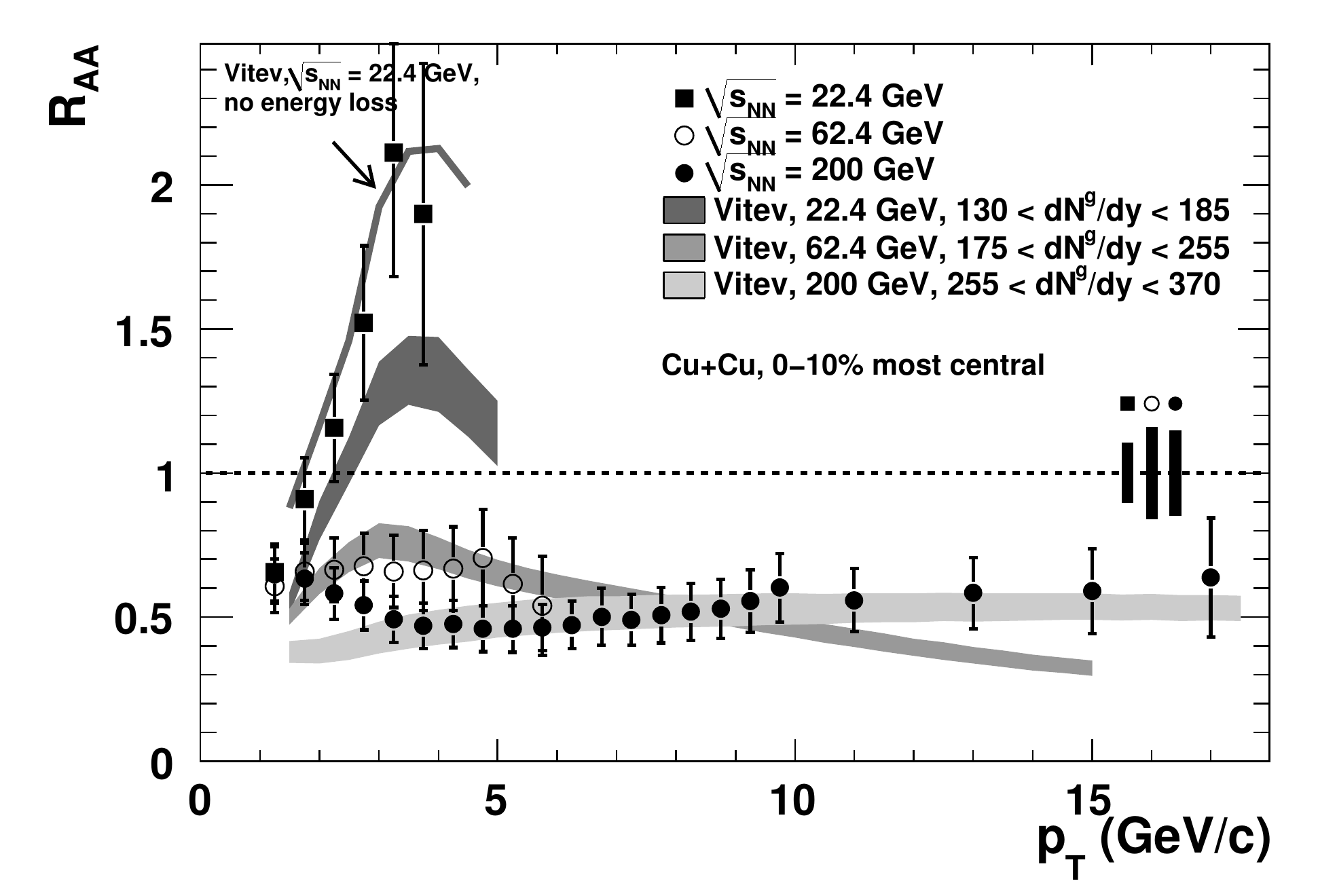} 
\caption[]{a) (left) Nuclear modification factor, $R_{AA}$ for direct-$\gamma$, $\pi^0$ and $\eta$ in Au+Au (0-10\%) central collisions at $\sqrt{s_{NN}}=200$ GeV~\cite{YAQM05}, together with GLV theory curve~\cite{GLV}. b) PHENIX $R_{AA}$ for $\pi^0$ in Cu+Cu central collisions at $\sqrt{s_{NN}}=200$, 62.4 and 22.4 GeV~\cite{ppg084}, together with Vitev theory curves~\cite{Vitev2}.}
\label{fig:QM05wowPXCu}
\end{figure}

In Fig.~\ref{fig:QM05wowPXCu}a, $R_{AA}(p_T)$ is shown for $\pi^0$, $\eta$ mesons and direct-$\gamma$ for $\sqrt{s_{NN}}=200$      
GeV Au+Au central (0-10\%) collisions. The $\pi^0$ and $\eta$ mesons, which are fragments of jets from outgoing partons are suppressed by the same amount while the direct-$\gamma$ which do not interact in the medium are not suppressed. This indicates a strong medium effect on outgoing partons. 
Fig.~\ref{fig:QM05wowPXCu}b shows that $R_{AA}$ for central (0-10\%) Cu+Cu collisions is comparable at $\sqrt{s_{NN}}=62.4$ and 200 GeV, but that there is no suppression, actually a Cronin enhancement~\cite{CroninEffect}, at $\sqrt{s_{NN}}=22.4$ GeV.  This indicates that the medium which suppresses jets is produced somewhere between $\sqrt{s_{NN}}=22.4$ GeV, the SpS Fixed Target highest c.m. energy, and 62.4 GeV. 

	The measurements at RHIC appear to be in excellent agreement with the theoretical curves~\cite{GLV,Vitev2}. The suppression can be explained by the energy loss of the outgoing partons in the dense color-charged medium due to coherent Landau-Pomeranchuk-Migdal radiation of gluons, predicted in QCD~\cite{BDMPS}, which is sensitive to properties of the medium. Measurements of two-particle correlations (discussed below, Sec.~\ref{sec:corrAA}) confirm the loss of energy of the away-jet relative to the trigger jet in Au+Au central collisions compared to p-p collisions. However, lots of details remain to be understood. 

\section{Direct photons at RHIC: thermal photons?}
\subsection{Internal Conversions---the first measurement anywhere of direct photons at low $p_T$}
\begin{figure}[!h]
\begin{center}
\begin{tabular}{cc}
\includegraphics[width=0.60\linewidth]{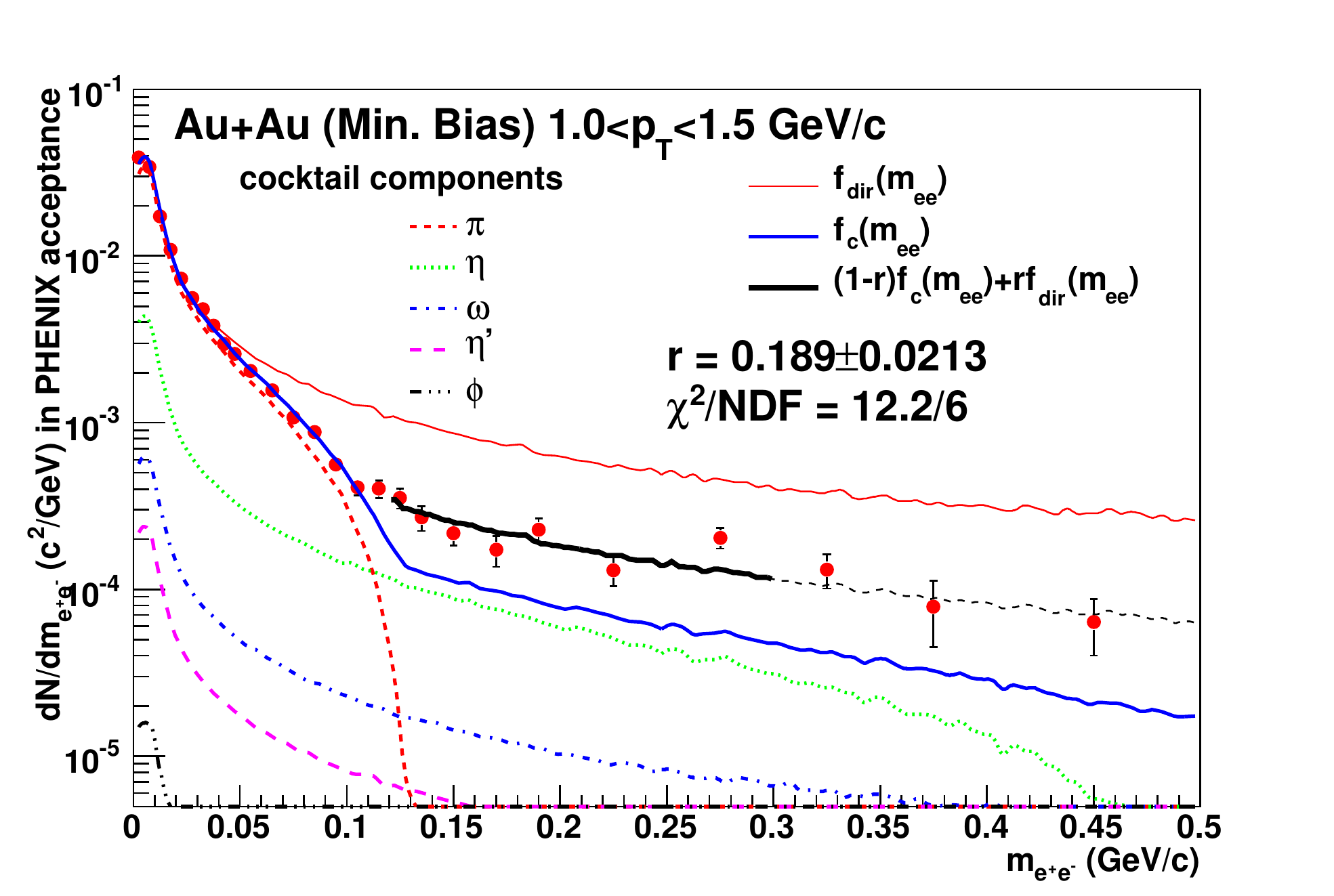}& 
\hspace*{-0.065\linewidth}\includegraphics[width=0.45\linewidth]{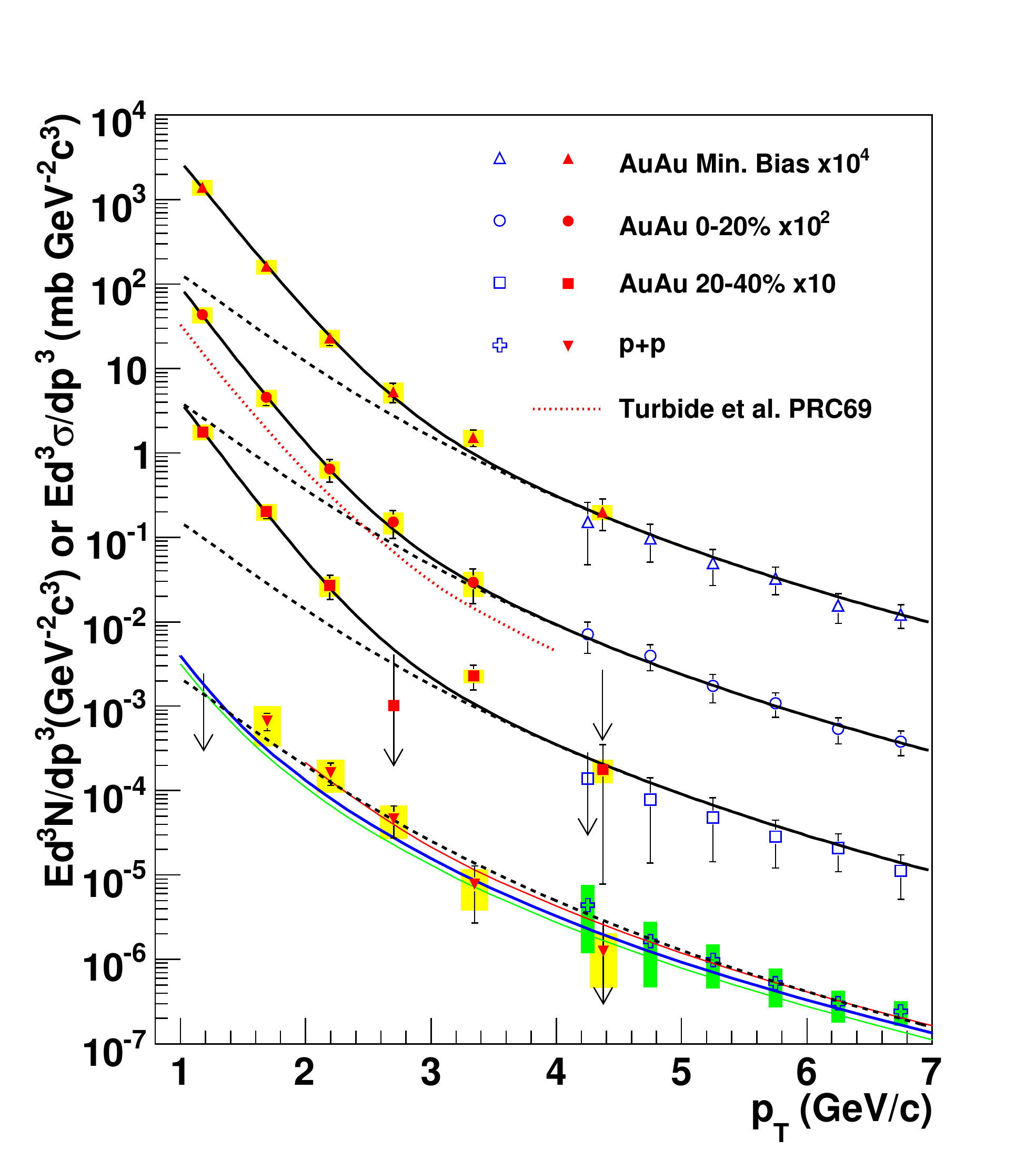} 
%\vspace*{-0.35in}
\end{tabular}
\end{center}
\caption[]{a) (left) Invariant mass ($m_{ee}$) distribution of $e^+ e^-$ pairs from Au+Au minimum bias events for $1.0< p_T<1.5$~GeV/c~\cite{ppg086}. Dashed lines are Eq.~\ref{eq:KW} for the mesons indicated. Blue solid line is $f_c(m)$, the total di-electron yield from the sum of contributions or `cocktail' of meson Dalitz decays; Red solid line is $f_{dir}(m)$ the internal conversion $m_{e e}$ spectrum from a direct-photon ($M>> m_{e e}$). Black solid line is a fit of the data to the sum of cocktail plus direct contributions in the range $80< m_{e e} < 300$ MeV/c$^2$. b) (right) Invariant cross section (p-p) or invariant yield (Au+Au) of direct photons as a function of $p_T$~\cite{ppg086}. Filled points are from virtual photons, open points from real photons.  
\label{fig:ppg086Figs} }
\end{figure}

   Internal conversion of a photon from $\pi^0$ and $\eta$ decay is well-known and is called Dalitz decay~\cite{egNPS}. Perhaps less well known in the RHI community is the fact that for any reaction (e.g. $q+g\rightarrow \gamma +q$) in which a real photon can be emitted, a virtual photon (e.g. $e^+ e^-$ pair of mass $m_{ee}\geq 2m_e$) can also be emitted. This is called internal-conversion and is generally given by the Kroll-Wada formula~\cite{KW,ppg086}:
   \begin{eqnarray}
   {1\over N_{\gamma}} {{dN_{ee}}\over {dm_{ee}}}&=& \frac{2\alpha}{3\pi}\frac{1}{m_{ee}} (1-\frac{m^2_{ee}}{M^2})^3 \quad \times \cr & &|F(m_{ee}^2)|^2 \sqrt{1-\frac{4m_e^2}{m_{ee}^2}}\, (1+\frac{2m_e^2}{m^2_{ee}})\quad ,
   \label{eq:KW}
   \end{eqnarray}
   where $M$ is the mass of the decaying meson or the effective mass of the emitting system. The dominant terms are on the first line of Eq.~\ref{eq:KW}:  the characteristic $1/m_{ee}$ dependence; and the cutoff of the spectrum for $m_{ee}\geq M$ (Fig.~\ref{fig:ppg086Figs}a)~\cite{ppg086}. Since the main background for direct-single-$\gamma$ production is a photon from $\pi^0\rightarrow \gamma +\gamma$, selecting $m_{ee} \gsim 100$ MeV/c$^2$  effectively reduces the background by an order of magnitude by eliminating the background from $\pi^0$ Dalitz decay, $\pi^0\rightarrow \gamma + e^+ + e^- $, at the expense of a factor $\sim 1000$ in rate. This allows the direct photon measurements to be extended (for the first time in both p-p and Au+Au collisions) below the value of $p_T\sim 4$~GeV/c, possible with real photons, down to $p_T=1$~GeV/c (Fig.~\ref{fig:ppg086Figs}b)~\cite{ppg086}, which is a real achievement. 
The solid lines on the p-p data are QCD calculations which work down to $p_T=2$~GeV/c. The dashed line is a fit of the p-p data to the modified power law $B (1+p_T^2/b)^{-n}$, used in the related Drell-Yan~\cite{Ito81} reaction, which flattens as $p_T\rightarrow 0$. 

	The relatively flat, non-exponential, spectra for the direct-$\gamma$ and Drell-Yan reactions as $p_T\rightarrow 0$ is due to the fact that there is no soft-physics production process for them, only production via the partonic subprocesses, $g+q\rightarrow \gamma+q$ and $\bar{q}+q\rightarrow e^+ + e^-$, respectively.  This is quite distinct from the case for hadron production, e.g. $\pi^0$, where the spectra are exponential  as $p_T\rightarrow 0$ in p-p collisions (Fig.~\ref{fig:pi0GamRHIC}a) due to soft-production processes, as well as in Au+Au collisions. 
 Thus, for direct-$\gamma$ in  Au+Au collisions, the exponential spectrum of excess photons above the $\mean{T_{AA}}$ extrapolated p-p fit is unique and therefore suggestive of a thermal source.   
\subsection{Low $p_T$ vs high $p_T$ direct-$\gamma$---Learn a lot from a busy plot}

\begin{figure}[!h]
\begin{center}
%\begin{tabular}{cc}
%\vspace*{-0.25in}
\includegraphics[width=0.90\linewidth]{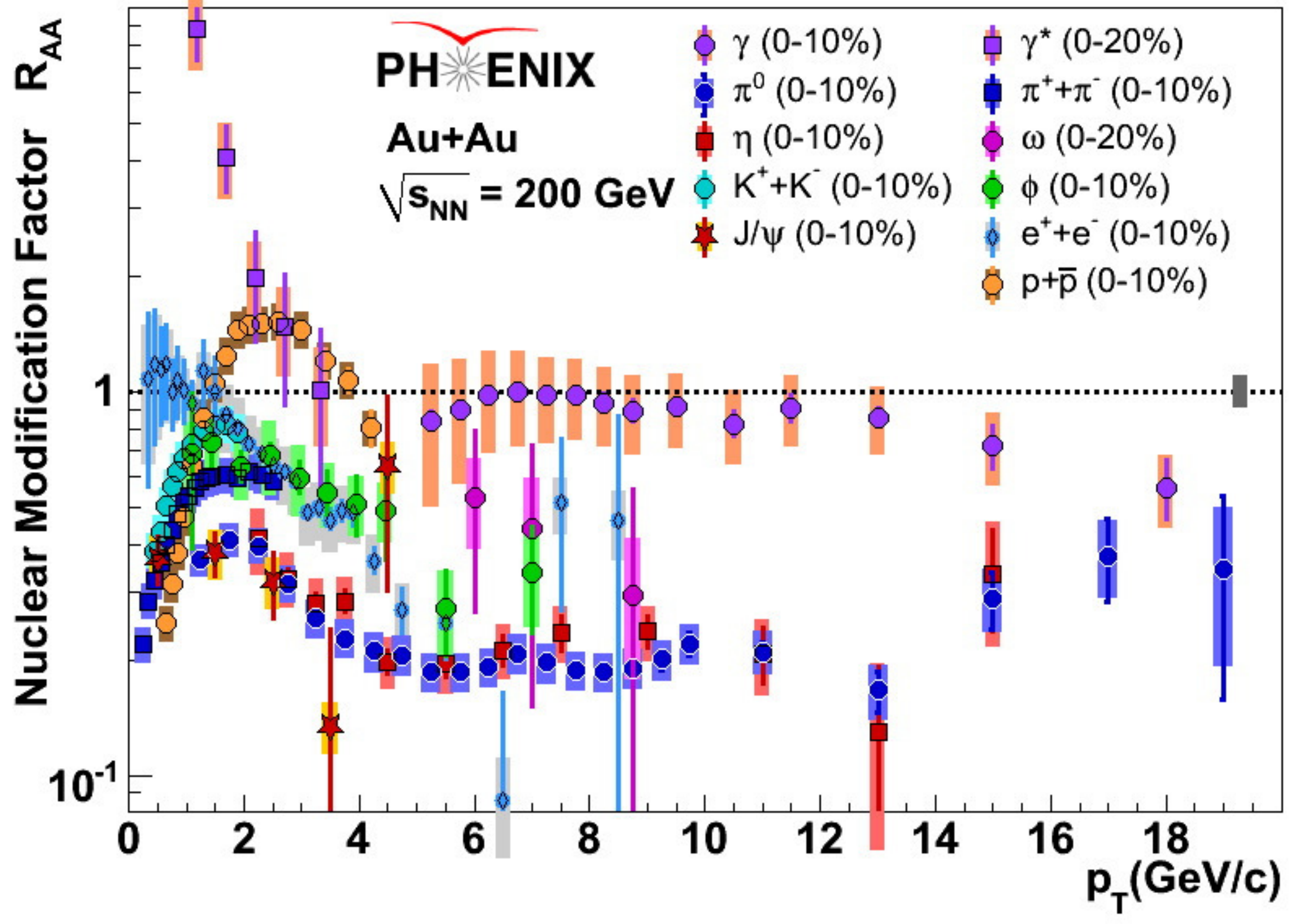} 
%\includegraphics[width=0.45\linewidth]{figs9/raa_Tshirt_v20g} 
%\vspace*{-0.35in}
%\end{tabular}
\end{center}
\caption[]{Nuclear Modification Factor, $R_{AA}(p_T)$ for all identified particles so far measured by PHENIX in central Au+Au collisions at $\sqrt{s_{NN}}=200$~GeV.~\cite{ThanksAM}   
\label{fig:Tshirt} }
\end{figure}

     The unique behavior of direct-$\gamma$ at low $p_T$ in Au+Au relative to p+p compared to any other particle is more dramatically illustrated by examining  the $R_{AA}$ of all particles measured by PHENIX in central Au+Au collisions at $\sqrt{s_{NN}}=200$~GeV (Fig.~\ref{fig:Tshirt})~\cite{ThanksAM}. For the entire region $p_T\leq 20$~GeV/c so far measured at RHIC, apart from the $p+\bar{p}$ which are enhanced in the region $2\leq p_T \lsim 4$~GeV/c (`the baryon anomaly'), the production of {\em no other particle} is enhanced over point-like scaling. 
The behavior of $R_{AA}$ of the low $p_T\leq 2$~GeV/c direct-$\gamma$ is totally and dramatically different from all the other particles, exhibiting an order of magnitude exponential enhancement as $p_T\rightarrow 0$. This exponential enhancement is certainly suggestive of a new production mechanism in central Au+Au collisions different from the conventional soft and hard particle production processes in p-p collisions and its unique behavior is attributed to thermal photon production by many authors (e.g. see citations in reference~\cite{ppg086}).

\subsubsection{Direct photons and mesons up to $p_T=20$~GeV/c}
   Other instructive observations can be gleaned from Fig.~\ref{fig:Tshirt}.  The $\pi^0$ and $\eta$ continue to track each other to the highest $p_T$. At lower $p_T$, the $\phi$ meson tracks the $K^{\pm}$ very well, but with a different value of $R_{AA}(p_T)$ than the $\pi^0$, while at higher $p_T$,the $\phi$ and $\omega$ vector mesons appear to track each other. Interestingly, the $J/\Psi$ seems to track the $\pi^0$ for $0\leq p_T\leq 4$~GeV/c; and it will be important to see whether this trend continues at higher $p_T$. 

\section{$J/\Psi$ suppression, still golden?}
\begin{figure}[!htb]
\begin{center}
\begin{tabular}{cc}
%\hspace*{-4cm}
\includegraphics[width=0.44\linewidth,height=0.54\linewidth]{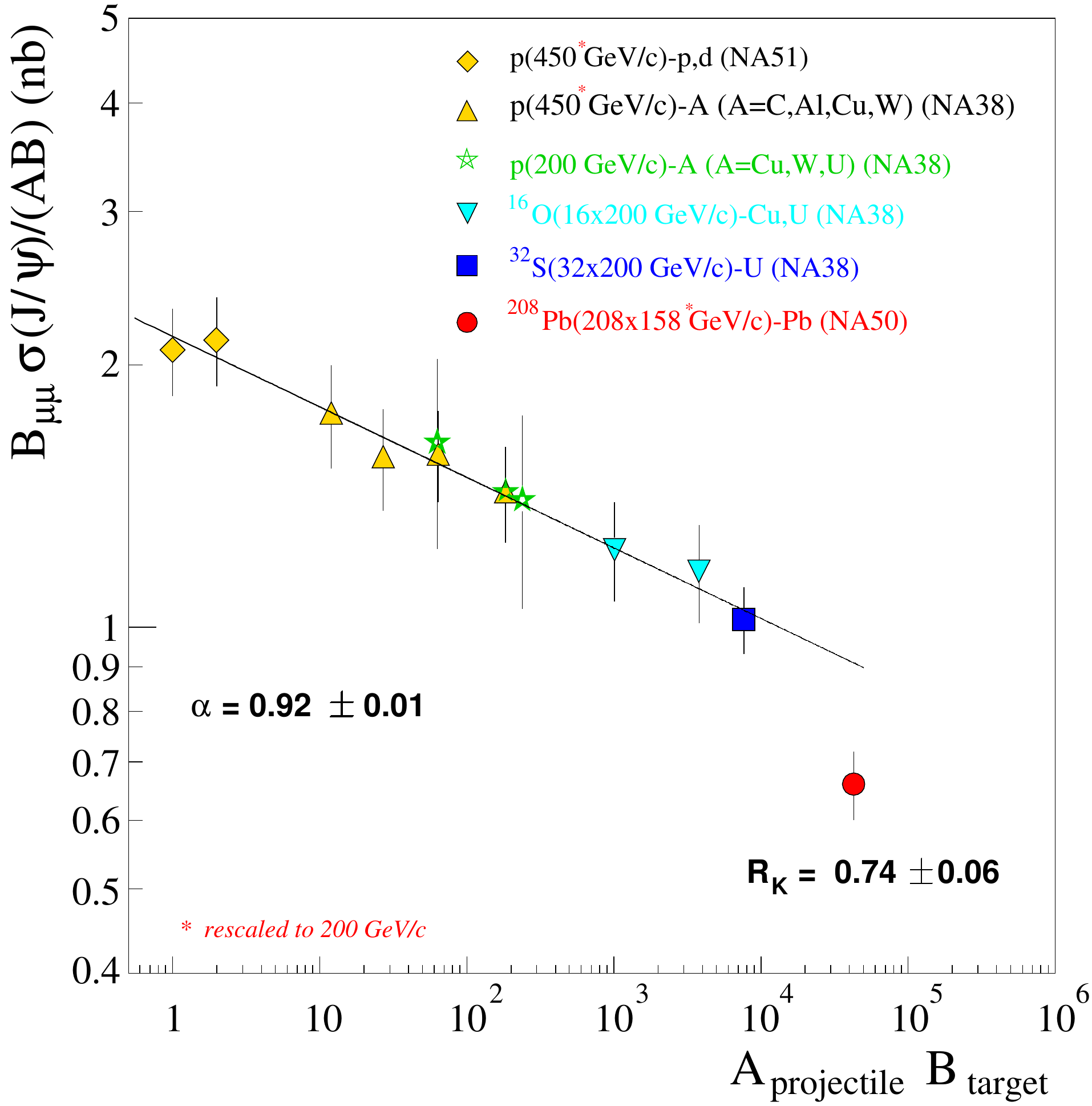}&\hspace*{0.2cm}
\includegraphics[width=0.43\linewidth]{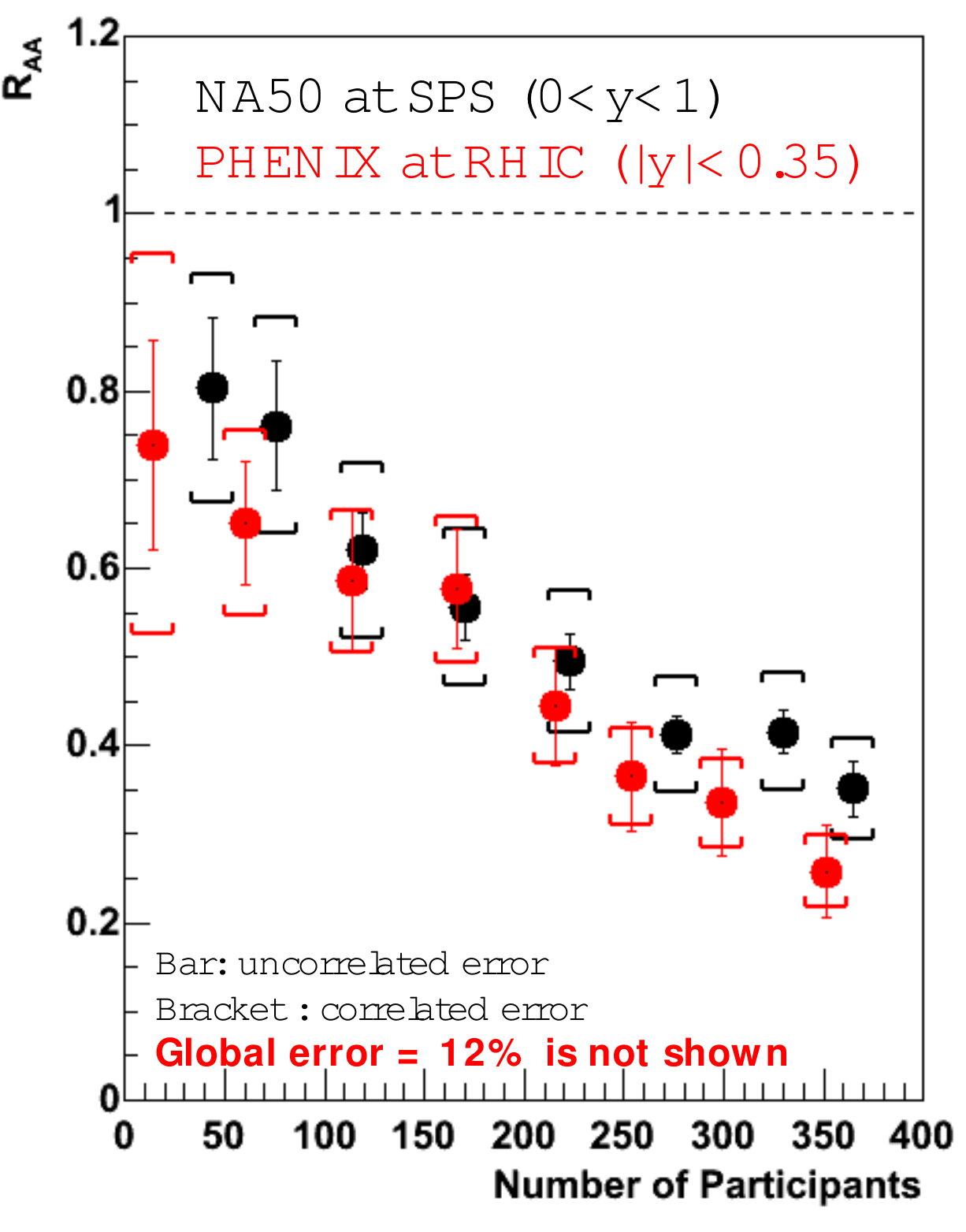}
\end{tabular}
\end{center}\vspace*{-0.25in}
\caption[]{\small a) (left) Total cross section for $J/\Psi$ production divided by $AB$ in A+B collisions at 158--200$A$ GeV~\cite{NA50EPJC39}. b) (right) $J/\Psi$ suppression relative to p-p collisions ($R_{AA}$) as a function of centrality ($N_{\rm part}$) at RHIC~\cite{PXJPsiAuAu200,GunjiQM06} and at the CERN/SPS~\cite{NA50EPJC39}. \label{fig:JPsiAB}}
\end{figure}

\begin{figure}[!t]
\begin{center}
\begin{tabular}{cc}
%\hspace*{-4cm}
\includegraphics[width=0.44\linewidth,height=0.40\linewidth]{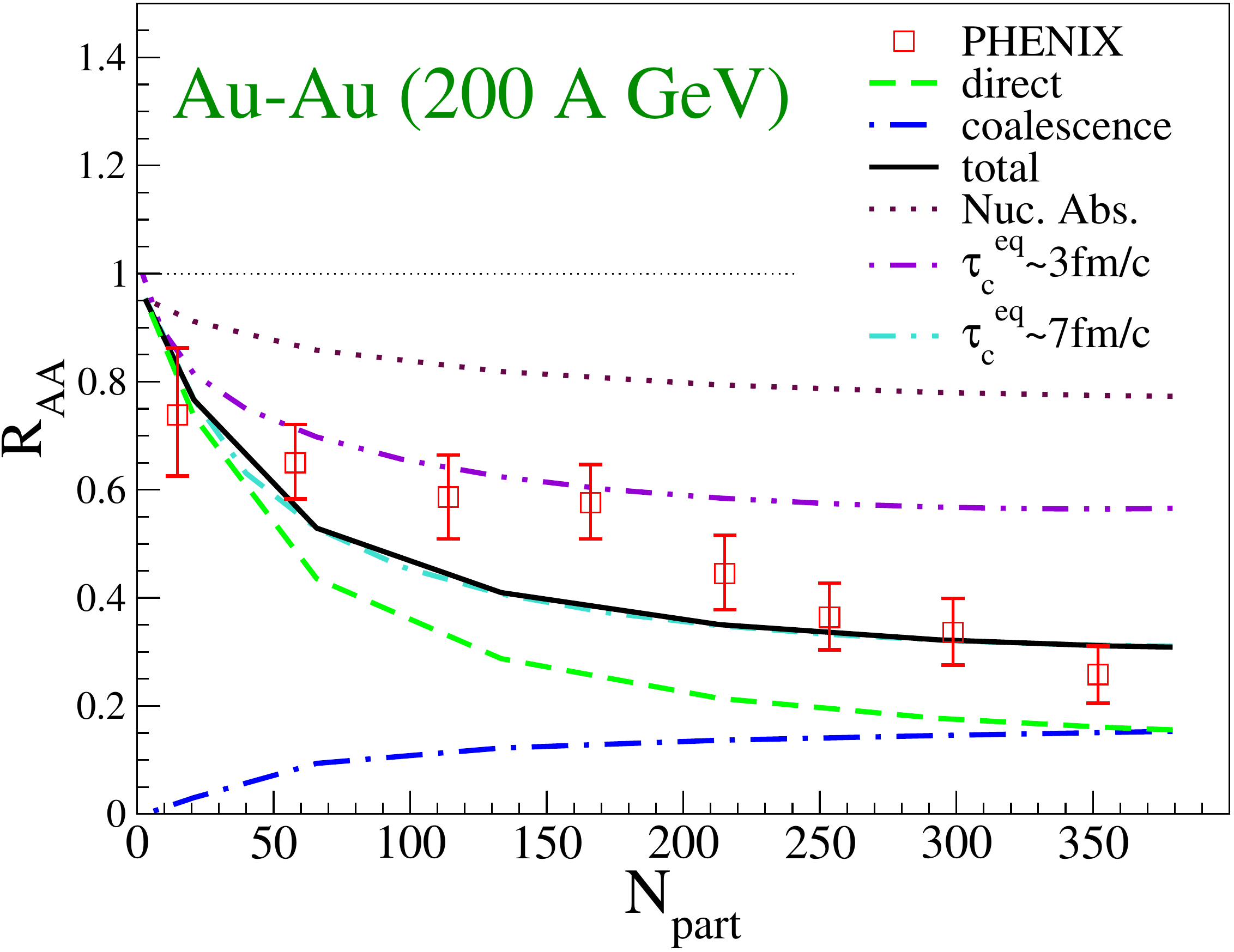}&\hspace*{0.2cm}
\includegraphics[width=0.43\linewidth]{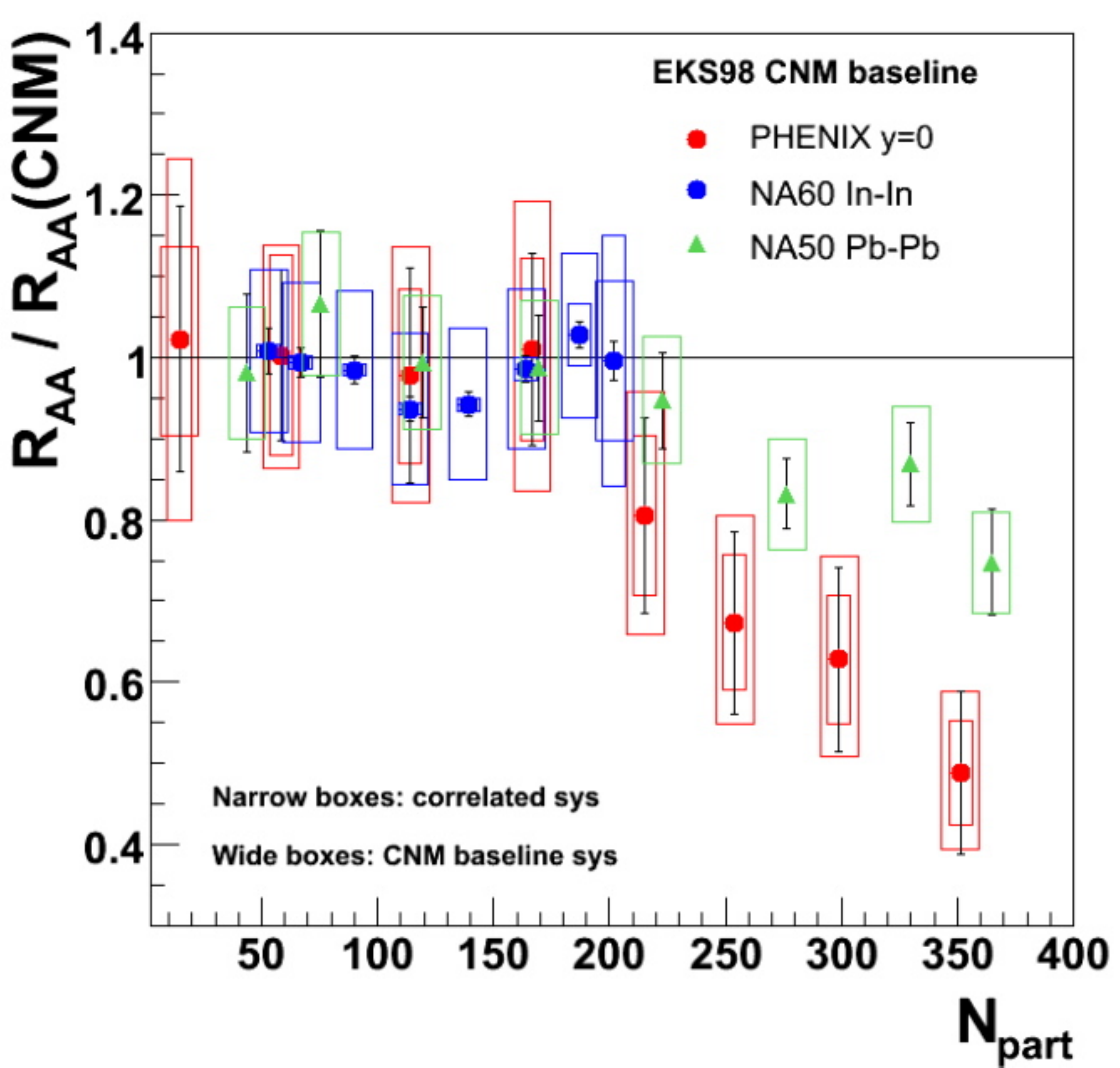}
\end{tabular}
\end{center}\vspace*{-0.25in}
\caption[]{\small a) (left) PHENIX measurement of $R_{AA}$ as a function of centrality from Fig.~\ref{fig:JPsiAB}b together with prediction from a coalescence model~\cite{RappPLB664}; b) (right) $R_{AA}$ of $J/\Psi$ at SpS and RHIC c.m. energies normalized to the measured $R_{AA}(CNM)$ from cold nuclear matter~\cite{ArnaldiECT09}\label{fig:JPsinew}}
\end{figure}

The dramatic difference in $\pi^0$ suppression from SpS to RHIC c.m. energy (Fig.~\ref{fig:QM05wowPXCu}b) is not reflected in $J/\Psi$ suppression, which is nearly identical at mid-rapidity at RHIC compared to the NA50 measurements at SpS (Fig.~\ref{fig:JPsiAB}b)~\cite{PXJPsiAuAu200,GunjiQM06}. 
This casts new doubt on the value of $J/\Psi$ suppression as a probe of deconfinement in addition to the previous complication that $J/\Psi$ are already suppressed (compared to point-like scaling) in p+A and B+A collisions (Fig.~\ref{fig:JPsiAB}a). One possible explanation is that $c$ and $\bar{c}$ quarks in the QGP recombine to regenerate $J/\Psi$, miraculously making the observed $R_{AA}$ equal at SpS and RHIC c.m. energies (Fig.~\ref{fig:JPsinew}a)~\cite{GunjiQM06,RappPLB664}. The good news is that such models predict the vanishing of $J/\Psi$ suppression or even an enhancement ($R_{AA}> 1$) at LHC energies~\cite{PBMStachelPLB490,ThewsPRC63,AndronicNPA79}, which would be spectacular, if observed.  

Even without the LHC startup, there has been progress this past year when, after $\sim 20$ years (!), p+A comparison data for the $J/\Psi$ from the CERN fixed target program at 158A GeV/c finally became available~\cite{ArnaldiQM09}. The cold nuclear matter effect of $J/\Psi$ suppression in p+A collisions is parameterized by an effective absorption cross section $\sigma^{J/\Psi}_{abs}$ which had been previously measured to be $4.3\pm 1.0$ mb at 400 GeV/c proton beam energy and {\em ``assumed to be independent of beam energy''}. The actual measurement for 158 GeV p+A collisions gives  $\sigma^{J/\Psi}_{abs}=7.6\pm 0.9$ mb which considerably reduces the ``anomalous suppression'' effect shown in Fig.~\ref{fig:JPsiAB}a to such an extent that there is now a clear difference between the CERN SpS and RHIC $J/\Psi$ suppression for the most central A+A collisions relative to the measured Cold Nuclear Matter effect (Fig.~\ref{fig:JPsinew}b)~\cite{ArnaldiECT09}. Maybe there is still some hope for $J/\Psi$ suppression as a QGP signature, but there is an important lesson for LHC. Comparison data for p-p and p+A MUST be taken and must be at the same $\sqrt{s_{NN}}$ as the A+A data.   

\section{Two-particle correlations}
\label{sec:corrAA}
\begin{figure}[!hbt]
\begin{center}
\begin{tabular}{cc}
\begin{tabular}[b]{c}
\hspace*{-0.03\linewidth}\includegraphics[width=0.509\linewidth]{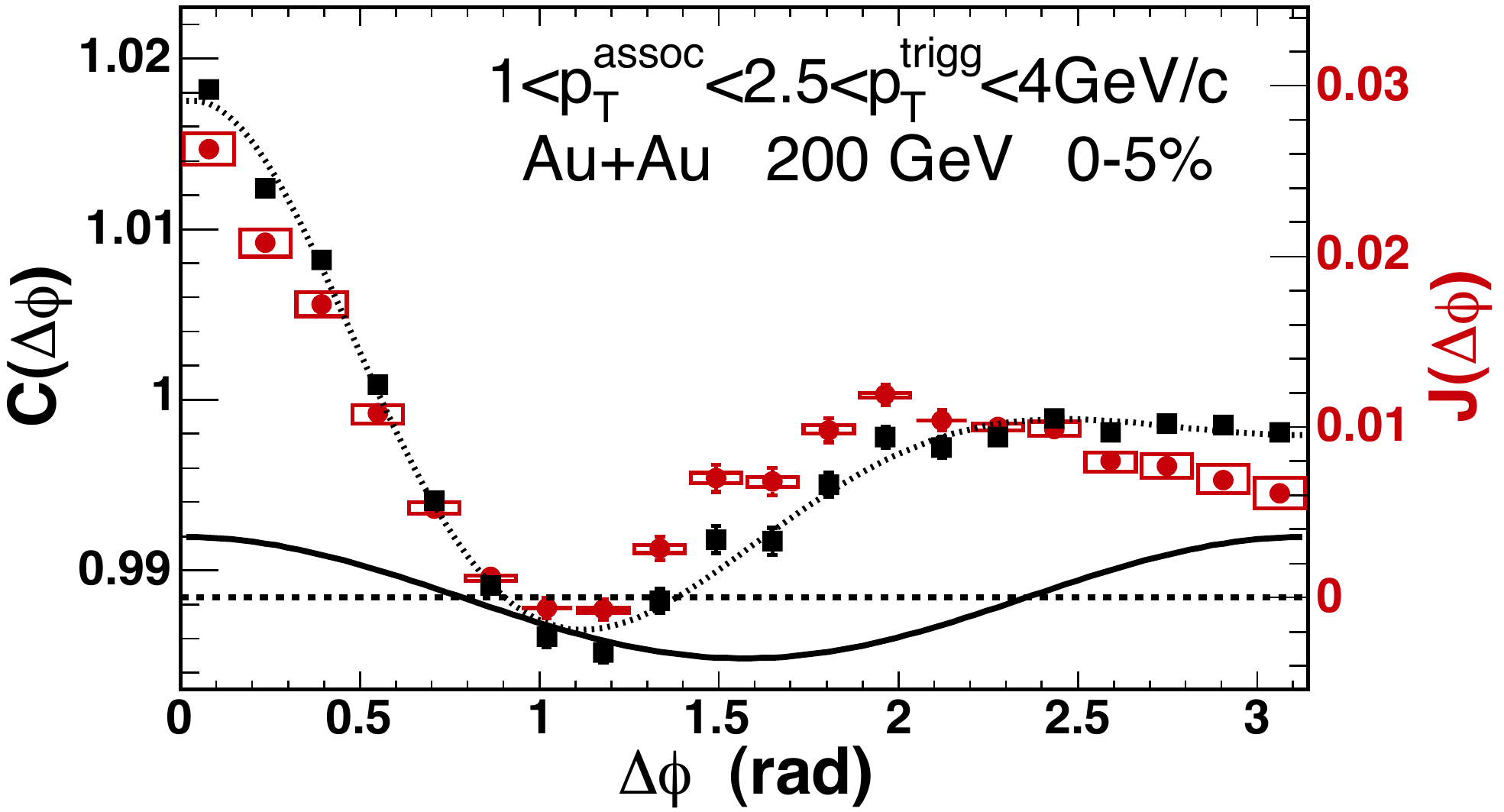}\cr
\hspace*{-0.03\linewidth}\includegraphics[width=0.435\linewidth]{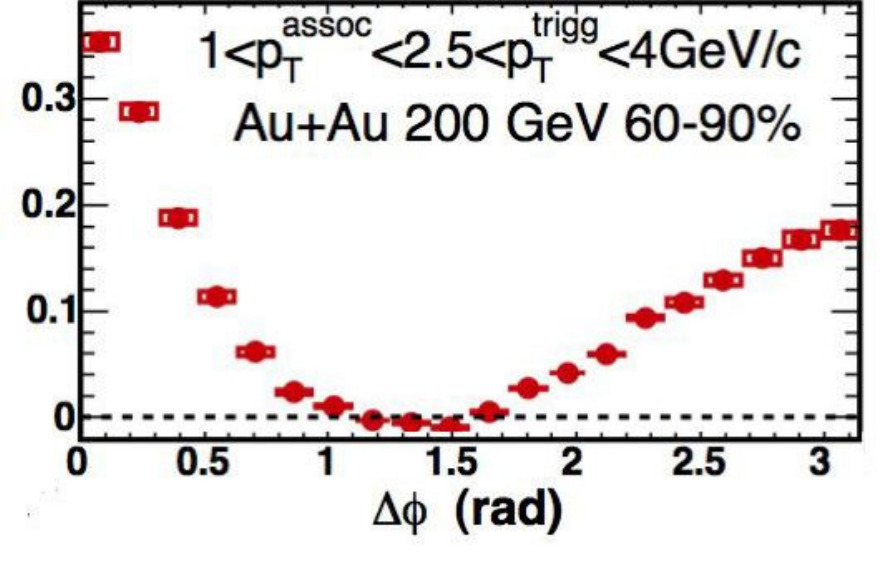}\hspace*{0.03\linewidth}
\end{tabular} 
\hspace*{-0.01\linewidth}\includegraphics[width=0.471\linewidth]{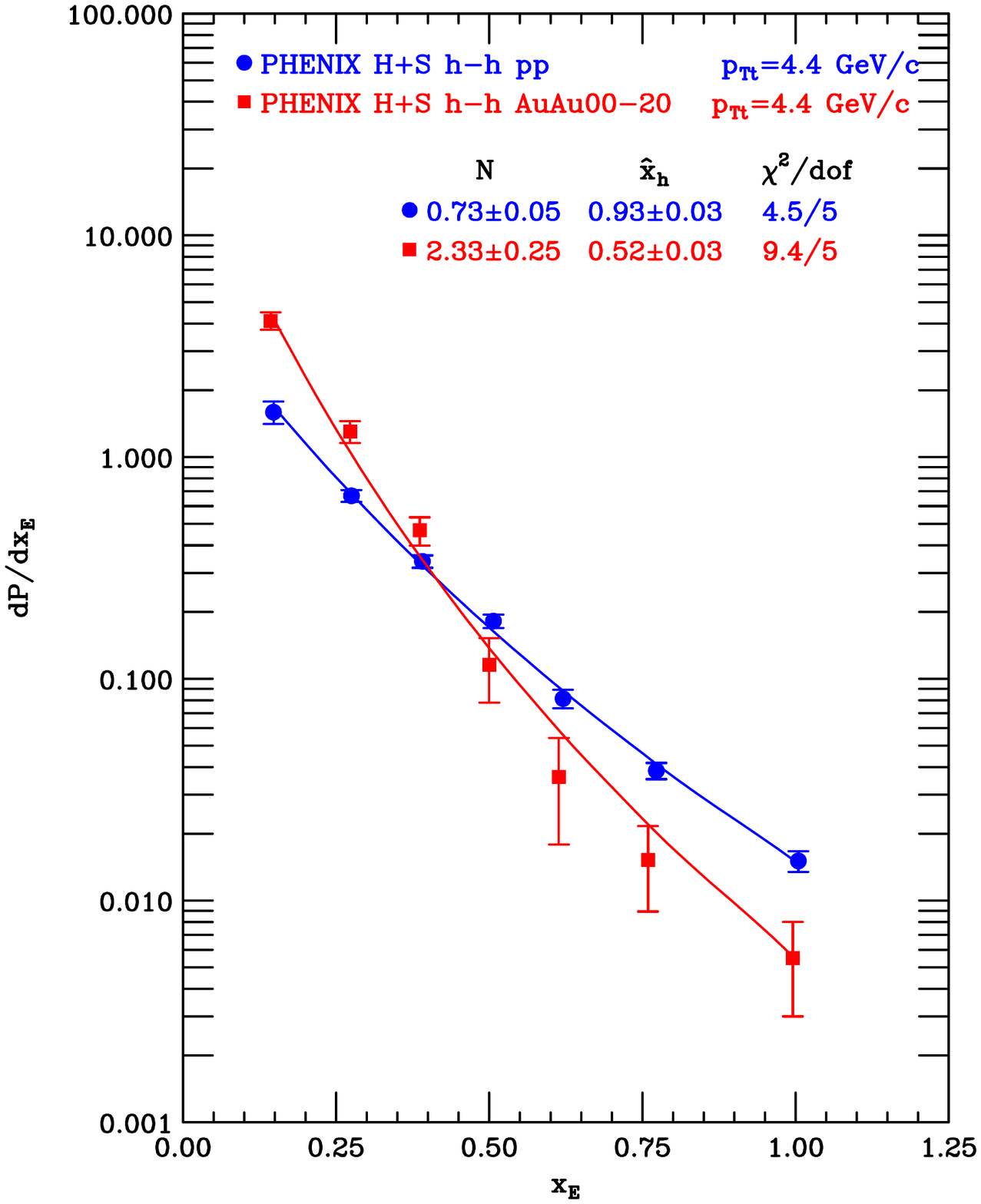}
\end{tabular}
\end{center}\vspace*{-1.7pc}
\caption[]{a) (left) Azimuthal correlation $C(\Delta\phi)$ of $h^{\pm}$ with $1\leq p_{T_a}\leq 2.5$ GeV/c with respect to a trigger $h^{\pm}$ with $2.5\leq p_{T_t} \leq 4$ GeV/c in Au+Au: (top) central collisions, where the line with data points indicates $C(\Delta\phi)$ before correction for the azimuthally modulated ($v_2$) background, and the other line is the $v_2$ correction which is subtracted to give the jet correlation function $J(\Delta\phi)$ (data points); (bottom)-same for peripheral collisions. b) (right) $x_E\approx p_{T_a}/p_{T_t}$ distribution for the Au+Au-central data compared to p-p. }
\label{fig:f8}
\end{figure}

    If the $\pi^0$ suppression shown in Fig.~\ref{fig:QM05wowPXCu} is in fact explained by the energy loss of the outgoing partons in the dense color-charged medium, this can be confirmed by measurements of two-particle correlations. These measurements are sensitive to the ratio of the energy of the away-jet to the trigger jet, which can be compared in Au+Au collisions and p-p collisions. 
    In analogy to Fig.~\ref{fig:mjt-ccorazi} (above), the two-particle correlations in Au+Au collisions (Fig.~\ref{fig:f8}a) show clear di-jet structure in both peripheral and central collisions. The away-side correlation in central Au+Au collisions is much wider than in peripheral Au+Au and p-p collisions and is further complicated by the large multiparticle background which is a modulated in azimuth by the $v_2$ collective flow of a comparable width to the jet correlation. After the $v_2$ correction, a double peak structure $\sim \pm 1$ radian from $\pi$ is evident, with a dip at $\pi$ radians. This may indicate a reaction of the medium to a passing parton in analogy to a ``sonic-boom''~\cite{ShuryakMach} and is under active study both theoretically and experimentally. 

%%%newin2011
    It is evident that $v_3$, a $\cos 3(\Delta\phi)$ term with lobes at $\Delta\phi=0, 2\pi/3$ and $4\pi/3 \approx 0, 2, 4$ radians, could explain the double peak structure at $\pi\pm D$ radian in the two-particle correlations.  There is presently lots of activity to confirm in detail whether taking account of the odd harmonics in addition to $v_2$ and $v_4$ in the background of Fig.~\ref{fig:f8}a  will result in narrower gaussian-like away-jet peaks in Au+Au central collisions like the peaks in peripheral Au+Au and p-p collisions.

    The energy loss of the away-parton is indicated by the fact that the $x_E$ distribution in Au+Au central collisions (Fig.~\ref{fig:f8}b) is steeper than that from p-p collisions. As noted above,   
we found in PHENIX~\cite{ppg029,mjtdeco} that the $x_E$ distribution did not measure the fragmentation function of the away-jet but is sensitive instead to $\hat{x}_h$, the ratio of the transverse momentum of the away-parton to that of the trigger parton, specifically~\cite{ppg029}:
\begin{equation}
\left . {dP\over dx_E}\right |_{p_{T_t}}=N (n-1) {1\over \hat{x}_h} {1 \over{(1+x_E/\hat{x}_h})^n}
\label{eq:xEdist} 
 \end{equation}
 where $N$ is a normalization factor, and $n$ (=8.1 at 200 GeV) is power of the inclusive invariant $p_{T_t}$ distribution.   

%%%

\section{A charming surprise}
\begin{figure}[!ht]
\begin{center} %\vspace*{-1pc}
\begin{tabular}{cc}
\hspace*{-0.04\linewidth}\includegraphics*[width=0.53\linewidth]{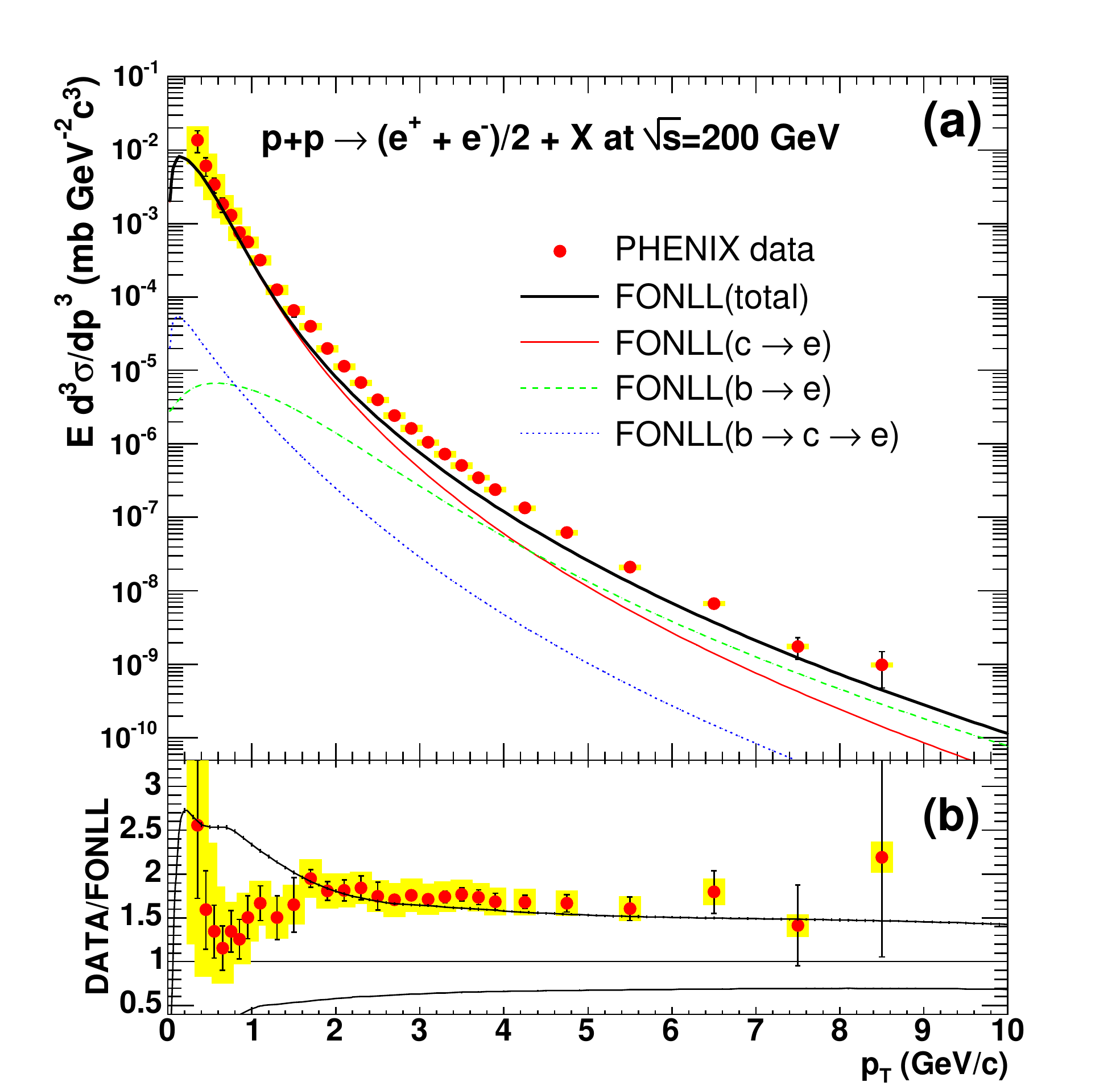} & 
%%\hspace*{-0.08\linewidth}\includegraphics[width=0.51\linewidth]{figs9/fig3se_data_only_logo} 
\hspace*{-0.06\linewidth}\includegraphics[width=0.51\linewidth]{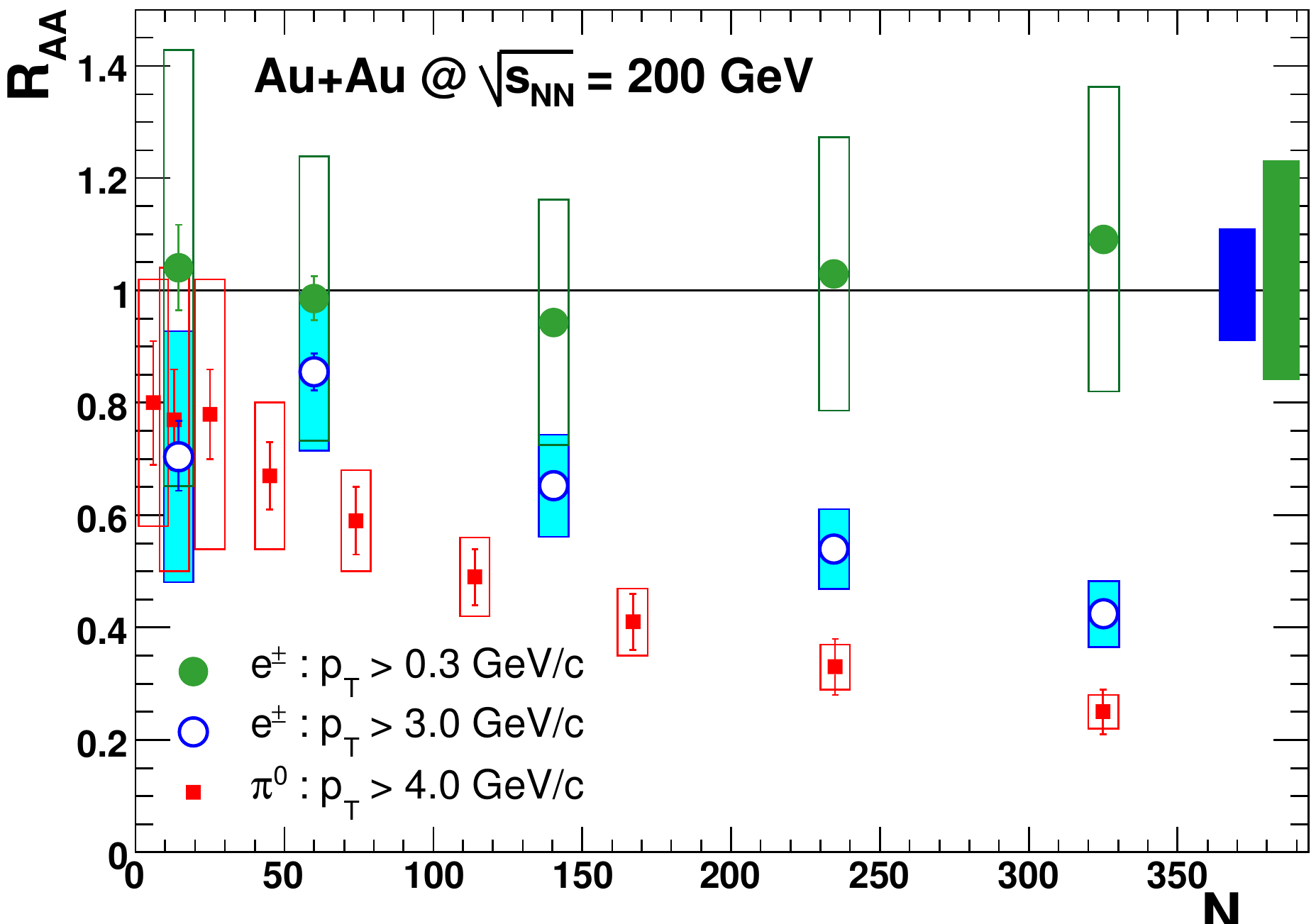} 
\end{tabular}
\end{center}\vspace*{-1.5pc}
\caption[]{a) (left) Invariant cross section of direct $e^{\pm}$ in p-p collisions ~\cite{PXcharmAA06} compared to theoretical predictions from $c$ and $b$ quark semileptonic decay. b) (right) $R_{AA}$ as a function of centrality ($N_{\rm part}$) for the total yield of $e^{\pm}$ from charm ($p_T > 0.3$) GeV/c, compared to the suppression of the $e^{\pm}$ yield at large $p_T>3.0$ GeV/c which is comparable to that of $\pi^0$ with ($p_T>4$ GeV/c)~\cite{PXcharmAA06}}
\label{fig:f7}
\end{figure}

   We designed PHENIX specifically to be able to detect charm particles via direct-single $e^{\pm}$ since this went along naturally with $J/\Psi\rightarrow e^+ +e^-$ detection and since the single particle reaction avoided the huge combinatoric background in Au+Au collisions. We thought that the main purpose of open charm production, which corresponds to a hard-scale ($m_{c\bar{c}}\gsim 3$ GeV/c$^2$), would be a check of our centrality definition and $\mean{T_{AA}}$ calculation since the total production of $c$ quarks should follow point-like scaling. In fact, our first measurement supported this beautifully~\cite{PXcharmPRL94}. However, our subsequent measurements proved to be much more interesting and even more beautiful. Figure~\ref{fig:f7}a shows our direct-single-$e^{\pm}$ measurement in p-p collisions at $\sqrt{s}=200$ GeV~\cite{PXcharmAA06} in agreement with a QCD calculation of $c$ and $b$ quarks as the source of the direct-single-$e^{\pm}$ (also called non-photonic $e^{\pm}$ at RHIC).   
The total yield of direct-$e^{\pm}$ for $p_T > 0.3$ GeV/c was taken as the yield of $c$-quarks in p-p and Au+Au collisions. The result, $R_{AA}=1$ as a function of centrality (Fig.~\ref{fig:f7}b), showed that the total $c-(\bar{c})$ production followed point-like scaling, as expected. The big surprise came at large $p_T$ where we found that the yield of direct-single-$e^{\pm}$ for $p_T>3$ GeV/c was suppressed nearly the same as the $\pi^0$ from light quark and gluon production. This strongly disfavors the QCD energy-loss explanation of jet-quenching because, naively, heavy quarks should radiate much less than light quarks and gluons in the medium; but opens up a whole range of new possibilities including string theory~\cite{egsee066}. 

\begin{figure}[!ht]
\begin{center} %\vspace*{-1pc}
\begin{tabular}{cc}
\hspace*{-0.02\linewidth}\includegraphics*[width=0.51\linewidth]{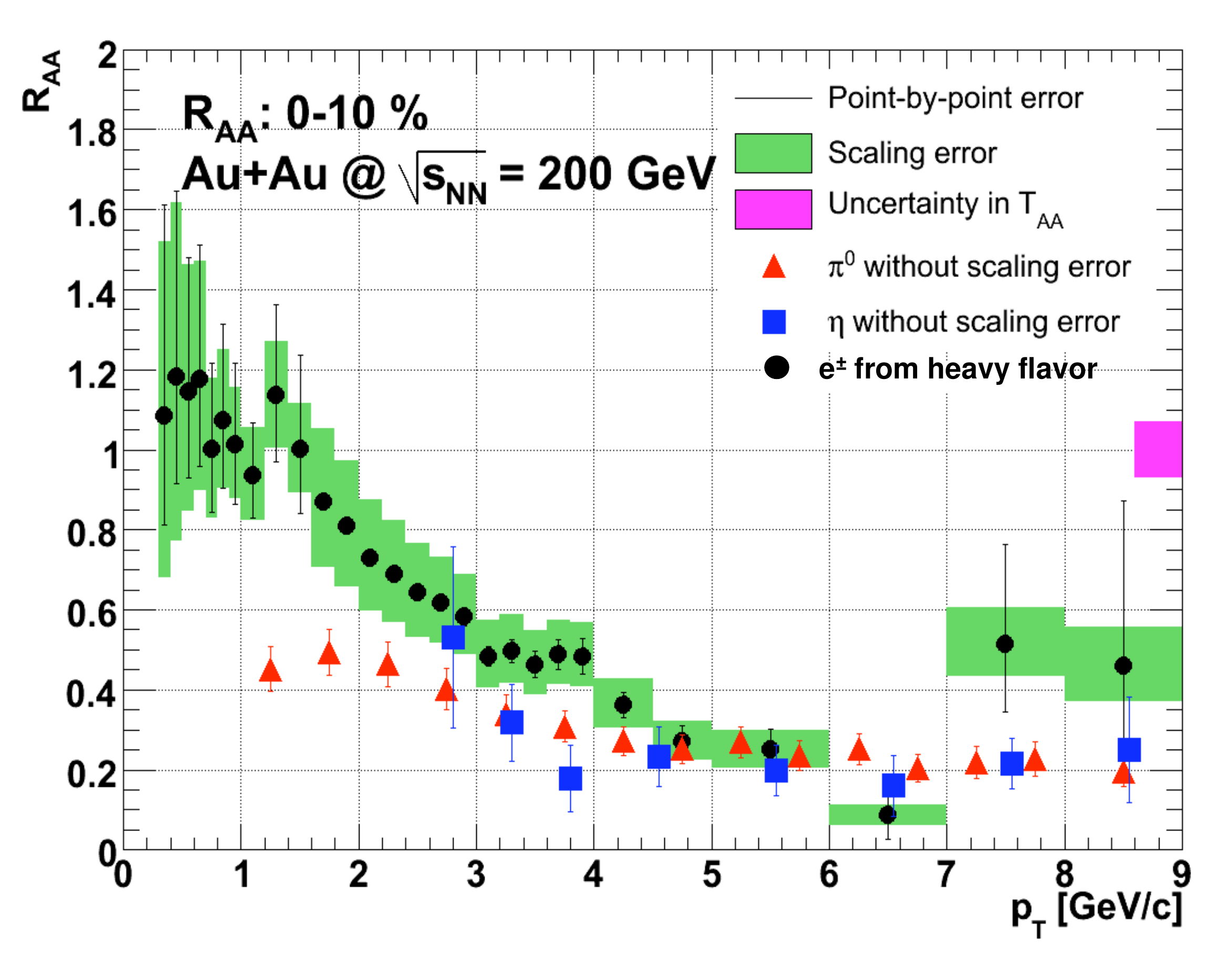} & 
%%\hspace*{-0.08\linewidth}\includegraphics[width=0.51\linewidth]{figs9/fig3se_data_only_logo} 
\hspace*{-0.02\linewidth}\includegraphics[width=0.50\linewidth]{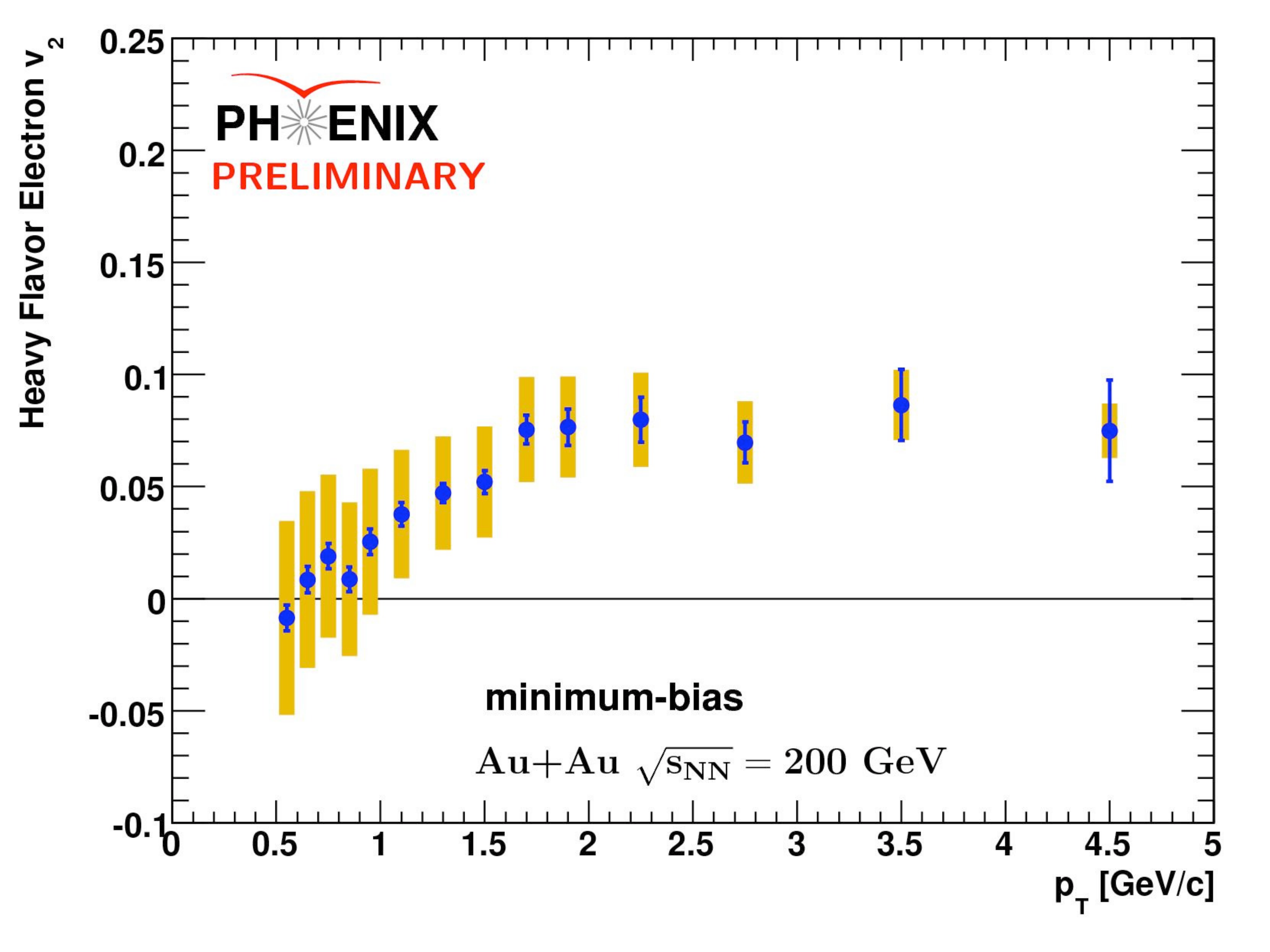} 
\end{tabular}
\end{center}\vspace*{-1.5pc}
\caption[]{a) (left) $R_{AA}$ (central Au+Au) b) (right) $v_2$ (minimum bias Au+Au) as a function of $p_T$ for direct-$e^{\pm}$ at $\sqrt{s_{NN}}=200$ GeV~\cite{PXcharmAA06}. }
\label{fig:fcrisis}
\end{figure}

The suppression of direct-single-$e^{\pm}$ is even more dramatic as a function of $p_T\gsim 5$ GeV/c (Fig~\ref{fig:fcrisis}a) which indicates suppression of heavy quarks as large as that for $\pi^0$ in the region where the $m\gsim 4$ GeV $b$-quarks dominate. Figure~\ref{fig:fcrisis}b  shows that heavy quarks exhibit collective flow ($v_2$), another indication of a very strong interaction with the medium. 
\section{Zichichi to the rescue?}
  In September 2007, I read an article by Nino, ``Yukawa's gold mine'' in the CERN Courier taken from his talk at the 2007 International Nuclear Physics meeting in Tokyo, Japan, in which he proposed:``We know that confinement produces masses of the order of a giga-electron-volt. Therefore, according to our present understanding, the QCD colourless condition cannot explain the heavy quark mass. However, since the origin of the quark masses is still not known, it cannot be excluded that in a QCD coloured world, the six quarks are all nearly massless and that the colourless condition is `flavour' dependent.'' 
  
  Nino's idea really excited me even though, or perhaps because, it appeared to overturn two of the major tenets of the Standard Model since it seemed to imply that: QCD isn't flavor blind;  the masses of quarks aren't given by the Higgs mechanism.  Massless $b$ and $c$ quarks in a color-charged medium would be the simplest way to explain the apparent equality of gluon, light quark and heavy quark suppression indicated by the equality of $R_{AA}$ for $\pi^0$ and direct single-$e^{\pm}$ in regions where both $c$ and $b$ quarks dominate. Furthermore RHIC and LHC-Ions are the only place in the Universe to test this idea. 
  
   It may seem surprising that I would be so quick to take Nino's idea so seriously. This confidence dates from my graduate student days when I checked the proceedings of the 12th ICHEP in Dubna, Russia in 1964 to see how my thesis results were reported and I found several interesting questions and comments by an ``A. Zichichi'' printed in the proceedings. One comment about how to find the $W$ boson in p+p collisions deserves a verbatim quote because it was exactly how the $W$ was discovered at CERN 19 years later: ``We would observe the $\mu$'s from W-decays. By measuring the angular and momentum distribution at large angles of K and $\pi$'s, we can predict the corresponding $\mu$-spectrum. We then see if the $\mu$'s found at large angles agree with or exceed the expected numbers.''
   
  Nino's idea seems much more reasonable to me than the string theory explanations of heavy-quark suppression (especially since they can't explain light-quark suppression). Nevertheless, just to be safe, I asked some distinguished theorists what they thought, with these results:
  \begin{itemize}
  \item Stan Brodsky:``Oh, you mean the Higgs field can't penetrate the QGP.''
 \item Rob Pisarski: `` You mean that the propagation of heavy and light quarks through the medium is the same.''
 \item Chris Quigg (Moriond 2008): ``The Higgs coupling to vector bosons $\gamma$, $W$, $Z$ is specified in the standard model and is a fundamental issue. One big question to be answered by the LHC is whether the Higgs gives mass to fermions or only to gauge bosons. The Yukawa couplings to fermions are put in by hand and are not required.'' ``What sets fermion masses, mixings?"
 \item Bill Marciano:``No change in the $t$-quark, $W$, Higgs mass relationship   if there is no Yukawa coupling: but there could be other changes.''
 \item Steve Weinberg: ``Lenny Susskind and I had a model, Technicolor (or Hypercolor),  that worked well in the vector boson sector but didn't give mass to the fermions.''
 \end{itemize}
	 
	 Nino proposed to test his idea by shooting a proton beam through a QGP formed in a Pb+Pb collision at the LHC and seeing the proton `dissolved' by the QGP. My idea is to use the new PHENIX VTX detector, installed in 2011, to  map out, on an event-by-event basis, the di-hadron correlations from identified $b-\overline{b}$ di-jets, identified $c-\overline{c}$ di-jets, which do not originate from the vertex, and light quark and gluon di-jets, which originate from the vertex and can be measured with $\pi^0$-hadron correlations. A steepening of the slope of the $x_E$ distribution of heavy-quark correlations as in Fig.~\ref{fig:f8}b will confirm in detail (or falsify) whether the different flavors of quarks behave as if they have the same energy loss (hence mass) in a color-charged medium. If Nino's proposed effect is true, that the masses of fermions are not given by the Higgs particle, and we can confirm the effect at RHIC or LHC-Ions, this would be a case where we Relativistic Heavy Ion Physicists may have something unique to contribute at the most fundamental level to the Standard Model, which would constitute a ``transformational discovery.'' Of course the LHC could falsify this idea by finding the Higgs decay to $b-\bar{b}$ at the expected rate in p-p collisions. Clearly, there are exciting years ahead of us! \\ 
\appendix 
\section {Appendix. Discussions} This appendix contains discussions among author and participants at the Erice 2009 International School of Subnuclear Physics\pagebreak
\begin{center}
\includegraphics[width=0.98\linewidth]{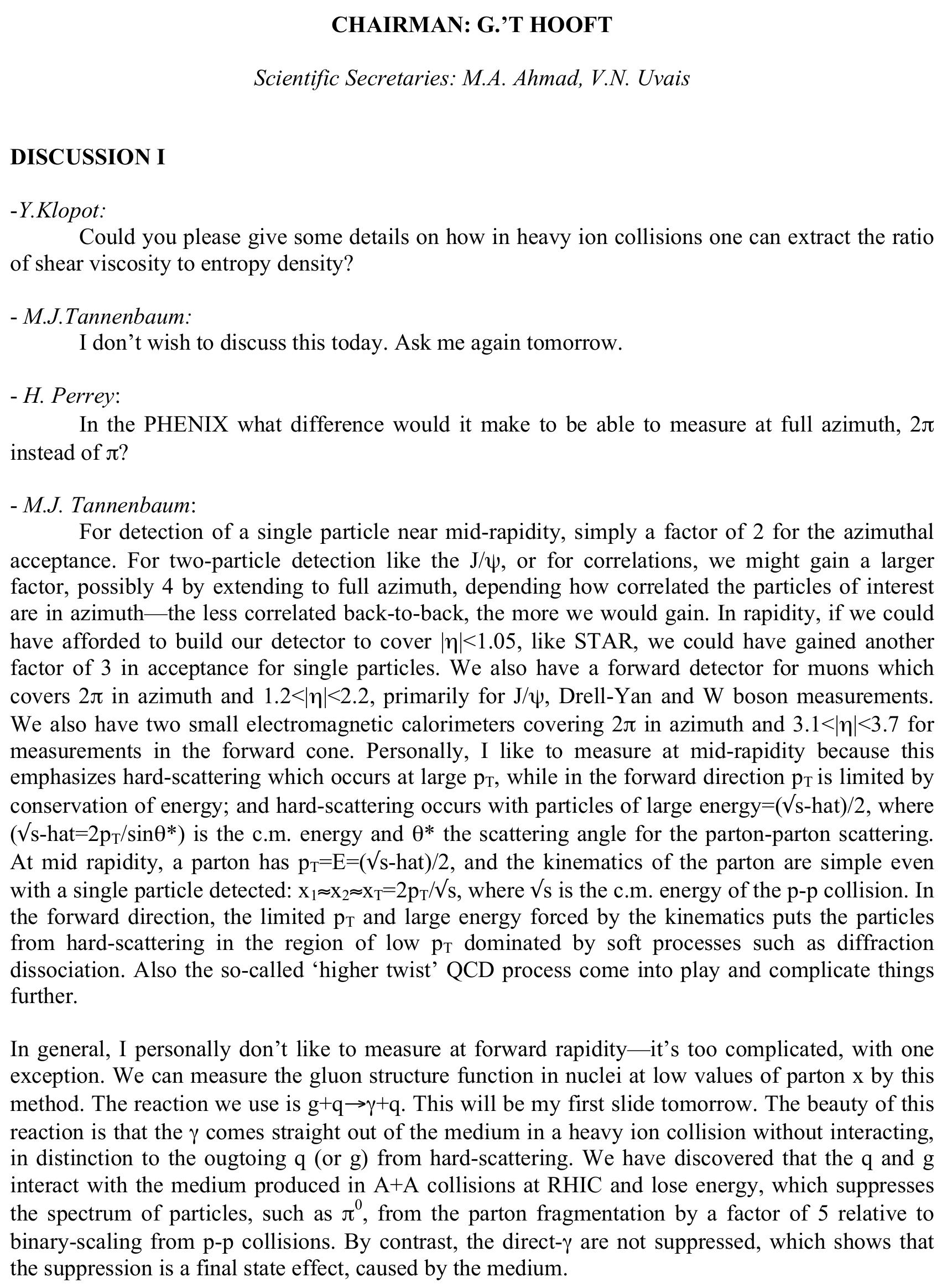}\newpage
\includegraphics[width=0.98\linewidth]{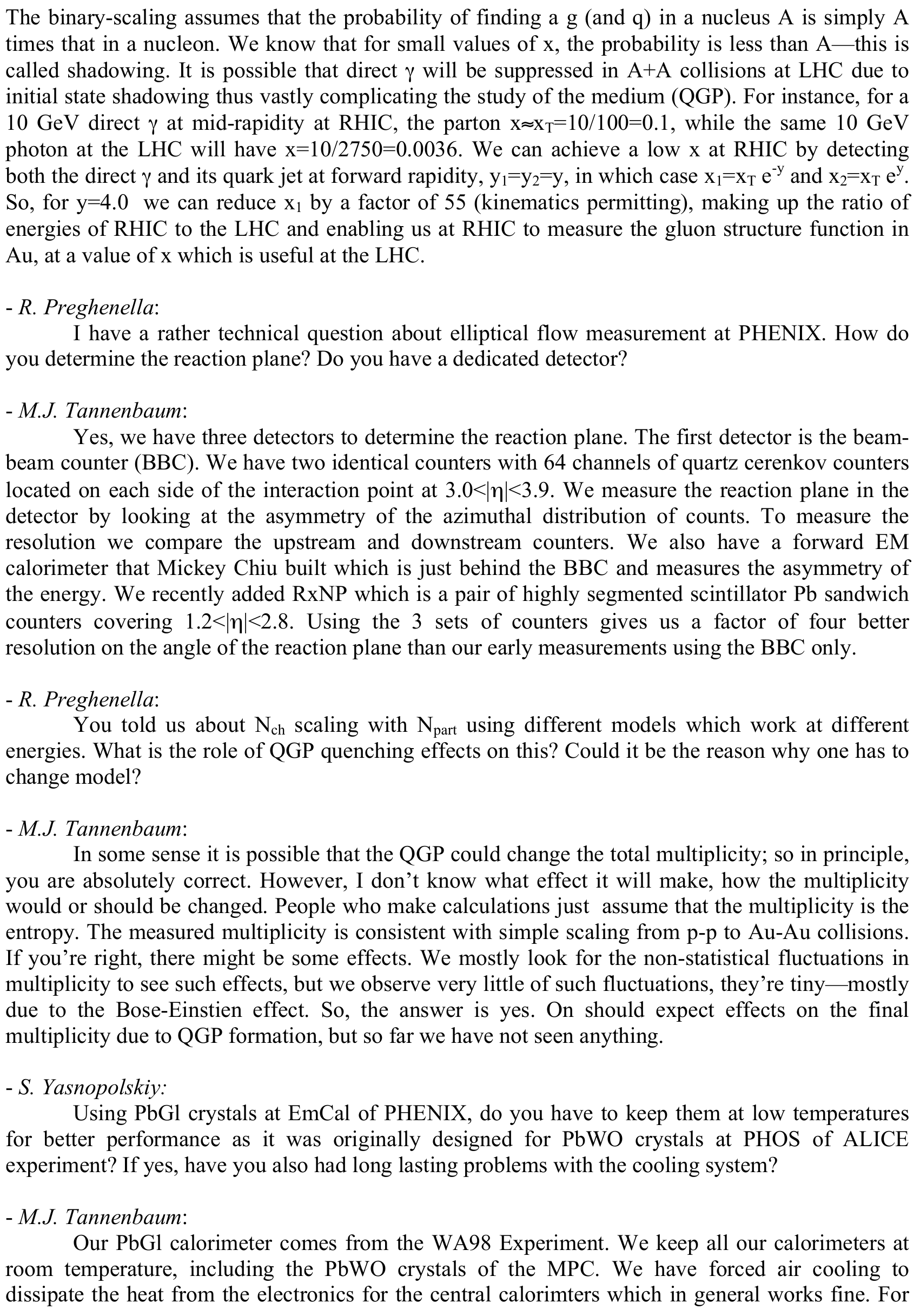}\newpage
\includegraphics[width=0.98\linewidth]{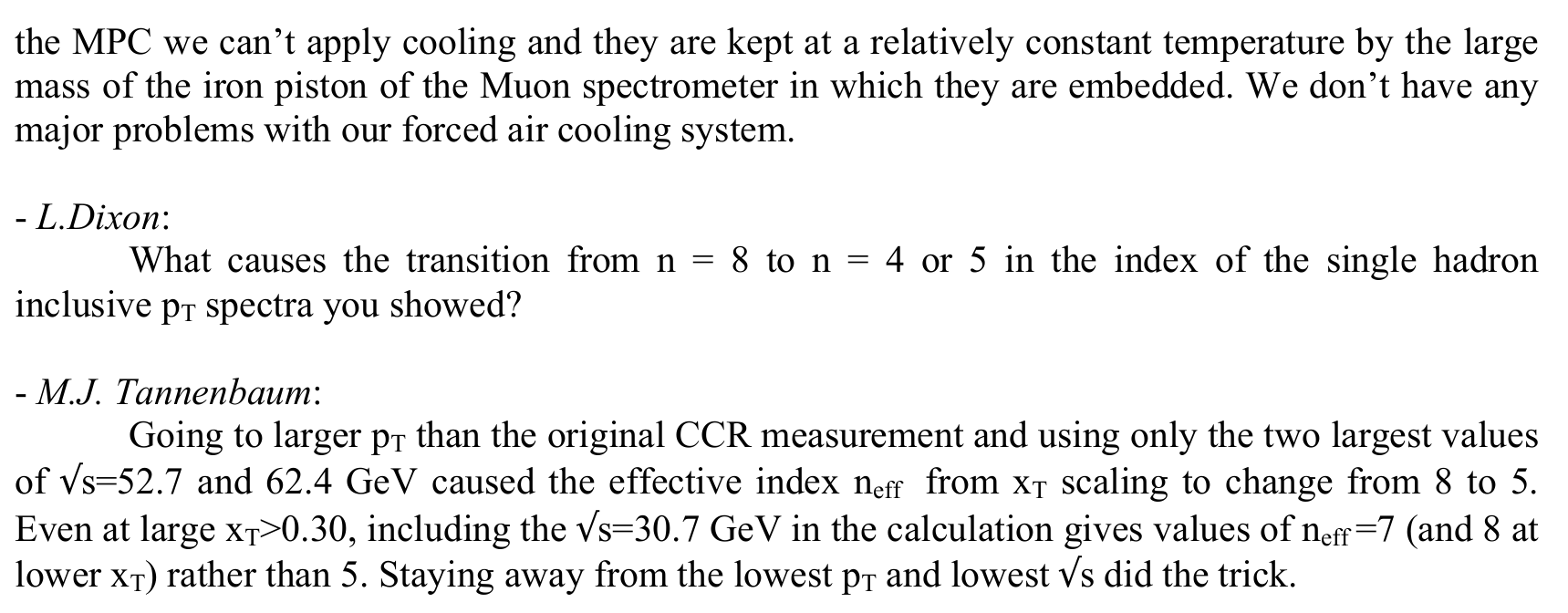}\newpage

%\vspace*{-2.0cm}
\includegraphics[width=0.98\linewidth]{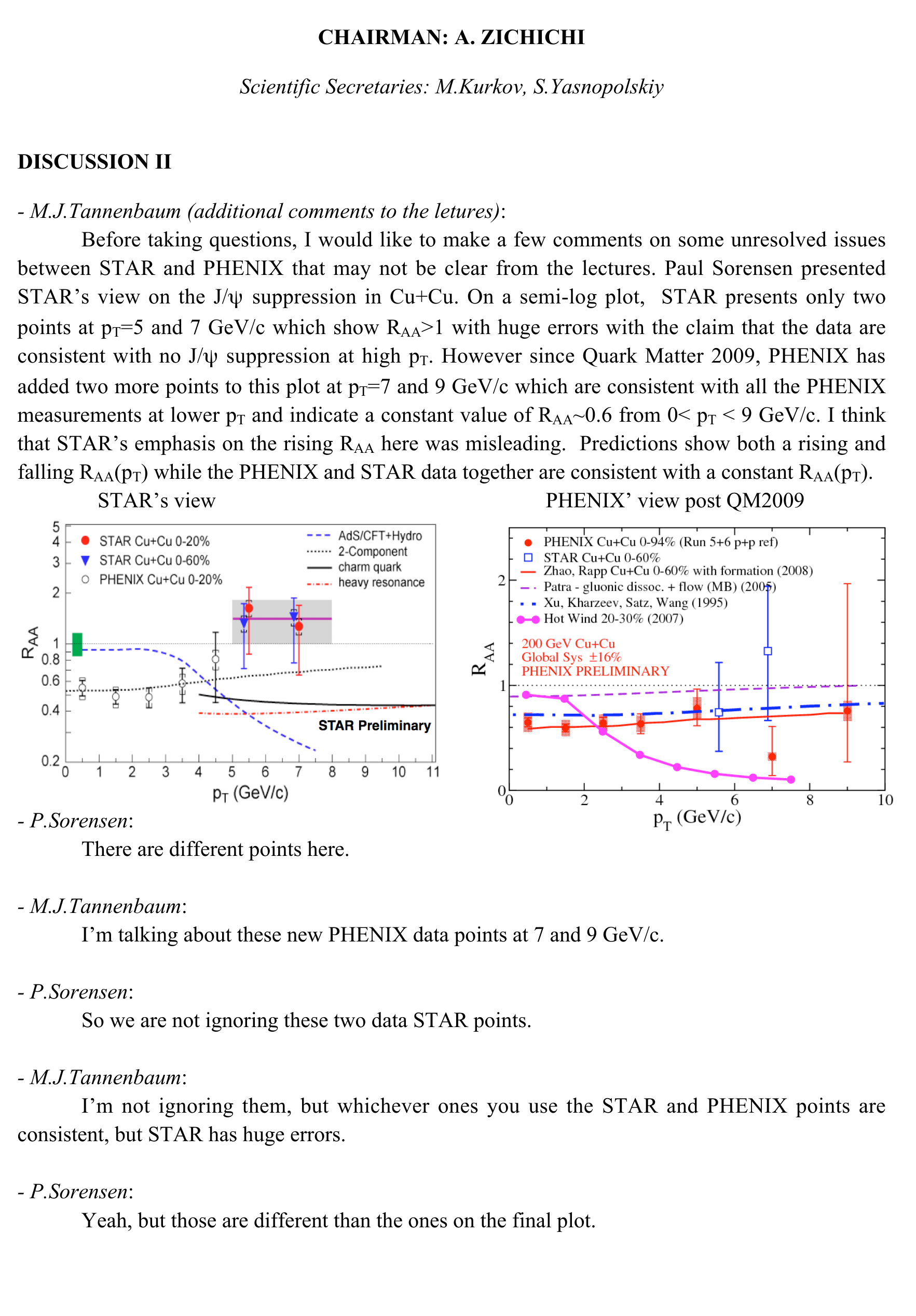}\newpage
\includegraphics[width=0.98\linewidth]{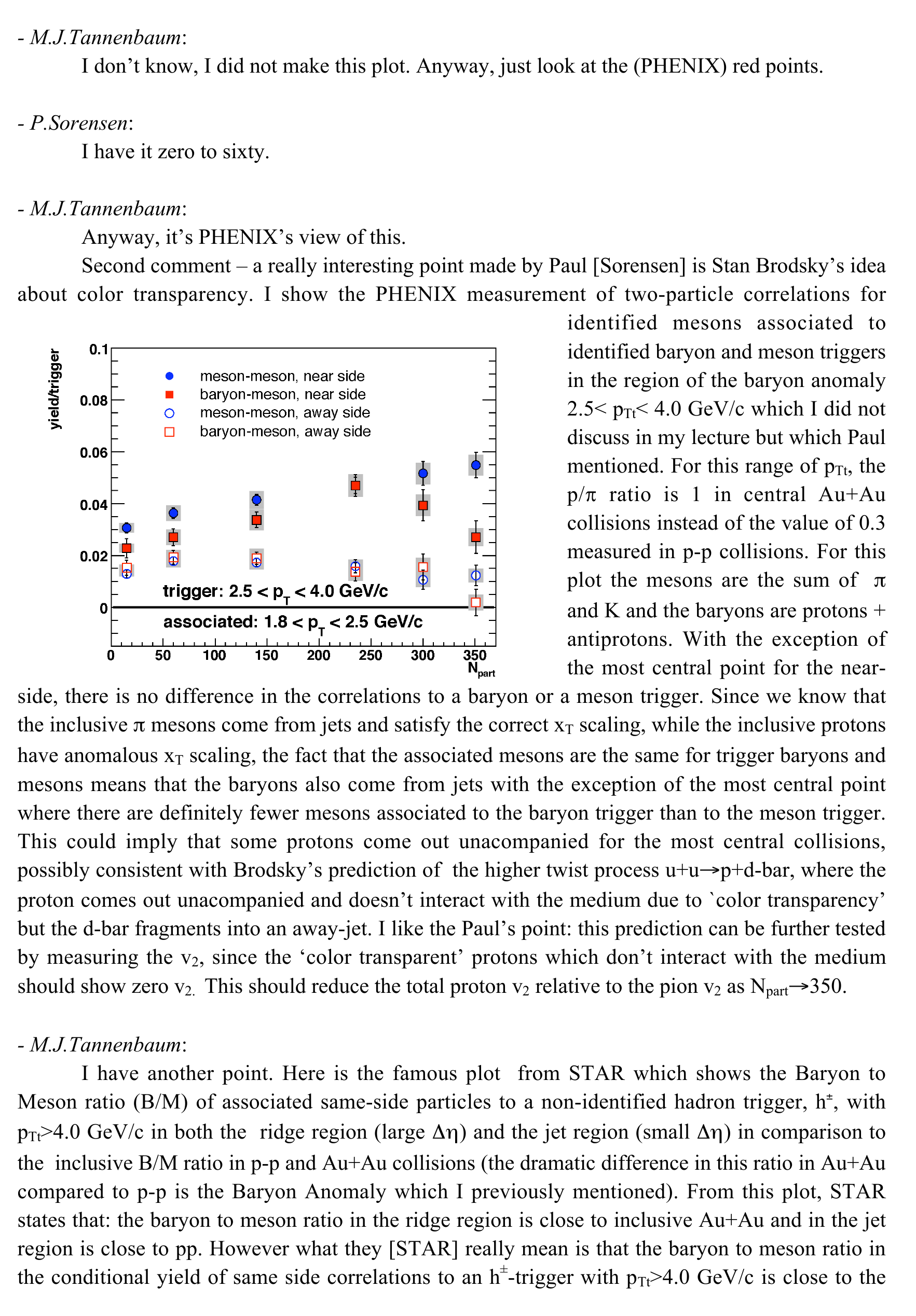}\newpage
\includegraphics[width=0.98\linewidth]{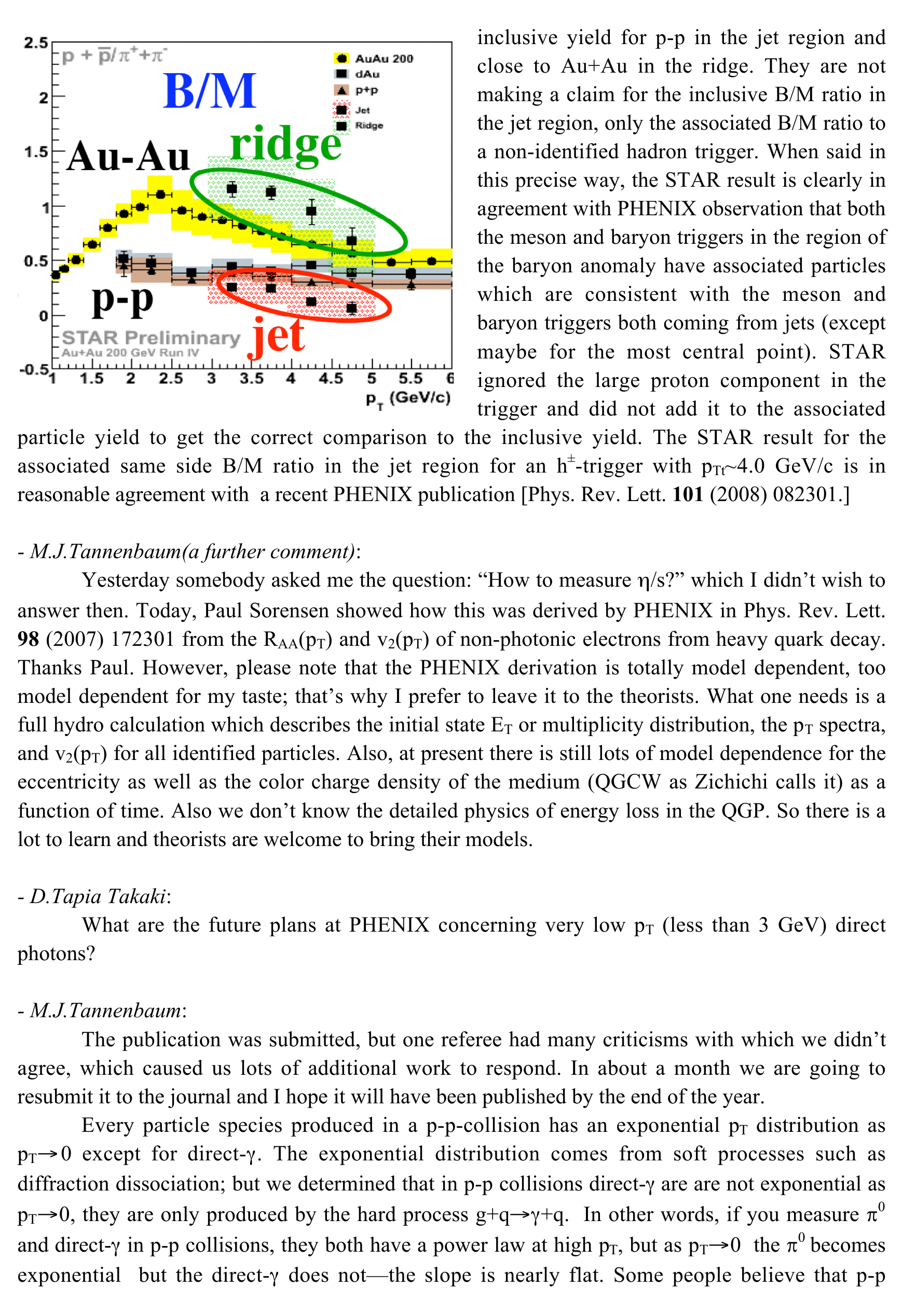}\newpage
\includegraphics[width=0.98\linewidth]{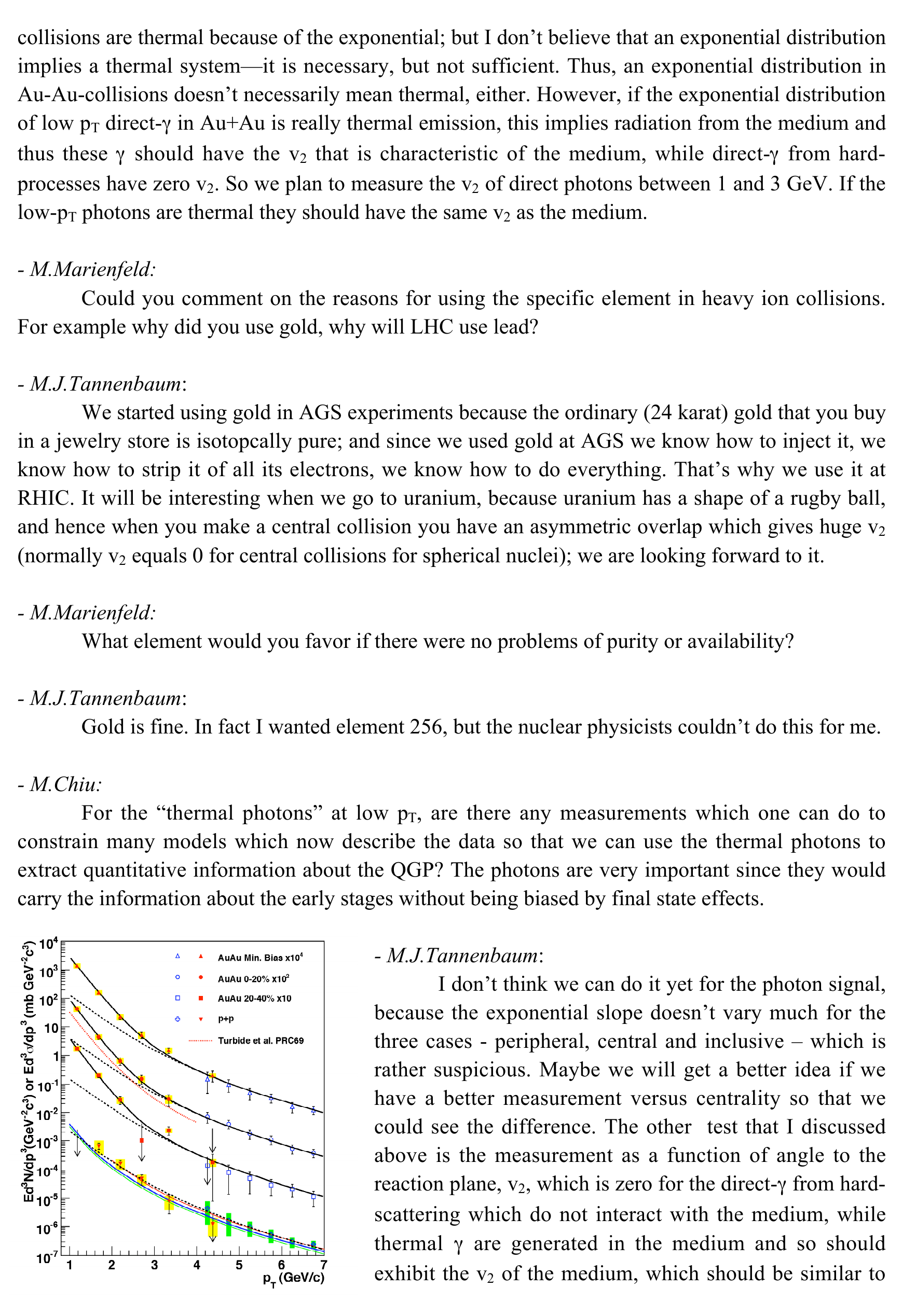}\newpage
\includegraphics[width=0.98\linewidth]{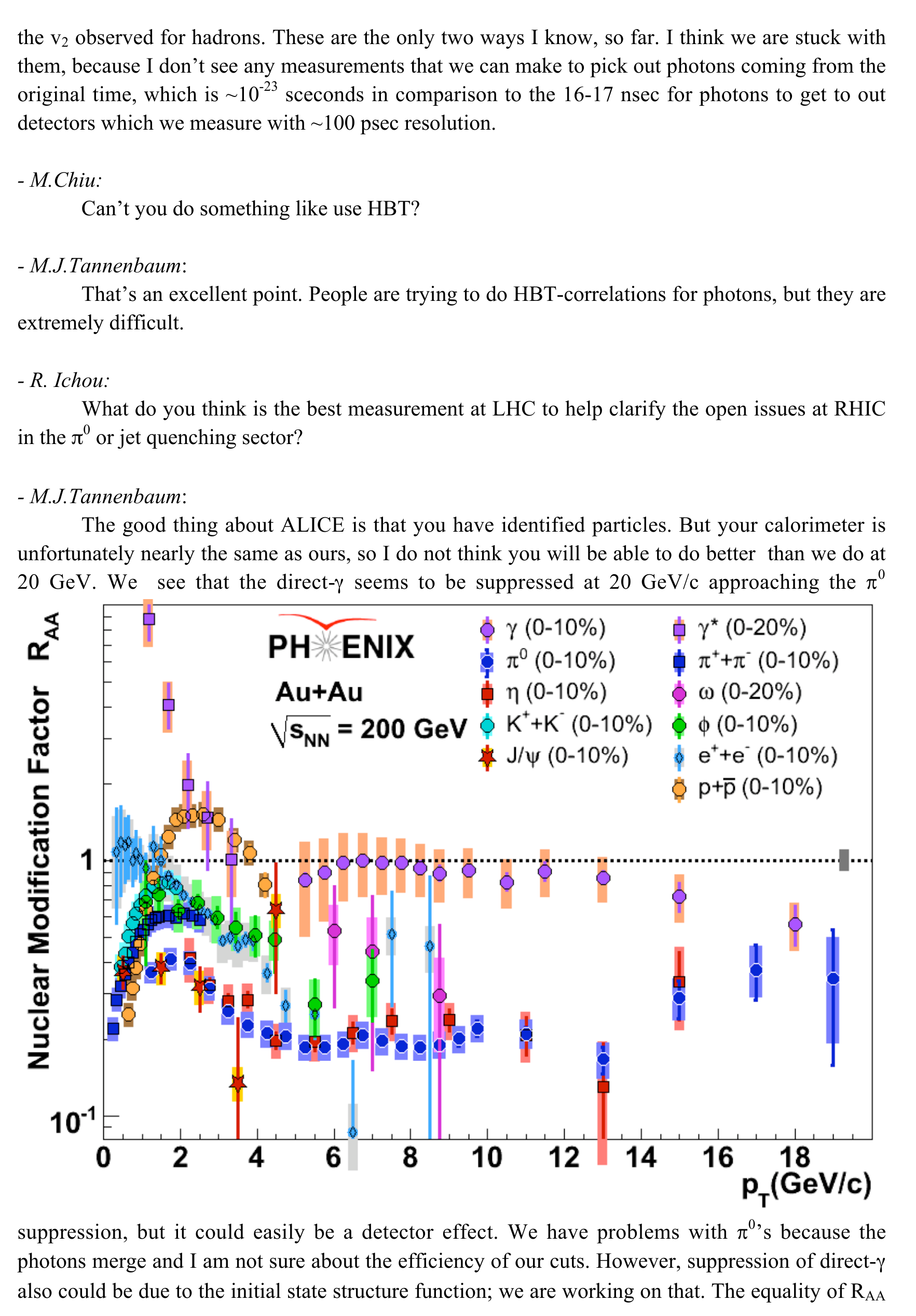}\newpage
\includegraphics[width=0.98\linewidth]{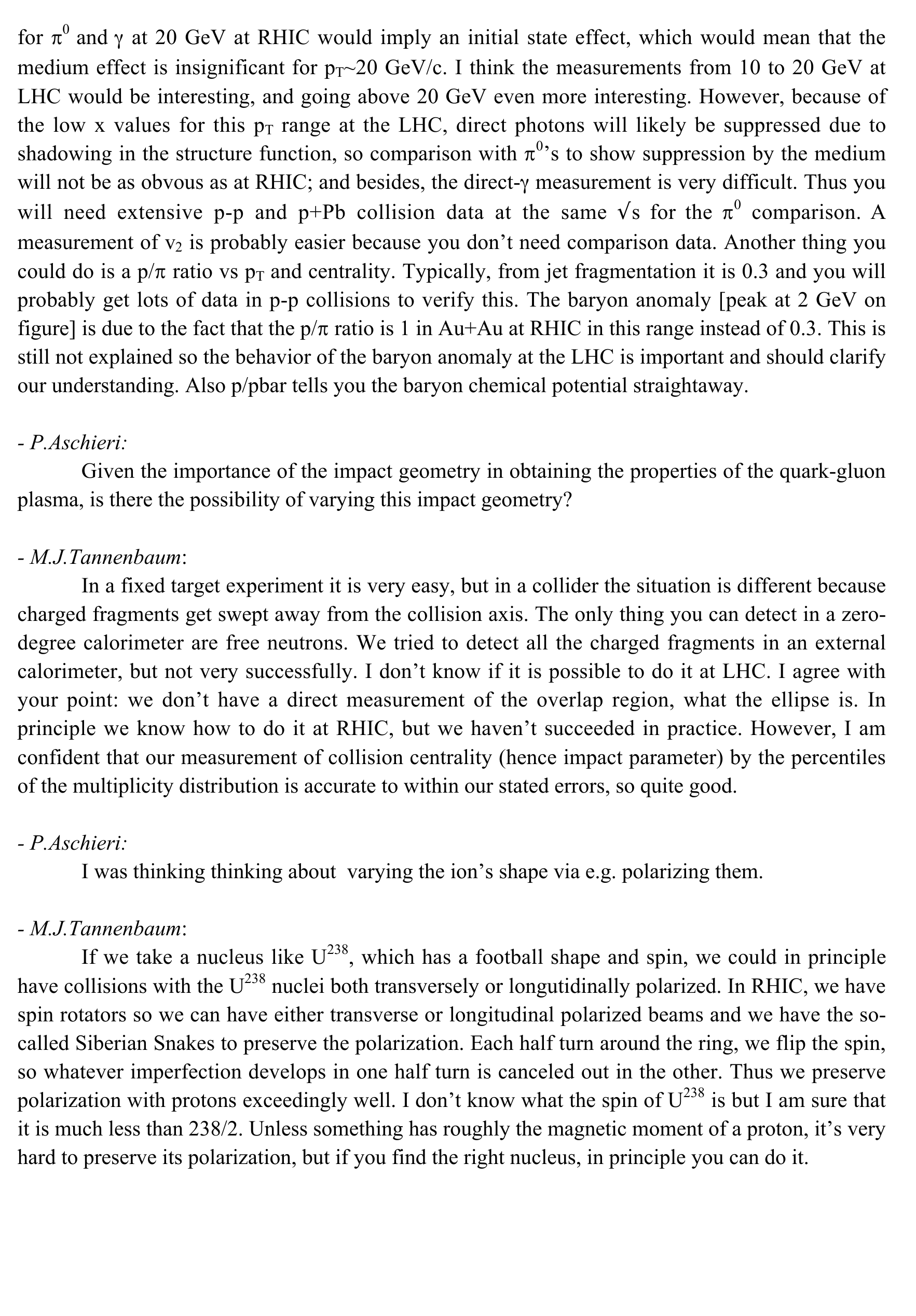}\newpage
\includegraphics[width=0.98\linewidth]{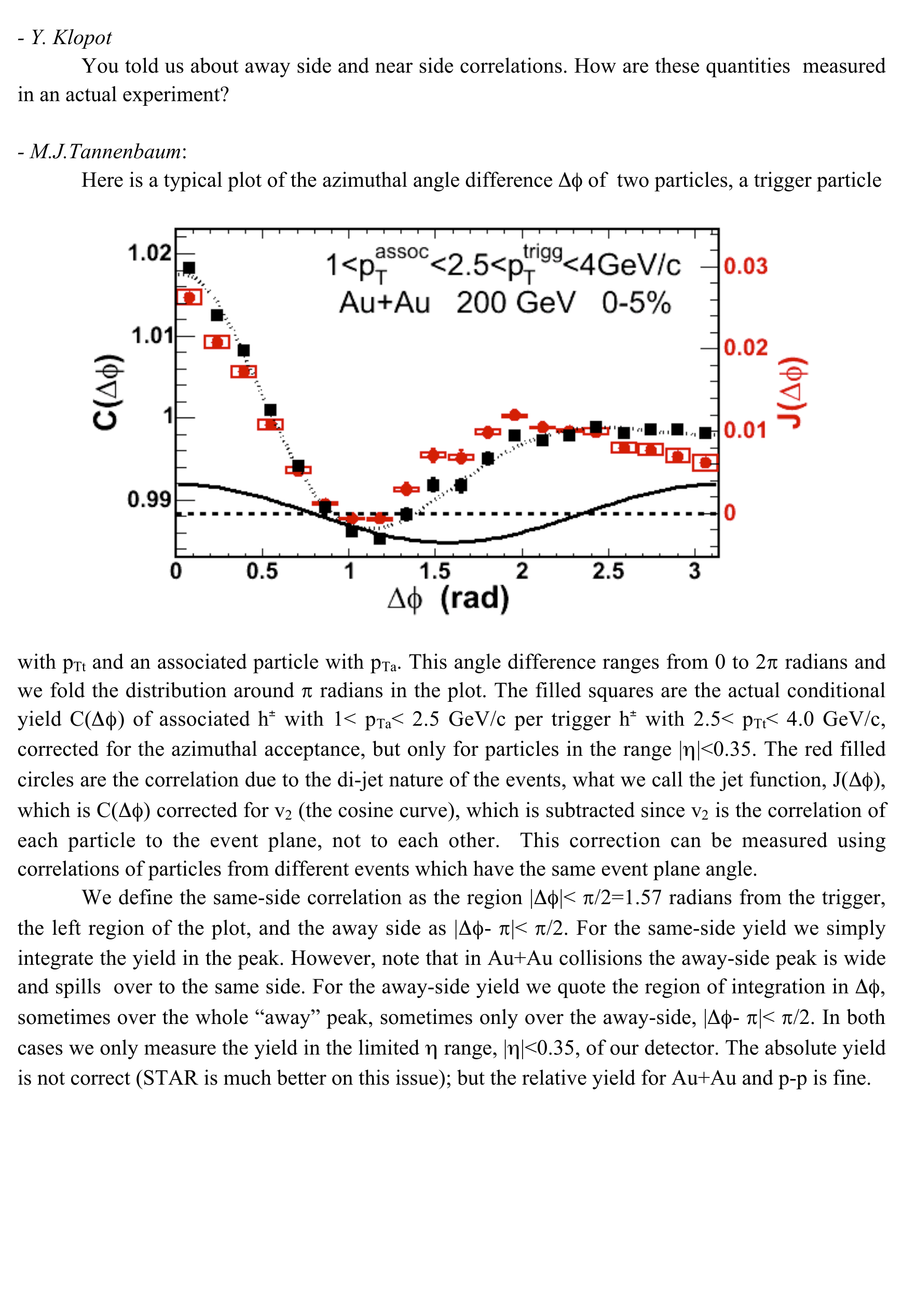}\newpage
\end{center}	  \


\begin{thebibliography}{99}
\bibitem{BearMountain} {\it Report of the Workshop on BeV/Nucleon Coliisions of Heavy Ions---How and Why}, Bear Mountain, NY, 29 November--1 December 1974. (BNL-50445, Upton NY, 1975). 
\bibitem{seeMJTROP} See Ref.~\cite{MJTROP} for a more extensive list of references. 
\bibitem{MJTROP} M.~J.~Tannenbaum, \Journal{\RPP}{69}{2005--2059}{2006}.
\bibitem{Shuryak80} E.~V.~Shuryak, \Journal{\PLC} {61} {71--158}{1980}. 
\bibitem{BRWP} I.~Arsene {\it et al.}, BRAHMS Collab.  \Journal{\NPA}{757}{1--27}{2005}.
\bibitem{PHWP} B.~B.~Back {\it et al.}, PHOBOS Collab. \Journal{\NPA}{757}{28--101}{2005}.
\bibitem{STWP} J.~Adams {\it et al.}, STAR Collab. \Journal{\NPA}{757}{102--183}{2005}.
\bibitem{PXWP} K.~Adcox {\it et al.}, PHENIX Collab. \Journal{\NPA}{757}{184--283} {2005}.
\bibitem{THWPS} D.~Rischke and G.~Levin, eds., \Journal{\NPA}{750}{1--171}{2005}. 
\bibitem{SatzRPP63} H.~Satz \Journal{\RPP}{63}{1511-1574}{2000}.
\bibitem{Krishna99} K.~Rajagopal, \Journal{\NPA}{661}{150c--161c}{1999}.
\bibitem{DAgostino05} M.~D'Agostino {\it et al.}, \Journal{\NPA}{749}{55c--64c}{2005}. 

\bibitem{Policastro} G.~Policastro, D.~T.~Son and A.~O.~Starinets \Journal{\PRL}{87}{081601}{2001}. 
\bibitem{Nastase-BlackHole} H.~Nastase {``The RHIC fireball as a dual black hole''} {\tt hep-th/0501068}.
\bibitem{specificity} It should be noted that the requirement of specificity to A+A collisions immediately rules out the QGP in p-p collisions, which is disputable. See Refs.~\cite{Weiner05,Alexopoulos02}. 
 
\bibitem{Weiner05} R.~M.~Weiner, \Journal{\IJMPE}{15}{37--70}{2006}.
\bibitem{Alexopoulos02} T.~Alexopoulos, {\it et al.}, \Journal{\PLB}{528}{43--48}{2002}.
\bibitem{MatsuiSatz86} 
T.~Matsui and H.~Satz, \Journal{\PLB}{178}{416}{1986}. 

\bibitem{NA50EPJC39} B.~Alessandro, {\it et al.}, NA50 Collab., \Journal{\EPJC}{39}{335--345}{2005}. See also, F.~Prino, Proc. XXX International Symposium on  Multiparticle Dynamics, {\tt hep-ex/0101052}.
\bibitem{E772} D.~M.~Alde, {\it et al.}, E772 Collab., \Journal{\PRL}{66}{2285--2288}{1991}. See, also, M.~J.~Leitch, \Journal{\EPJC}{43}{157--160}{2005} and references therein.  
\bibitem{UA1} C.~Albajar, {\it et al.}, UA1 Collab., \Journal{\PLB}{186}{237}{1987}. 

\bibitem{ppg019} S.~S.~Adler, {\it et al.} PHENIX Collab., \Journal{\PRC}{71}{034908}{2005}.
\bibitem{RHICNIM} {\it The Relativistic Heavy Ion Collider Project: RHIC and its Detectors}, \Journal{\NIMA}{499}{235--880}{2003}.
\bibitem{egseePT} The detector is so non-conventional that it made the cover of Physics Today, October 2003. 
\bibitem{LaceyQM05} R.~A.~Lacey, \Journal{\NPA}{774}{199--214}{2006}.
\bibitem{KanetaQM04} M.~Kaneta {\it et al.}, PHENIX Collab., \Journal{\JPG}{30}{S1217--S1220}{2004}.
\bibitem{PXArkadyQM06} A.~Adare, {\it et al.}, PHENIX Collab., \Journal{\PRL}{98}{162301}{2007}.

\bibitem{Ollitrault} J.-Y.~Ollitrault, \Journal{\PRD}{46}{229--245}{1992}; \Journal{\NPA}{638}{195c--206c}{1998}.
\bibitem{HeiselbergLevy} H.~Heiselberg and A.-M.~Levy, \Journal{\PRC}{59}{2716--2727}{1999}.
\bibitem{VoloshinQM02} S.~A.~Voloshin, \Journal{\NPA}{715}{379c--388c}{2003}.
\bibitem{TeaneyPRC68} D.~Teaney, \Journal{\PRC}{68}{034913}{2003}.
\bibitem{Kovtun05} P.~K.~Kovtun, D.~T.~Son, and A.~O.~Starinets \Journal{\PRL}{94}{111601}{2005}. 
%%%v3here
\bibitem{AlverOllitrault} B.~H.~Alver, C.~Gombeaud, M.~Luzum, and J.~Y.~Ollitrault, \Journal{\PRC}{82}{034913}{2010}.
\bibitem{AlverRoland} B.~Alver and G.~Roland, \Journal{\PRC}{81}{054905}{2010}.
\bibitem{BrazilNuXuv3} J.~Takahashi, {\it et al.}, \Journal{\PRL}{103}{242301}{2009}. Also, see A.~P.~Mishra, {\it et al.}, \Journal{\PRC}{77}{064902}{2008}.  
\bibitem{EsumiQM11} S.~Esumi, {\it et al.} (PHENIX Collab.), \href{http://arxiv.org/abs/1110.3223v1}{arXiv:1110.3223v1 [nucl-ex]}.
\bibitem{alsoMJT95} Also, see, for example, M.~J.~Tannenbaum, Proc. Adriatico Research Conference on Trends in Collider Spin Physics, ICTP, Trieste, Italy, Dec. 5--8, 1995, Eds. Y. Onel, N. Paver, A. Penzo (World Scientific, Singapore, 1997) pp 31-53. 
\bibitem{PXpi0PRD} S.~S.~Adler, {\it et al.} PHENIX Collab., \Journal{\PRD}{76}{051006(R)}{2007}.
\bibitem{Aurenche} P.~Aurenche, {\it et al.}, \Journal{\PRD}{73}{094007}{2006}.
\bibitem{Cocconi1961} G. Cocconi, L.~J. Koester, and D.~H. Perkins, {Technical Report No.~UCRL-10022 (1961)}, 
Lawrence Radiation Laboratory,  (unpublished), p. 167.
\bibitem{BBK} S.~M.~Berman, J.~D.~Bjorken and J.~B.~Kogut, 
\Journal{\PRD}{4}{3388}{1971}.
\bibitem{CCR} F.~W.~B\"usser, et al. (CCR), \Journal{\PLB}{46}{471}{1973}.
\bibitem{CCOR} A.~L.~Angelis, et al. (CCOR), \Journal{\PLB}{79}{505}{1978}.
\bibitem{BBKBBGCGKS} See M.~J.~Tannenbaum, \Journal{\NPA}{749}{219c}{2005} for references. 
\bibitem{Owens} J.~F.~Owens, \Journal{\RMP}{59}{465}{1987}.
\bibitem{CutlerSivers} R.~Cutler and D.~Sivers, \Journal{\PRD}{17}{196}{1978}; \Journal{\PRD}{16}{679}{1977}.
\bibitem{Combridge:1977dm} B.~L.~Combridge, J.~Kripfganz and J.~Ranft, \Journal{\PLB}{70}{234}{1077}.
\bibitem{Angelis79} A.~L.~S.~Angelis, {\it et al.}, Physica Scripta {\bf 19}\ (1979) 116.
\bibitem{JacobEPS79} Data from Ref.~\cite{Angelis79} shown by M. Jacob, Proc. EPS Int. Conf. High Energy Physics, Geneva, 27 June--4 July, 2009, Ed. A.~Zichichi (CERN, Geneva, 1980), p 512.
\bibitem{ppg029} S.~S.~Adler, {\it et al.} PHENIX Collab., \Journal{\PRD}{74}{072002}{2006}. 
\bibitem{Paris82} { Proc. 21st Int'l Conf. HEP}, Paris, 1982, eds 
P.~Petiau, M.~Porneuf, J. Phys. C{\bf 3}\ (1982): see J.~P.~Repellin, p.  
C3-571; also see M.~J.~Tannenbaum, p. C3-134, G.~Wolf, p. C3-525.  
\bibitem{CCOR82NPB} A.~L.~S.~Angelis, {\it et al.}, 
\Journal{\NPB}{209}{284}{1982}.
\bibitem{QCDCompton} H.~Fritzsch and P.~Minkowski, \Journal{\PLB}{69}{316}{1977}.
\bibitem{CMOR}A.~L.~S.~Angelis, et al (CMOR), \Journal{\NPB}{327}{541-568}{1989}.

\bibitem{CCRS}F.~W.~B\"usser, et al. (CCRS), \Journal{\NPB}{113}{189-245}{1976}.
\bibitem{HLLS}I.~Hinchliffe and C.~H.~Llewellyn Smith, \Journal{\NPB}{114}{45-60}{1976}. 

\bibitem{CCRSPLB56}F.~W.~B\"usser, et al. (CCRS), \Journal{\PLB}{56}{482-486}{1976}.
\bibitem{Clark}A.~G.~Clark, et al. (CSZ), \Journal{\NPB}{142}{29-52}{1978}.

\bibitem{PXpi062pp} A.~Adare, {\it et al.} PHENIX Collab., \Journal{\PRD}{79}{012003}{2009}.
\bibitem{YAQM05} Y.~Akiba, {\it et al.} PHENIX Collab., \Journal{\NPA}{774}{403--408}{2006}.
\bibitem{GLV} M.~Gyulassy, P.~Levai and I.~Vitev, \Journal{\PRL}{85}{5535-5538}{2000}; I.~Vitev and M.~Gyulassy, \Journal{\PRL}{89}{252301}{2002}.
\bibitem{ppg084} A.~Adare, {\it et al.} PHENIX Collab., \Journal{\PRL}{101}{162301}{2008}.

\bibitem{Vitev2} I.~Vitev, \Journal{\PLB}{639}{38--45}{2006}, and private communication to Ref.~\cite{ppg084}. 
\bibitem{CroninEffect} D.~Antreasyan, {\it et al.}, \Journal{\PRD}{19}{764-778}{1979}.
\bibitem{BDMPS} See R.~Baier, D.~Schiff and B.~G.~Zakharov, \Journal{\ARNPS}{50}{37--69}{2000}, and references therein. 
\bibitem{egNPS} e.g. see N.~P.~Samios, \Journal{\PR}{121}{275--281}{1961}.
\bibitem{KW}N.~M.~Kroll and W.~Wada, \Journal{\PR}{98}{1355--1359}{1955}.
\bibitem{ppg086} A.~Adare, {\it et al.} PHENIX Collab., \Journal{\PRL}{104}{132301}{2010}.
\bibitem{Ito81} A.~S.~Ito, {\it et al.}, \Journal{\PRD}{23}{604--633}{1981}.
\bibitem{ThanksAM} Thanks to Sasha Milov for the plot of $R_{AA}(p_T)$ for all PHENIX published and preliminary measurements. With the exception of the internal-conversion direct-$\gamma$ where the fit to the p-p data is used to compute $R_{AA}$, all the other values of $R_{AA}$ are computed from Eq.~\ref{eq:RAA} using the measured Au+Au and p-p data points.

\bibitem{PXJPsiAuAu200} A.~Adare, {\it et al.}, PHENIX Collab., \Journal{\PRL}{98}{232301}{2007}.
\bibitem{GunjiQM06} T.~Gunji, {\it et al.}, PHENIX Collab., \Journal{\JPG}{34}{S749--S752}{2007},   {\tt http://www.sinap.ac.cn/qm2006/parallel.htm\#p21} .

\bibitem{RappPLB664} X.~Zhao and R.~Rapp, \Journal{\PLB}{664}{253--257}{2008}
 
\bibitem{PBMStachelPLB490} P.~Braun-Munzinger and J.~Stachel, \Journal{\PLB}{490}{196--202}{2000}.
\bibitem{ThewsPRC63} R.~L.~Thews, M.~Schroedter and J.~Rafelski, \Journal{\PRC}{63}{054905}{2001}.
\bibitem{AndronicNPA79} A.~Andronic, P.~Braun-Munzinger, K.~Redlich and J.~Stachel, \Journal{\NPA}{789}{334--356}{2007}.
\bibitem{ArnaldiQM09} R.~Arnaldi, {\it et al.}, NA60 Collab., \Journal{\NPA}{830}{345c--352c}{2009}.
\bibitem{ArnaldiECT09} R.~Arnaldi, presented at Heavy Quarkonia Production in Heavy Ion Collisions, Trento, Italy, 25-29 May 2009, {\tt http://www.ect.it/Meetings/ConfsWksAndCollMeetings/ ConfWksDocument/2009/Talks/WORKSHOP\_25May09/arnaldi.ppt} 
\bibitem{ShuryakMach} J.~Casalderrey-Solana, E.~V.~Shuryak and D.~Teaney, \Journal{\JPCS}{27}{22-31}{2005}.
\bibitem{mjtdeco} M.~J.~Tannenbaum, PoS(CFRNC2006)001 (2006)
\bibitem{PXcharmPRL94} S.~S.~Adler {\it et al.}, PHENIX Collab., \Journal{\PRL}{94}{082301}{2005}.
\bibitem{PXcharmAA06} A.~Adare, {\it et al.}, PHENIX Collab., \Journal{\PRL}{98}{172301}{2007}.
\bibitem{egsee066} e.g. see Ref.~\cite{PXcharmAA06} for a list of references. 


\end{thebibliography}
\end{document}